\documentclass[a4paper,fleqn,usenatbib,usedcolumn,useAMS]{mnras}

\hypersetup{pdfauthor={C. P. Coughlan, D.C. Gabuzda},
               pdftitle={High resolution VLBI polarisation imaging of AGN with the Maximum Entropy Method},
               pdfkeywords={astronomical instrumentation, methods, and techniques, techniques: image processing -- galaxies: active -- galaxies: magnetic fields -- galaxies, techniques: high angular resolution},
               bookmarksnumbered=true}
\volume{{\rm in press}}

\usepackage{mathptmx}
\usepackage{amsmath}
\usepackage{txfonts}

\usepackage[T1]{fontenc}
\usepackage{ae,aecompl}

\usepackage{subfig,graphicx,overpic}

\def\reff@jnl#1{{\rm#1\/}}

\def\aj{\reff@jnl{AJ}}                  
\def\araa{\reff@jnl{ARA\&A}}            
\def\apj{\reff@jnl{ApJ}}                        
\def\apjl{\reff@jnl{ApJ}}               
\def\apjs{\reff@jnl{ApJS}}              
\def\ao{\reff@jnl{Appl.Optics}}         
\def\apss{\reff@jnl{Ap\&SS}}            
\def\aap{\reff@jnl{A\&A}}               
\def\aapr{\reff@jnl{A\&A~Rev.}}         
\def\aaps{\reff@jnl{A\&AS}}             
\def\azh{\reff@jnl{AZh}}                        
\def\baas{\reff@jnl{BAAS}}              
\def\jrasc{\reff@jnl{JRASC}}            
\def\memras{\reff@jnl{MmRAS}}           
\def\mnras{\reff@jnl{MNRAS}}            
\def\pra{\reff@jnl{Phys. Rev. A}}         
\def\prb{\reff@jnl{Phys. Rev. B}}         
\def\prc{\reff@jnl{Phys. Rev. C}}         
\def\prd{\reff@jnl{Phys. Rev. D}}         
\def\prl{\reff@jnl{Phys. Rev. Lett}}      
\def\pasp{\reff@jnl{PASP}}              
\def\pasj{\reff@jnl{PASJ}}              
\def\qjras{\reff@jnl{QJRAS}}            
\def\rmxaa{\reff@jnl{RMxAA}}		
\def\skytel{\reff@jnl{S\&T}}            
\def\solphys{\reff@jnl{Solar~Phys.}}    
\def\sovast{\reff@jnl{Soviet~Ast.}}     
\def\ssr{\reff@jnl{Space~Sci.Rev.}}     
\def\zap{\reff@jnl{ZAp}}                        
\def\nat{\reff@jnl{Nature}}             

\def\p#1by#2{{\partial{#1} \over \partial{#2}}}
\def\pp#1by#2#3{{\partial^2{#1} \over \partial{#2}\partial{#3}}}
\def\d#1by#2{{{\rm d}{#1} \over {\rm d}{#2}}}
\def\dd#1by#2#3{{{\rm d}^2{#1} \over {\rm d}{#2}{\rm d}{#3}}}

\title[VLBI polarisation imaging of AGN with MEM]{High resolution VLBI polarisation imaging of AGN with the Maximum Entropy Method}
\author[Colm P. Coughlan, Denise C. Gabuzda]{Colm P. Coughlan$^{1}$\thanks{Email: coughlan@cp.dias.ie}, Denise C. Gabuzda$^{2}$\\
$^{1}$School of Cosmic Physics, Dublin Institute for Advanced Studies, Ireland\\
$^{2}$Department of Physics, University College Cork, Ireland}

\date{Accepted XXX. Received YYY; in original form ZZZ}

\pubyear{2016}

\begin{document}
\label{firstpage}
\pagerange{\pageref{firstpage}--\pageref{lastpage}}
\maketitle

\begin{abstract}
Radio polarisation images of the jets of Active Galactic Nuclei (AGN) can provide a deep insight into the launching and collimation mechanisms of relativistic jets. However, even at VLBI scales, resolution is often a limiting factor in the conclusions that can be drawn from observations. The Maximum Entropy Method (MEM) is a deconvolution algorithm that can outperform the more common CLEAN algorithm in many cases, particularly when investigating structures present on scales comparable to or smaller than the nominal beam size with ``super-resolution''. A new implementation of the MEM suitable for single- or multiple-wavelength VLBI polarisation observations has been developed and is described here. Monte Carlo simulations comparing the performances of CLEAN and MEM at reconstructing the properties of model images are presented; these demonstrate the enhanced reliability of MEM over CLEAN when images of the fractional polarisation and polarisation angle are constructed using convolving beams that are appreciably smaller than the full CLEAN beam. The results of using this new MEM software to image VLBA observations of the AGN 0716+714 at six different wavelengths are presented, and compared to corresponding maps obtained with CLEAN. MEM and CLEAN maps of Stokes $I$, the polarised flux, the fractional polarisation and the polarisation angle are compared for convolving beams ranging from the full CLEAN beam down to a beam one-third of this size. MEM's ability to provide more trustworthy polarisation imaging than a standard CLEAN-based deconvolution when convolving beams appreciably smaller than the full CLEAN beam are used is discussed. 
\end{abstract}


\section{Introduction}

The radio imaging of Active Galactic Nuclei (AGN) with Very Long Baseline Interferometry (VLBI) is a complex process, involving careful calibration and computational processing of the visibility data. A critical part of this process is the use of a deconvolution algorithm to make an image of the data that is as free as possible from the effects of unsampled visibilities, thus undoing the convolution of the original intensity and polarisation distributions with the dirty beam corresponding to the point response function of the observing array. No deconvolution algorithm can operate with perfect accuracy, as the finite visibility sampling function of any interferometric array means that the data required to do so simply have not been observed. However, the choice of a suitable deconvolution algorithm can effectively model the missing visibilities using the observed data and result in a reliable deconvolved map which accurately reproduces the real intensity distribution.

Imaging the distributions of the intensity and polarisation of AGN with VLBI requires a deconvolution algorithm which works with all of the Stokes parameters commonly used to describe the intensity and polarisation of the radio emission.  The most popular deconvolution algorithm used for such imaging is CLEAN \citep{Hogbom1974}. Originally developed by H\"{o}gbom in 1974, the CLEAN algorithm can be used to deconvolve both intensity (Stokes $I$) and polarisation (Stokes $Q$, $U$, $V$) data, treating the polarisation Stokes parameters independently.

The CLEAN algorithm has evolved over the years since it was first introduced and today many different variants of it are in use, including the standard Clark CLEAN and multiscale CLEAN \citep{Clark1980,Wakker1988}.  The Clark CLEAN algorithm is the most widely used deconvolution algorithm in VLBI polarisation studies of AGN. However, while CLEAN is popular, fast and effective, limitations related to its underlying modeling of the source as a series of $\delta$ functions (or in the case of multiscale CLEAN, Gaussian components) can curb the accuracy of the deconvolved maps when imaging multi-component structures with characteristic sizes appreciably smaller than the full CLEAN beam (a Gaussian fitted to the central lobe of the dirty beam). Alternative deconvolution algorithms with more realistic intrinsic models of the flux distribution and polarisation properties of the source can give better performance when probing such compact structures.

The development of a new generation of radio imaging algorithms is an active field of research, with many different approaches being considered to improve the accuracy, sensitivity and resolution of deconvolved radio images. Much of this development has been focussed on the re-formulation of radio imaging as a compressed sensing problem, an approach which has yielded a new understanding of how an accurate image can be recovered via an extremely sparse sampling visibility sampling function (see, for example, \citet{Candes2006}). \citet{Carrillo2014}, \citet{Dabbech2015} and \citet{Garsden2015} have developed algorithms utilising this new formulation in the context of the sparse sampling common in radio interferometry and have achieved notable improvements over standard CLEAN imaging.

Further improvements in the field have also come from the formulation of radio imaging as a problem of Bayesian inference, where a model of the observing process can be used to derive original sky model as the posterior distribution of the observed radio data. \citet{Sutter2014} and \citet{Junklewitz2016} have developed algorithms which, although not yet suitable for polarisation imaging, show substantial improvements over CLEAN. Additional work centred around the Maximum Entropy Method (MEM) seeks to use its high resolution properties to achieve better results than CLEAN. \citet{Chael2016} have recently extended a form of polarimetric MEM suitable at higher frequency mm wavelengths for the Event Horizon Telescope.

The Maximum Entropy Method is an established deconvolution algorithm which was first used to deconvolve radio images in the 1970s\citep{Wernecke1977,Frieden1978}. The connection between MEM and the principle of maximum entropy as developed by \citet{Jaynes1957} is a somewhat contentious one, with some parties maintaining a strong link between the two, while others suggest that MEM is a more heuristic algorithm with only a tenuous connection to Jaynes' maximum entropy principle. \citet{Gull1984} offers a discussion as to why MEM might be considered a ``true'' application of the maximum entropy principle, while \citet{Cornwell1984} and \citet{OSullivan1984} disagree and suggest alternative formulations of the maximum entropy principle for application to image reconstruction. \citet{Nityananda1982}, \citet{Narayan1984} and \citet{numericalc} suggest a practical outlook, whereby MEM can be considered a useful regularisation parameter without any deeper interpretation. This approach is also followed in the simulations and observational results presented in this work.

\citet{Cornwell1985} outlined an efficient implementation of MEM as a constrained optimisation method based on a consideration of the function
\begin{equation}
\begin{aligned}
\label{mem-jeqn}
J =\ & H(I_{\mathrm{m}},P_{\mathrm{m}}) - \alpha \chi_{\mathrm{I}}^{2}(\widehat{V_{\mathrm{I}}}, \widetilde{V_{\mathrm{I}}})\\
&-\beta (\chi_{\mathrm{Q}}^{2} (\widehat{V_{\mathrm{Q}}}, \widetilde{V_{\mathrm{Q}}})+\chi_{\mathrm{U}}^{2} (\widehat{V_{\mathrm{U}}}, \widetilde{V_{\mathrm{U}}})) - \gamma G
\end{aligned}
\end{equation}
\noindent
where $H$ is the entropy of a continuous model map of the source, $I_\mathrm{m}$ and $P_\mathrm{m}$ are the intensity and polarisation properties of the model map, $\chi^{2}$ is a measure of the difference between the model ($\widehat{V}$) and the observed ($\widetilde{V}$) visibilities (there are three $\chi^{2}$ terms, one for Stokes $I$ and two for Stokes $Q$ and $U$), $\alpha$, $\beta$ and $\gamma$ are Lagrangian optimisation parameters and $G$ is a function equal to the difference between the total Stokes $I$ model flux and the zero-spacing flux. The zero-spacing flux is an estimate of the integrated flux, which can be obtained by extrapolating the amplitudes of the visibilities with the shortest baselines. A form of entropy suitable for polarisation emission developed by \citet{Gull1984} and used by \citet{Holdaway1990} and \citet{Sault1999} is
\begin{equation}
\begin{aligned}
\label{mem-heqn}
H=\ & -\sum_{\mathrm{k}} I_{\mathrm{k}}(\log(\frac{2 I_{\mathrm{k}}}{I_{B_{\mathrm{k}}}e}) +\frac{1+m_{\mathrm{k}}}{2}\log(\frac{1+m_{\mathrm{k}}}{2})\\
&+\frac{1-m_{\mathrm{k}}}{2}\log(\frac{1-m_{\mathrm{k}}}{2}))
\end{aligned}
\end{equation}
\noindent
where $I_{B_{\mathrm{k}}}$ is the flux of a bias map (normally chosen to be a flat map with a total flux equal to the flux estimated for the source)  at pixel $k$ and $I_{\mathrm{k}}$ and $m_{\mathrm{k}}$ are the Stokes $I$ flux and the fractional polarisation, respectively, of pixel $\mathrm{k}$. This exact form of entropy was suggested in \citet{Sault1999}, though a very similar form was used by \citet{Holdaway1990Thesis}.

Following the initial formulation of a form of entropy suitable for polarisation by \citet{Ponsonby1973} and \citet{Nityananda1983},  \citet{Gull1984} derived Equation \ref{mem-heqn} guided by the formal mathematical analysis of \citet{Shore1980}, which used the axioms of transformation invariance and system and subset independence to suggest that the MEM entropy functional should be of the form $H = - \Sigma p_{\mathrm{i}} log(p_{\mathrm{i}} /B_{\mathrm{i}})$, where $p_{\mathrm{i}}$ is the probability associated with configuration $i$ and $B_{i}$ is a bias or initial estimate. \citet{Gull1984} then extended this form to polarised emission by representing the probability distribution of the polarisation in a single pixel as a diagonalised probability density matrix $p$, the eigenvalues of which were found to be $\frac{1}{2}(1 \pm m$). This gives rise to a polarisation contribution to the entropy of the form
\begin{equation}
Tr(-p ~\log p) =  -\frac{1+m_{\mathrm{k}}}{2}\log(\frac{1+m_{\mathrm{k}}}{2})-\frac{1-m_{\mathrm{k}}}{2}\log(\frac{1-m_{\mathrm{k}}}{2}),
\end{equation}

which was then generalised to Equation \ref{mem-heqn} for multiple pixels.

An examination of the form of $H$ gives an indication as to how it will react to different types of source structures.
The Gull and Skilling entropy of a source that is described well by the bias map is high -- the data do not require a meaningful model at all, and the amount of useful information which can be gained from the model is minimal. A source which has low fractional polarisation (i.e. disordered magnetic field) will also have high Gull and Skilling entropy.

The Gull and Skilling entropy is thus maximised for an unpolarised source that is identical to the bias map. This is the map that MEM will produce in the absence of any data that force it to make a more complicated model. If data are provided to the MEM model, the $\chi^{2}$ terms in Equation \eqref{mem-jeqn} force MEM to make a model that maximises the Gull and Skilling entropy while reproducing the data and observed flux to within the noise levels. In this way, the MEM can be thought of as a `tug of war' between the Gull and Skilling entropy, favouring disorder, and the $\chi^{2}$ terms and flux condition in Equation \eqref{mem-jeqn}, favouring fidelity to the observed data.

As noted by \citet{Holdaway1990Thesis}, the convergence of the MEM can be studied by examining the derivative of the $\chi^{2}$ term in Equation \ref{mem-jeqn} as follows:
\begin{equation}
\chi^{2} = \sum_{\mathrm{i}=0}^{k}\omega_{\mathrm{k}}(V_{\mathrm{m,k}} - V_{\mathrm{obs,k}})^{2}
\end{equation}
\begin{equation}
\frac{\partial \chi^{2}}{\partial I_{\mathrm{i}}} = 2 \sum_{\mathrm{i}=0}^{\mathrm{k}} \omega_{\mathrm{k}} \mathbb{R}(V_{\mathrm{m,k}} - V_{\mathrm{obs,k}}) \cos( 2 \pi (u_{\mathrm{k}}x_{\mathrm{i}} + v_{\mathrm{k}}y_{\mathrm{i}})),
\label{eqn-mem-res}
\end{equation}
\noindent
where $V_{\mathrm{m}}$ are the model visibilities, $V_{\mathrm{obs}}$ the observed visibilities, $\omega$ the weights of the observed visibilities, $u$ and $v$ the visibility coordinates and $x$ and $y$ the coordinates of the resulting image. For a particular visibility, $\chi^{2}$ reaches an extremum where either the model and observed visibilities agree completely, or
\begin{equation}
2 \pi (u_{\mathrm{k}}x_{\mathrm{i}} + v_{\mathrm{k}}y_{\mathrm{i}}) = \frac{n \pi}{2}.
\label{eqn-mem-res2}
\end{equation}
\noindent

Equation \ref{eqn-mem-res2} shows that the effective resolution of a MEM map varies across the map. Further examining the resolution for $y=0$ gives the smallest non-zero value of $x$ probed as
\begin{equation}
x_{min} = \frac{1}{4 u_{\mathrm{max}}},
\end{equation}
\noindent
with an analogous result in the $y$ direction. Thus, while the resolution of the MEM model map varies across the map, a MEM model converges to a resolution higher than the Nyquist sampling limited resolution (corresponding roughly to the CLEAN beam). While this is a result of the norm chosen to describe the distance between the current MEM model and the data, this feature works together with the continuous nature of the MEM model to give rise to a set of model maps that can be a better and more consistent overall image of the source. In an implementation of MEM with polarisation as described above, MEM's simultaneous awareness of all Stokes parameters can also lead to a more consistent overall picture. Thus, even though super-resolved MEM maps (i.e., MEM maps made using a convolving beam that is appreciably smaller than the nominal CLEAN beam obtained by fitting a Gaussian to the central lobe of the dirty beam) are not any more ``real'' than super-resolved CLEAN maps, it is arguably reasonable to expect them to exhibit a higher degree of fidelity to the true structure (it should be noted that the CLEAN algorithm itself can model data on scales smaller than the CLEAN beam -- but its intrinsic delta function model can be less realistic than MEM's continuous model, and it does not have the advantage of a multi-Stokes parameter model). \citet{Gull1984} suggest that while the result of a MEM super-resolution may not be more likely than other approaches, it may be ``preferred'' in that it is maximally non-committal about measured parameters. Mathematically however, it remains difficult to judge the accuracy of any super-resolution achieved with either MEM or CLEAN and their respective performances may be better investigated by empirical modelling.

\section{The PMEM Software}

\label{sec-pmem}

The potential advantages of MEM over the CLEAN algorithm come at a cost of increased complexity and demand for computational resources. Implementing MEM and successfully using it to deconvolve a VLBI image can be more challenging than using CLEAN -- however the rewards can be correspondingly greater. Versions of MEM suitable for the imaging of Stokes $I$ data alone have been implemented in many popular astronomical software suites, including the NRAO's {\sc AIPS} and {\sc CASA} \citep{Greisen2002,McMullin2007}. While these implementations of MEM can deconvolve VLBI Stokes $I$ emission, they have no support for deconvolving Stokes $Q$ and $U$ maps (this is due to the choice of the standard Shannon entropy instead of the Gull and Skilling entropy shown in Equation \ref{mem-heqn}). Conversely, a version of MEM suitable for imaging polarised emission is present in CSIRO's {\sc MIRIAD} data reduction package \citep{Sault1995}, however it is incompatible with data from the Very Long Baseline Array (VLBA).

We have written new deconvolution software,  ``Polarised Maximum Entropy Method'' ({\sc PMEM}), to address these issues and implement a version of MEM based on the Cornwell--Evans algorithm with support for polarisation \cite{Coughlan2014}. This implementation builds on work done by \citet{Holdaway1990}. It is intended for studies of the polarisation properties of AGN on VLBI scales and enables the creation of VLBI polarisation and Faraday rotation measure maps. The following section describes some of the innovations present in PMEM and gives some details of the implementation.

\subsection{Optimization method}

PMEM offers several algorithmic improvements over the previous implementations in AIPS and MIRIAD and in particular aims to provide the polarisation user extra functionality to get the highest performance from their data. By default PMEM uses the polarisation extensions to the diagonalised Newton-Raphson method based Cornwell--Evans algorithm \citep{Cornwell1985} developed by \citet{Holdaway1990} and \citet{Sault1999}. Additional solvers have also been incorporated and tested, including gradient only solvers such as steepest descent and conjugate gradient approaches, as well as Davidson--Fletcher--Powell (DFP) and Broyden--Fletcher--Goldfarb--Shanno (BFGS) solvers which do not use the the exact diagonal Hessian identified by \citet{Cornwell1985}, but instead construct one iteratively from the data and the first derivatives.

While the gradient only solvers are extremely slow for this problem, the DFP and BFGS solvers perform very similarly to the Cornwell--Evans implementation and may offer more robust performance in cases where the exact Hessian does not give the fastest or most accurate solution. However as the Newton-Raphson method was found to give the best results in general, it has been used to reduce all the simulations and observations presented in this paper. It should be noted that while multiple methods of calculating the Hessian are used, the Hessian is approximated by a diagonal matrix in each case. This means that the correlation between adjacent pixels remains difficult to describe within the framework -- in keeping with MEM, but unlike frameworks such as RESOLVE \citep{Junklewitz2016} where correlation is accounted for. Whether using the Newton--Raphson, DFP or BFGS solvers, PMEM is efficient and processes maps of up to $10^{6}$ pixels in a matter of seconds.

\subsection{Improving polarisation performance}

MEM iteratively creates optimal model maps in Stokes $I$, $Q$ and $U$ by maximising Equation \ref{mem-jeqn}, balancing maximising the Gull and Skilling entropy of the model maps and minimising the disagreement between the convolved models and the observed data. A marked improvement in the resulting Stokes $Q$ and $U$ MEM maps can be achieved by including a term up-weighting the visibility disagreement term in Equation \ref{mem-jeqn}. This weighting accounts for the difference in magnitude between the total intensity and polarisation terms resulting from the fact that the maximum polarisation that can be expected from synchrotron radiation is 75\%, and in most sources the actual detected polarisation is far less -- often of the order of 10\% or so \citep{Pacholczyk1970}. This results in the misfit for the Stokes $I$ intensity almost always being much greater than the polarisation misfit when both are measured in the same scale. While this is strictly true, it can result in an optimal MEM model that agrees very well with the Stokes $I$ intensity, but quite poorly with the polarisation data. As a major purpose of PMEM is to provide increased resolution for polarisation observations of AGN jets, this behaviour is undesirable. To overcome this issue a new parameter $w_{\mathrm{p}}$ is introduced to increase the weight of the polarisation misfits relative to the Stokes $I$ misfit. This results in the total polarisation misfit being calculated in image space as
\begin{equation}
\begin{aligned}
\label{mem-main-eqn-wp}
F =\ & w_{\mathrm{p}}(\sum_{\mathrm{i}}^{N_{\mathrm{pix}}}(\sum_{\mathrm{j}=0}^{N_{\mathrm{pix}}}P_{\mathrm{i,j}}Q_{\mathrm{j}}-\mathrm{DMQ}_{\mathrm{i}})^{2}\\
& + \sum_{\mathrm{i}}^{N_{\mathrm{pix}}}(\sum_{\mathrm{j}=0}^{N_{\mathrm{pix}}}P_{\mathrm{i,j}}U_{\mathrm{j}}-\mathrm{DMU}_{\mathrm{i}})^{2}),
\end{aligned}
\end{equation}
\noindent
where the indices $\mathrm{i}$ and $\mathrm{j}$ sum over pixels, $\mathrm{DMQ}_{\mathrm{i}}$ is the $\mathrm{i}^{th}$ pixel of the Stokes $Q$ dirty map and $\sum_{\mathrm{j}=0}^{N_{\mathrm{pix}}}P_{\mathrm{i,j}}Q_{\mathrm{j}}$ indicates the convolution of the Stokes $Q$ model map with the dirty beam $P_{\mathrm{i,j}}$.

The most suitable value for $w_{\mathrm{p}}$ can differ for different datasets, depending on the relative complexities of the intensity and polarisation structures and the noise present in each. As a rough guide, finding the average fractional polarisation $m_{\mathrm{avg}}$ by dividing the total polarised flux of the source being deconvolved by its total intensity, then setting $w_{\mathrm{p}} = 1/m_{\mathrm{avg}}$ would approximately equalise the total intensity and polarisation terms in Equation \eqref{mem-main-eqn-wp}. The final Stokes $I$ map is relatively insensitive to changes in $w_{\mathrm{p}}$, however the use of an appropriate value can significantly improve the final Stokes $Q$ and $U$ maps. A value of 2.0 was found empirically to be a good compromise between weighting up the importance of good convergence in the polarisation Stokes parameters while not causing an appreciable change to the Stokes $I$ map, though higher values were often appropriate. A value of 1.0 is equivalent to the standard Cornwell--Evans implementation.

An alternative method of implementing this feature would be to weight the Stokes $Q$ and $U$ maps differently. This would be of use in sources where there was detectable flux in both Stokes $Q$ and $U$ with a significant difference in magnitude between them. In practice it was found that a single parameter was sufficient to allow good convergence in all three Stokes parameters.

\subsection{Convergence and aliasing}

Different observations have different dirty beam profiles and different levels of noise in the observed visibilities. While the unaltered algorithm performs well for the majority of VLBI jet observations, observations of structure on the same scale as the main lobe of the dirty beam can prove challenging for both MEM and CLEAN to image successfully. The inclusion of a stepping factor $\Delta_{\mathrm{step}}$ and an edge pixel exclusion option $N_{\mathrm{exclude}}$ gives the user some flexibility in responding to challenging sources. More often, they allow the optimisation of maps made for better performing sources.

$\Delta_{\mathrm{step}}$ is applied as a multiplicative factor to the maximum size of a step allowed during the maximisation of Equation \eqref{mem-jeqn} using the Newton--Raphson method. \citet{Cornwell1985} describe how the gain of the MEM, i.e. the change in $J$ for a change in the Stokes $I$ model map, can be calculated as $\nabla J \cdot \nabla J / 1\cdot1$, where the inner product is taken using the metric $(-\nabla \nabla J)^{-1}$. If the gain is high, $J$ may change rapidly, while if the gain is low $J$ may be difficult to change. Applying $\Delta_{\mathrm{step}}$ to the maximum step limit allowed in the Newton-Raphson maximisation as
\begin{equation}
\Delta_{\mathrm{max}} \propto \Delta_{\mathrm{step}} \frac{\nabla J \cdot \nabla J}{1 \cdot 1}
\label{eqn-dstep}
\end{equation}
\noindent
allows the user to manually slow down the changes in a single iteration of the algorithm in the case of poor performance or increase the changes (reducing the compute time) in the case of well-performing sources. This parameter was especially useful in optimising the performance of PMEM for the Monte Carlo simulations detailed further on in this Section. Setting $\Delta_{\mathrm{step}} = 1$ performs well in most cases.

$N_{\mathrm{exclude}}$ is a parameter which forces the flux in the outer $N_{\mathrm{exclude}}$ pixels of the model map to zero. This parameter is useful in reducing problems caused by aliasing due to the Fast Fourier Transforms (FFTs) performed on the model map. Setting $N_{\mathrm{exclude}}$ to a value equal to around 5\% of the image size is often sufficient, though for some combinations of source structure and image size this can be as high as 25\%. A PMEM mask may also be used to exclude pixels in a similar way to a CLEAN mask. See Appendix \ref{app-pmem} for further details.

\subsection{Diagonalising the Hessian}
\label{sec-diag-hess}

The maximisation of Equation \ref{mem-jeqn} requires the calculation and inversion of the Hessian matrix $\nabla \nabla J$. For an image with a total number of pixels equal to $N$ this would normally require the storage and inversion of a matrix of size $N \times N$ -- a calculation which challenges both the processing power and memory of modern computers. Luckily, the breakthrough approximation of the Cornwell and Evans algorithm \citep{Cornwell1985} was a description of how the Hessian of $J$ of can be diagonalised with the approximation
\begin{equation}
P_{\mathrm{ij}} \approx A
\end{equation}
\noindent
where $P_{\mathrm{ij}}$ is an $N \times N$ dimensional matrix representing convolution with the dirty beam and $A$ is a factor which should represent the power in the main lobe of the primary beam. The main information lost by making the Cornwell-Evans approximation is the correlation between neighbouring pixels in the model maps represented by the off-diagonal terms in the Hessian. In practice, with the choice of a suitable value of $A$, the approximation leads to a successful convergence of the MEM, however there are cases where it can lead to poor convergence. Generally these cases occur when there are many pixels across a beam and neglecting the off-diagonal elements of the Hessian has a significant effect on the inversion. This can be avoided by insuring the ratio of beam-size to pixel size does not get too high (3-4 cells across each beam is usually adequate) and by employing a linear search in the step size suggested by the MEM at each iteration as described in \citet{Cornwell1985}.\\

\citet{Sault1990} suggested setting $A$ equal to the following expression
\begin{equation}
A = \sqrt{\sum_{\mathrm{i}}^{N_{\mathrm{pix}}}P_{i}^{2}},
\end{equation}
\noindent
where $P_{\mathrm{i}}$ is the dirty beam. This represents the gain of the array for white noise and usually leads to a successful convergence of the MEM. However, it was found useful to introduce a manual term, $a_{\mathrm{factor}}$ to modify the factor $A$ as follows
\begin{equation}
A = a_{\mathrm{factor}}\sqrt{\sum_{\mathrm{i}}^{N_{\mathrm{pix}}}P_{i}^{2}}.
\end{equation}
\noindent

Varying $a_{\mathrm{factor}}$ in conjunction with $\Delta_{\mathrm{step}}$ as defined in Equation \eqref{eqn-dstep} is a useful technique in optimising the deconvolution process for poorly performing sources, or for sources where the ratio of beam-size to pixel size invalidates the diagonalisation of $\nabla \nabla J$ as discussed above. A value of $a_{\mathrm{factor}} = 1$ is equivalent to the suggestion of \citet{Sault1999}, however the choice of a higher value leads to smaller steps in the MEM and can allow the algorithm to recover from the effects of ignoring significant correlation present in the off-diagonal elements of $\nabla \nabla J$. Conversely, a smaller value of $a_{factor}$ can lead to faster convergence in cases where the Cornwell-Evans approximation performs well. In practice, choice of a value for $a_{\mathrm{factor}}$ depends on the $UV$ coverage of the observation, the structure of the source and the imaging parameters (primarily the number of pixels across a beam). 

\subsection{Implementation}

The C++ programming language was chosen due to its speed and efficiency. It also has a wide range of external libraries available which were used to further reduce the computational time for a typical MEM execution.

OpenMP was used to parallelise the computation where possible. The external library FFTW was used to perform the Fast Fourier Transforms (FFTs) needed in the convolutions. LAPACK, a high performance linear algebra library, was used to perform the matrix calculations needed for MEM. The FITS (Flexible Image Transport System) file format was used at all times, ensuring compatibility with all of the major astronomical software packages. This was implemented with the CFITSIO library via an easy to use interface layer (quickfits). The open-source codes of AIPS's `VTESS' task and MIRIAD's `PMOSMEM' task \citep{Greisen2002,Sault1995} were also of great help in writing PMEM. Both tasks contain excellent implementations of the Cornwell-Evans algorithm and were of immense help in avoiding numerical problems and deciding how to structure PMEM.

PMEM is open-source and availble at \url{https://github.com/colmcoughlan/pmem}. It makes use of the quickfits interface to CFITSIO, also available at \url{https://github.com/colmcoughlan/quickfits}. It should be noted that in principle this current implementation of PMEM should perform well for some non-VLBI observations. In any situation where the field is narrow enough that the standard dirty beam is a good description of the Fourier transform of the UVW sampling function (i.e. the w-term can be neglected) and spectral modelling of the emission is not required within the bandwidth, PMEM should achieve similar results to the VLBI results presented in this paper. Further information on PMEM, including recommended parameters, is available in Appendix \ref{app-pmem}.

\section{Monte Carlo Testing}
\label{section-montecarlo}

\begin{figure}
\begin{center}
\includegraphics[width=1.0 \columnwidth,trim={1cm 5cm 0cm 4.5cm}]{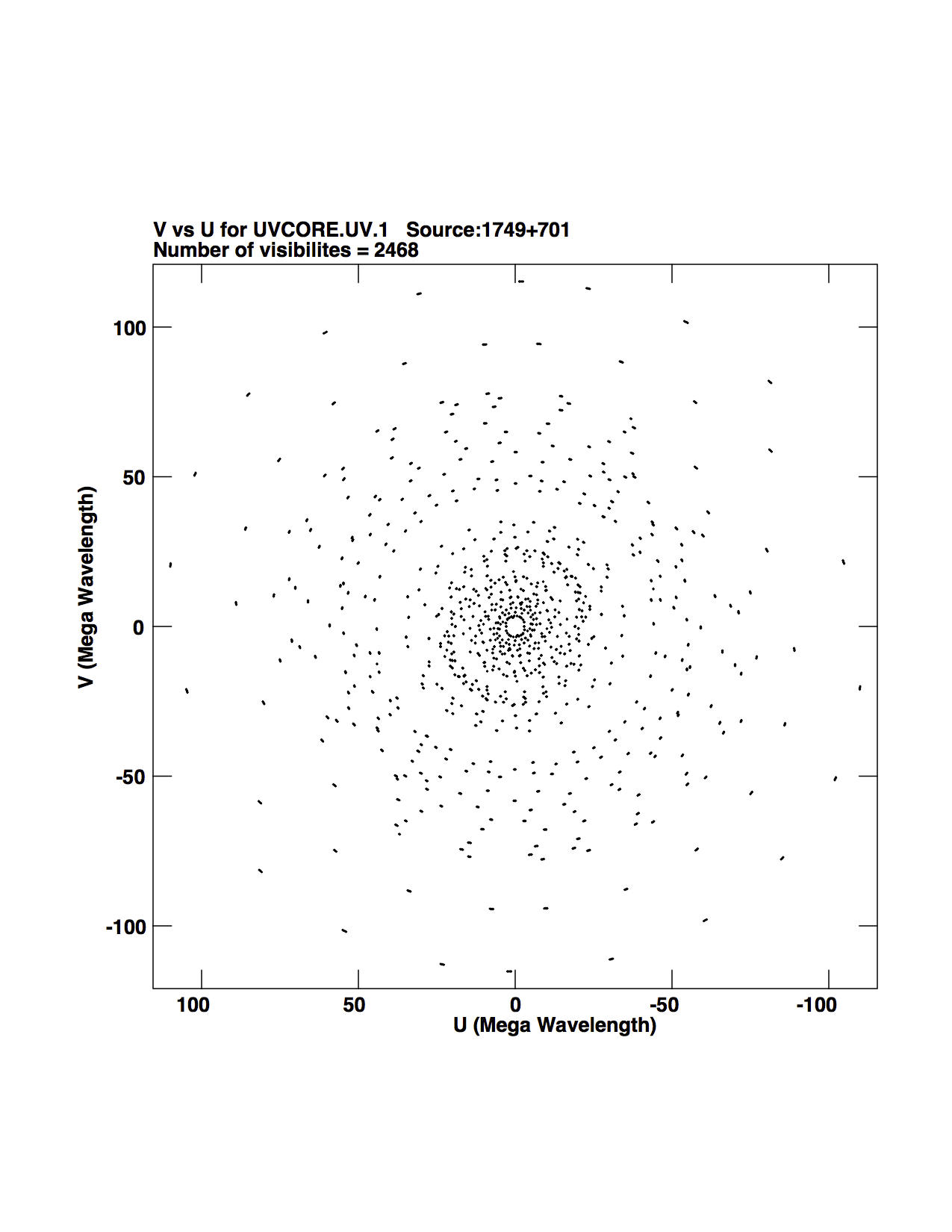}
\caption[UV distribution used in Monte Carlo Simulations.]{The UV distribution used in the Monte Carlo simulations, which corresponds to ``snapshot'' VLBA observations of a source at a declination of $+70^{\circ}$ at 4.6 GHz.}
\label{mem-main-fig-uvcore}
\end{center}
\end{figure}

To test the performance of MEM as implemented in {\sc PMEM}, we carried out a series of Monte Carlo simulations using the $UV$ distribution shown in Fig.~\ref{mem-main-fig-uvcore} and a two simple model sources -- a triple source consisting of 3 circular Gaussian components (Table~\ref{mem-main-tg-table}), and a continuous bent jet with a complex polarisation structure. Both of these sources are plotted in Figure ~\ref{mem-main-fig-mctg-model}.

\subsection{Methodology}

\begin{figure*}
\begin{center}
\subfloat{
	\includegraphics[width=0.8 \columnwidth,trim={0 2cm 0 1cm}]{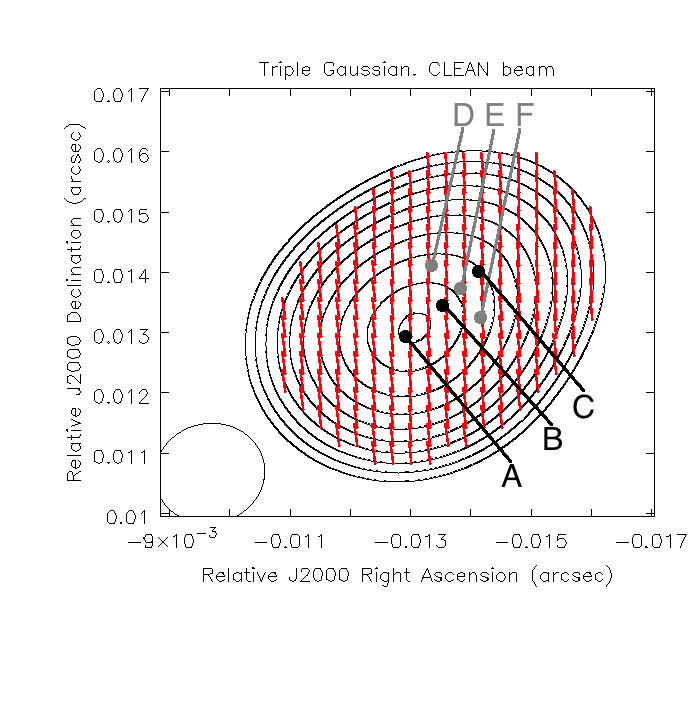}
}
\subfloat{
	\includegraphics[width=0.8 \columnwidth,trim={0 2cm 0 1cm}]{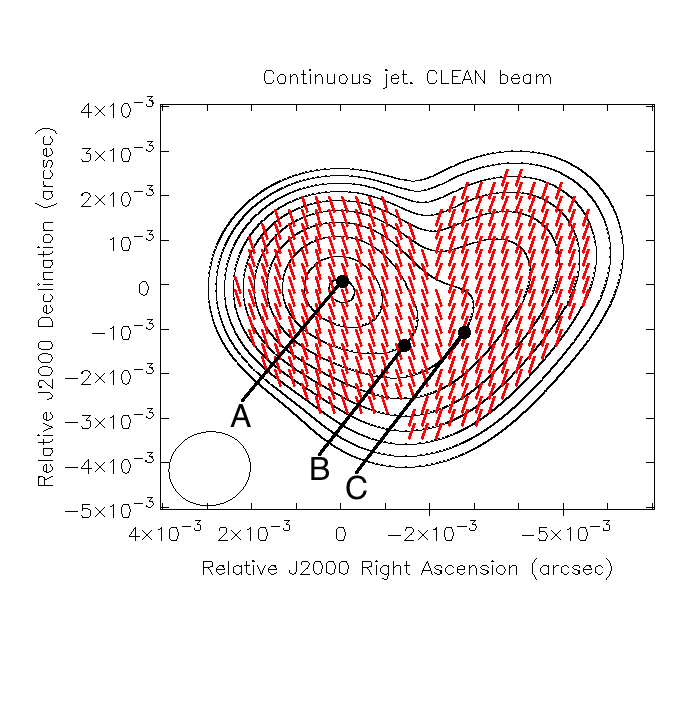}
}

\subfloat{
	\includegraphics[width=0.8 \columnwidth,trim={0 2cm 0 1cm}]{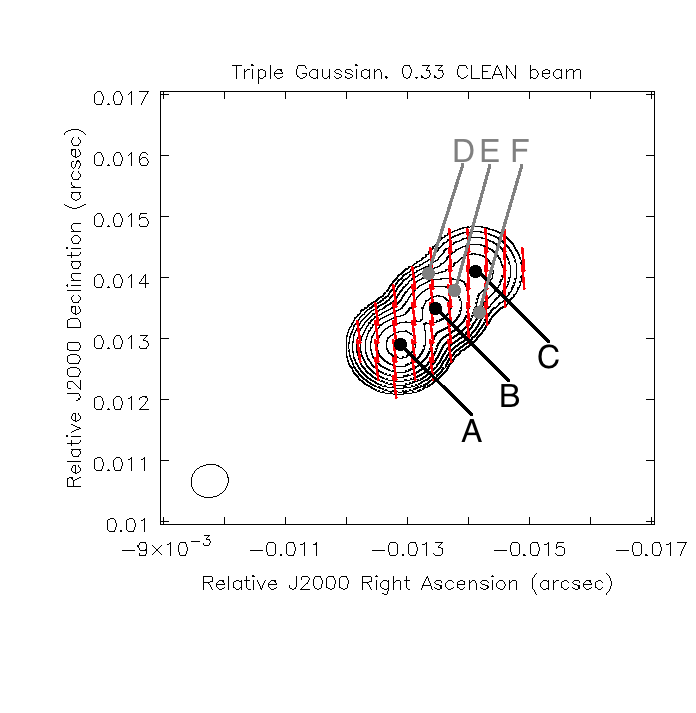}
}
\subfloat{
	\includegraphics[width=0.8 \columnwidth,trim={0 2cm 0 1cm}]{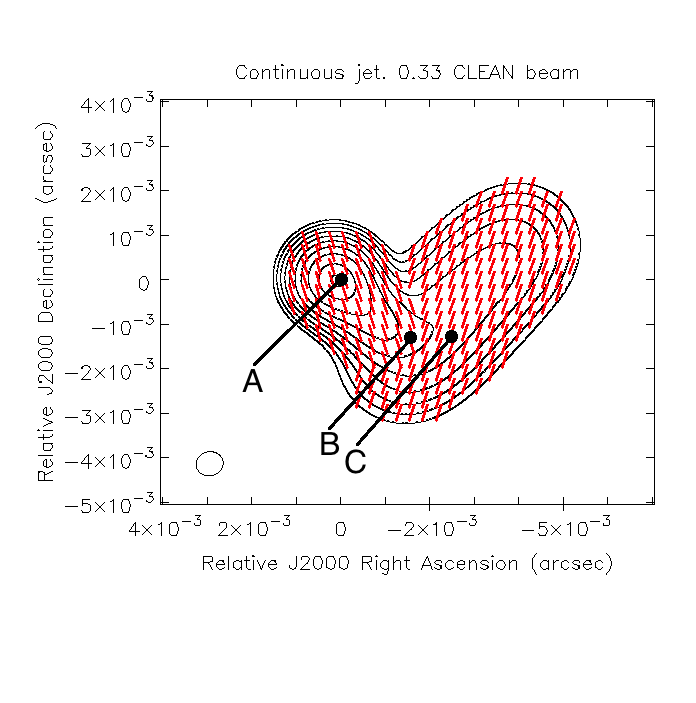}
}
\caption[Convolved model sources]{The model triple Gaussian and continuous bent jet sources convolved with the full CLEAN beam corresponding to the $UV$ distribution in Figure \ref{mem-main-fig-uvcore} and with a beam with a FHWM one third the size of the standard CLEAN FWHM. Points A-F indicate positions sampled by the Monte Carlo analysis. In the case of the triple Gaussian points A, B and C are also the location of the Gaussian components. Contours increase in powers of two with a final contour at 95\% of peak. The lowest contours for the triple Gaussian are: 3.3~mJy (CLEAN beam), and 2.4~mJy ($\frac{1}{3}$ CLEAN beam). The lowest contours for the continuous bent jet model are: 0.96~mJy (CLEAN beam), and 0.36~mJy ($\frac{1}{3}$ CLEAN beam). The red lines indicate the direction of the polarisation angle.}
\label{mem-main-fig-mctg-model}
\end{center}
\end{figure*}

The parameters of the three Gaussians in the first source were chosen to be reasonably typical of a fairly strong $\simeq 1$-Jy, compact AGN; the separation between the ``core'' and outer jet component is comparable to the major axis of the full CLEAN beam (see Table \ref{mem-main-tg-table}). The continuous bent jet source is similar, but is comprised of continuous emission instead of three Gaussian components. It has an integrated flux of 0.7~Jy and a increasing fractional polarisation from 10\% in the core to 20\% in the jet region. The polarisation angle also switches from -45$^{\circ}$ to +45$^{\circ}$ across the jet. Figure.~\ref{mem-main-fig-mctg-model} shows maps of both sources made by convolving the intrinsic model with (a) the full CLEAN beam and (b) a beam with a FWHM equal to one-third of the CLEAN beam's. The positions of the three components (A, B and C) which were considered for comparing CLEAN and MEM are indicated, along with a further three components (D, E and F) in the triple Gaussian which were also considered to examine performance in measuring values transverse to the jet direction.

The CLEAN algorithm as implemented in {\sc CASA}'s ``clean'' was also used to provide a benchmark and establish whether any additional information could be gained with the MEM deconvolution.

The images produced using both algorithms were tested for fidelity using five different restoring beams, corresponding to the full CLEAN beam and $\frac{3}{4}$, $\frac{1}{2}$, $\frac{1}{3}$ and $\frac{1}{4}$ of the CLEAN beam (the full CLEAN beam was $1.84 \times 1.64$~mas$^{2}$ in $-78^{\circ}$). It is important to note that these factors were applied to both the major and minor axes of the beam, thus $\frac{3}{4}$ of both CLEAN beam parameters corresponds to a beam with an area equal to $\frac{9}{16}$ of the original CLEAN beam. We carried out the MEM and CLEAN deconvolutions for 
Stokes $I$, $Q$ and $U$. We assumed negligible Stokes $V$, due to the typically weak circularly polarised flux expected (and observed) from AGN jets (e.g., \citep{Homan2006}), however a Stokes $V$ map would be treated in MEM the same was as Stokes $Q$ or $U$ (indeed, internally PMEM does not distinguish between Stokes $Q$, $U$ and $V$ in a meaningful way). The ability of MEM and CLEAN to restore the total flux and local fluxes at various locations was tested. Maps of the polarised flux, fractional polarisation and polarisation angle were also generated and used for these comparisons.

\begin{table}
\begin{center}
\begin{tabular}{| c | c | c | c | c | c | c | c |}
	\hline
	Comp & r & $\sigma$ & $I$ & $Q$ & $U$ & m & $\chi$\\
	& (mas) & (mas) & (Jy) & (Jy) & (Jy) &  & \\
	\hline                 
	A & 0.00 & 0.05 & 1.00 & 0.035 & 0.035 & 0.049 & $22.5^{\circ}$ \\
	\hline 
	B & 0.85 & 0.10 & 0.50 & 0.018 & 0.018 & 0.051 & $22.5^{\circ}$\\
	\hline 
	C & 1.70 & 0.40 & 0.10 & 0.031 & 0.016 & 0.35 & $13.6^{\circ}$\\
	\hline 
\end{tabular}
\caption[Model Triple Gaussian Source.]{Components A, B and C of the model triple Gaussian source. r is the separation from the strongest component (the ``core''); $\sigma$ the circular Gaussian FWHM; $I$, $Q$ and $U$ the corresponding Stokes parameters; $m$ the fractional polarisation; and $\chi$ the polarisation angle.}
\label{mem-main-tg-table}
\end{center}
\end{table}

\subsection{Simulation Results}

\begin{figure*}
\begin{center}
\subfloat[TG - point B. Stokes I]{
	\includegraphics[width=  1.0\columnwidth]{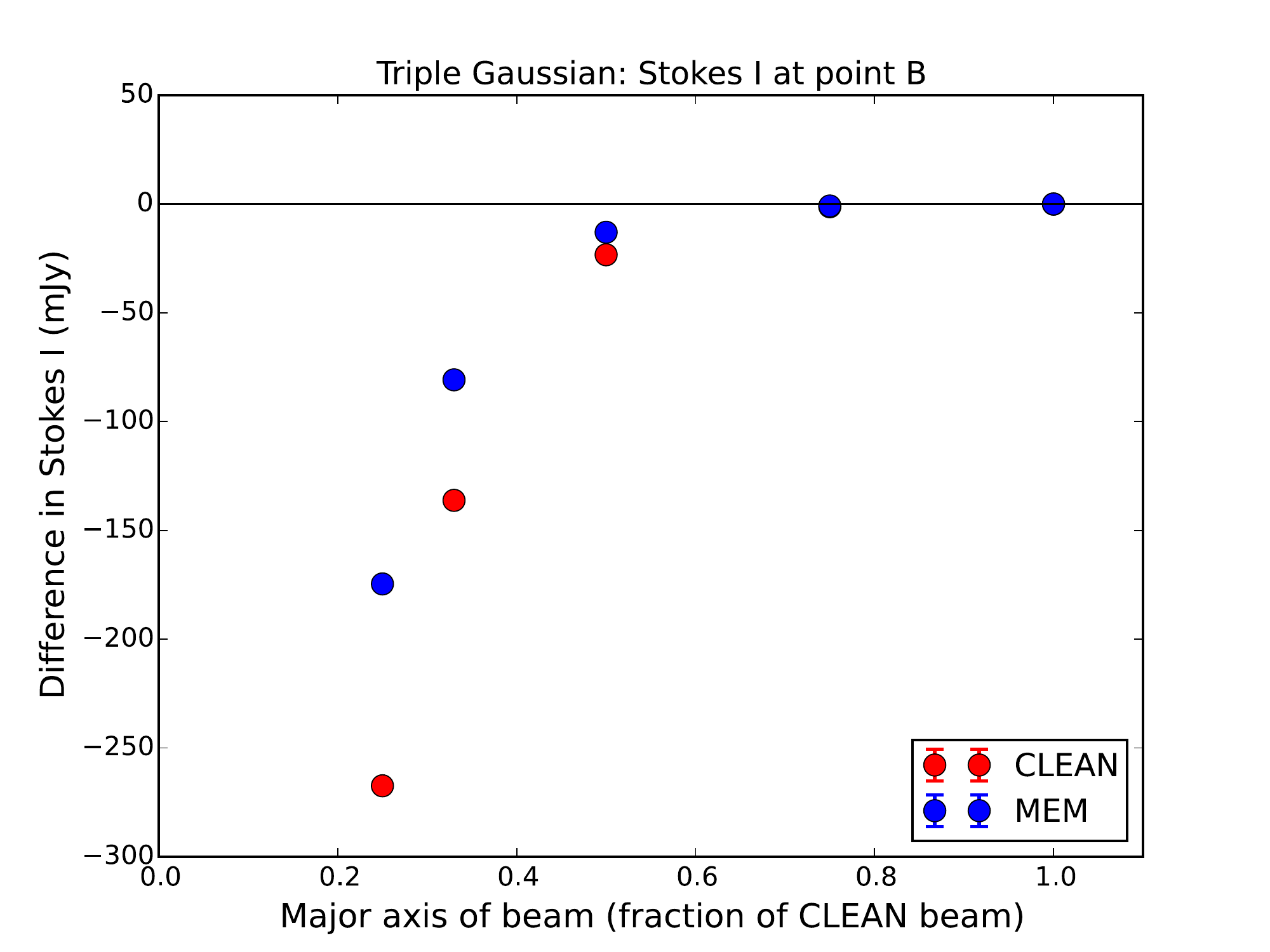}
}
\subfloat[CJ - point B. Stokes I]{
	\includegraphics[width=  1.0\columnwidth]{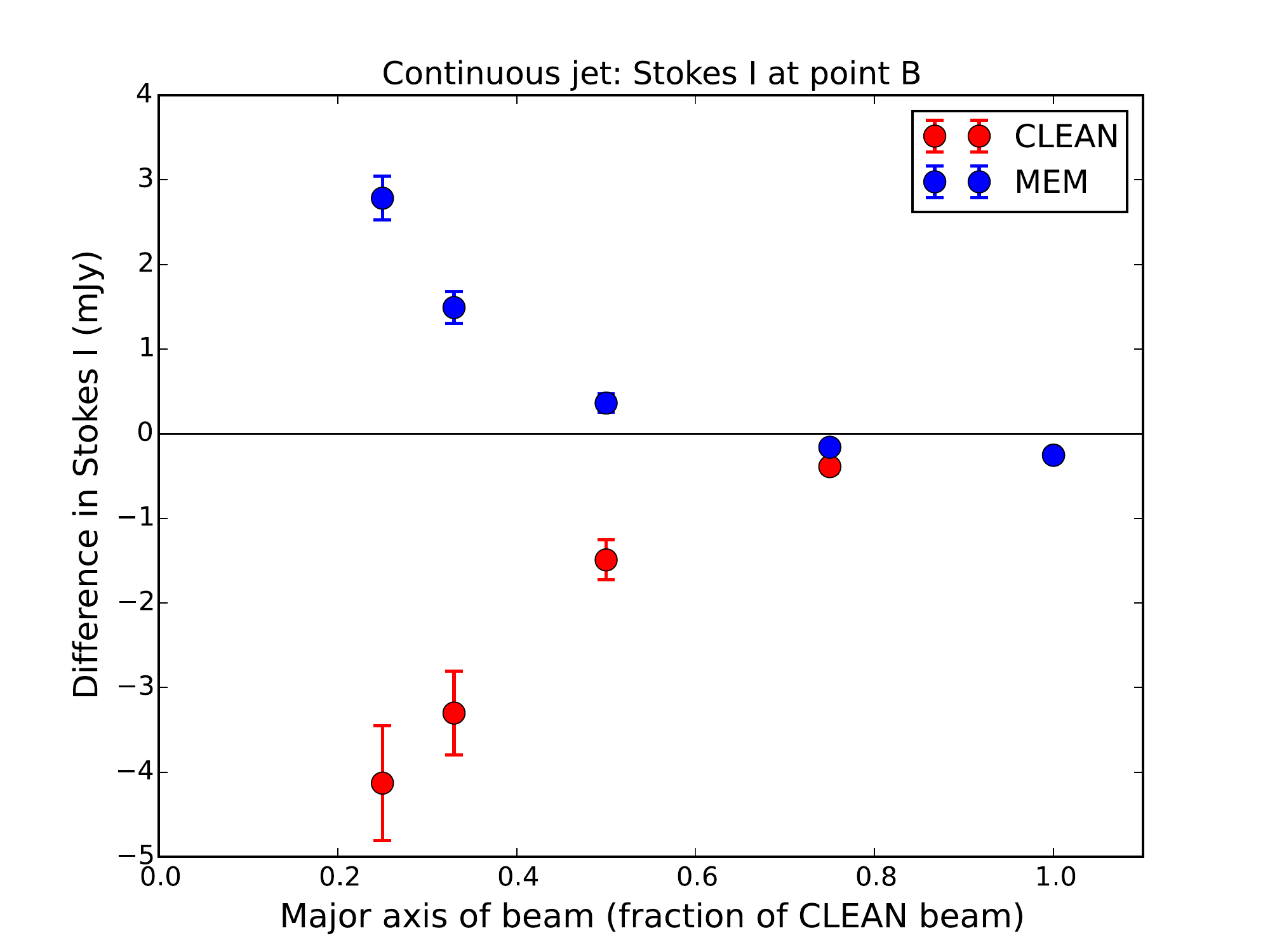}
}

\subfloat[TG - point B. $m$]{
	\includegraphics[width=  1.0\columnwidth]{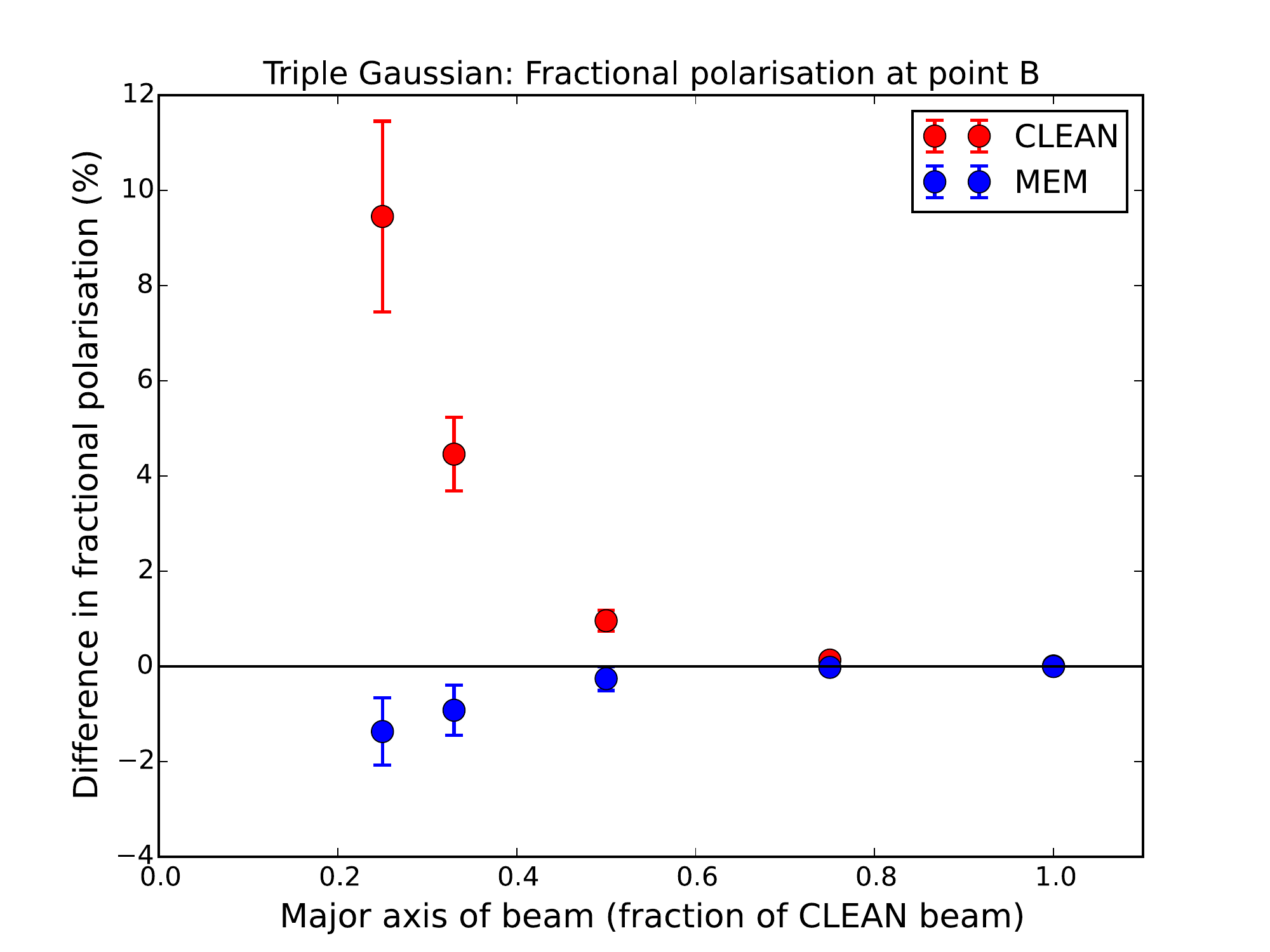}
}
\subfloat[CJ - point B. $m$]{
	\includegraphics[width=  1.0\columnwidth]{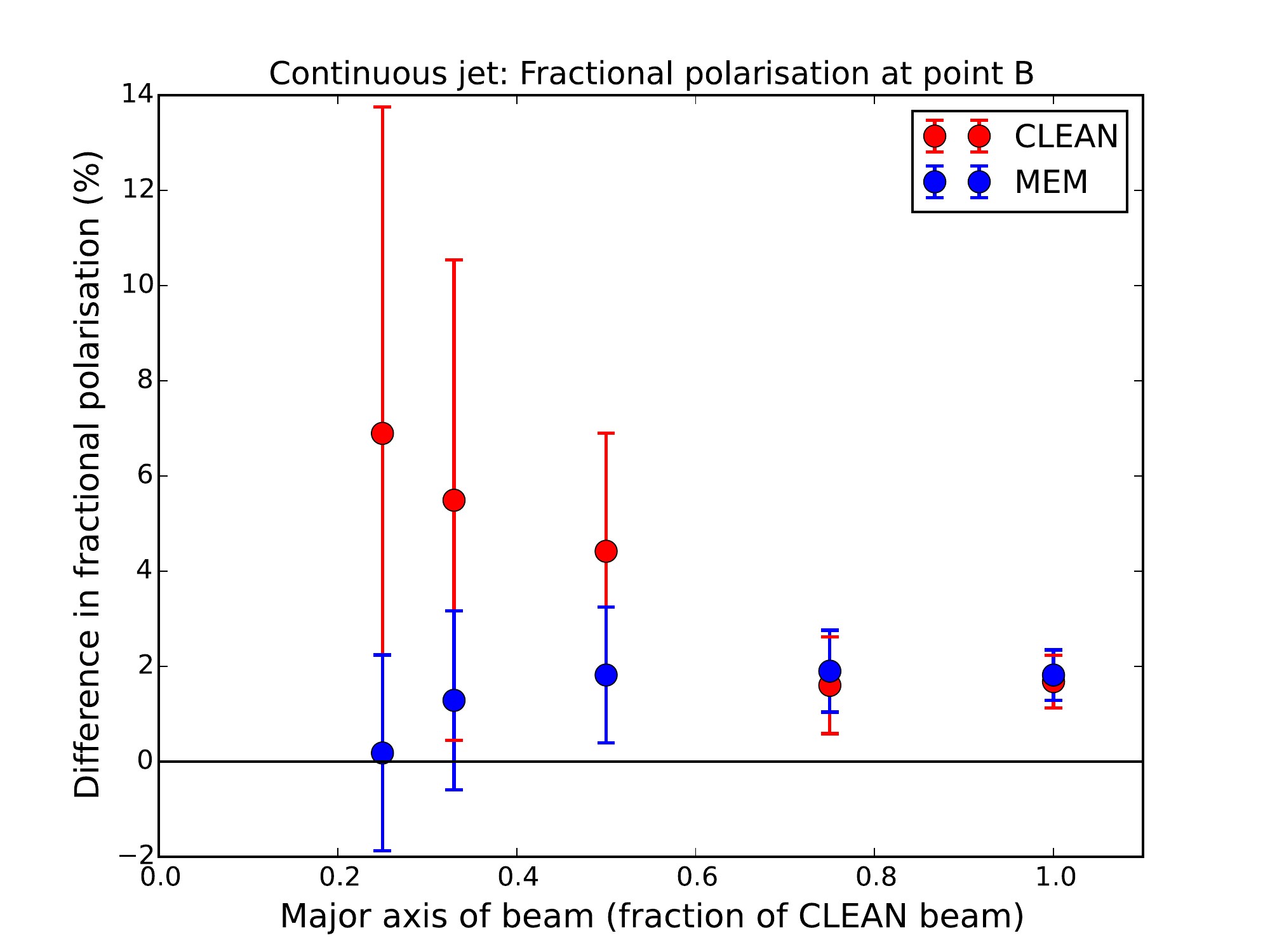}
}

\subfloat[TG - point B. $\chi$]{
	\includegraphics[width=  1.0\columnwidth]{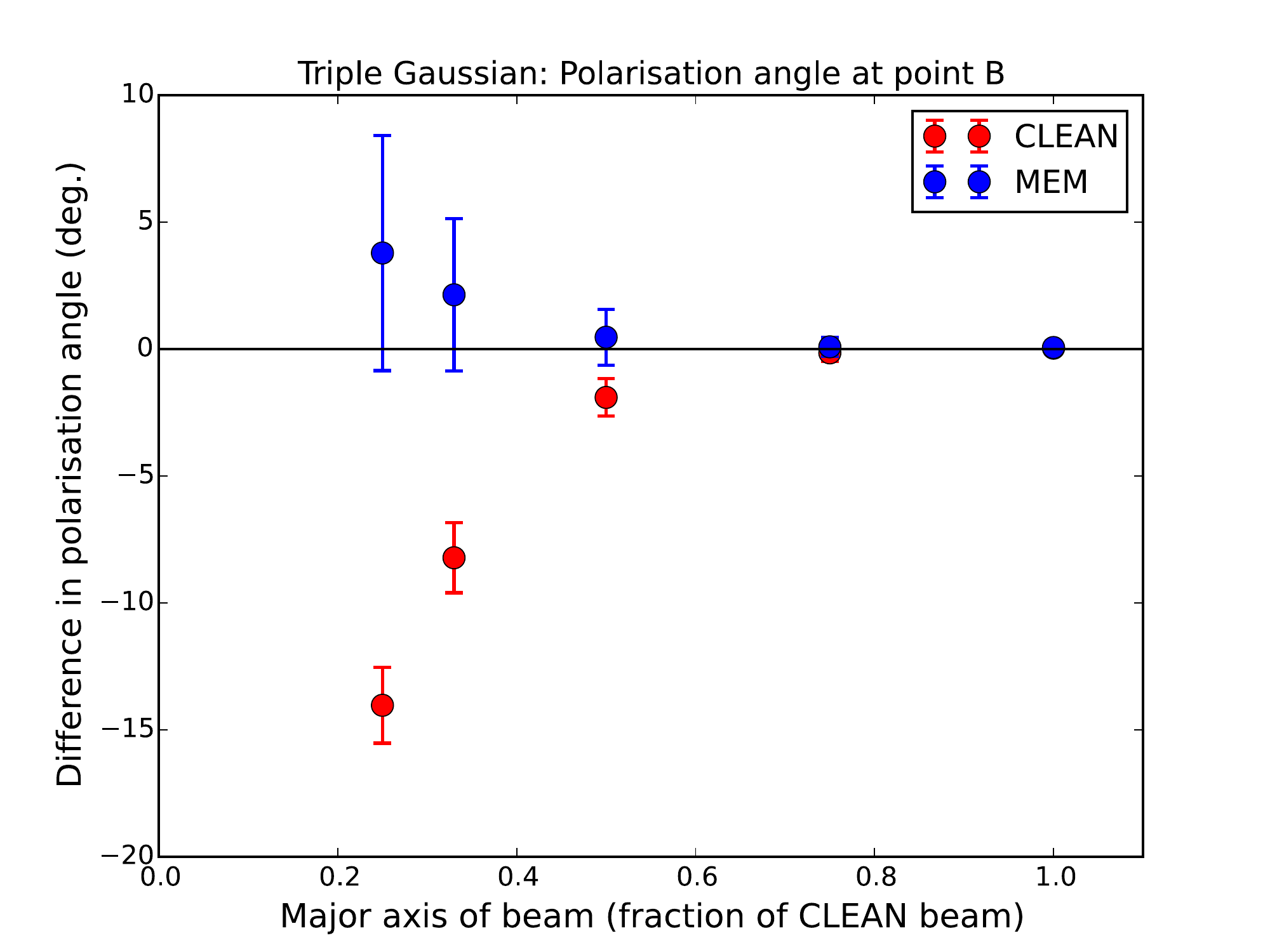}
}
\subfloat[CJ - point B. $\chi$]{
	\includegraphics[width=  1.0\columnwidth]{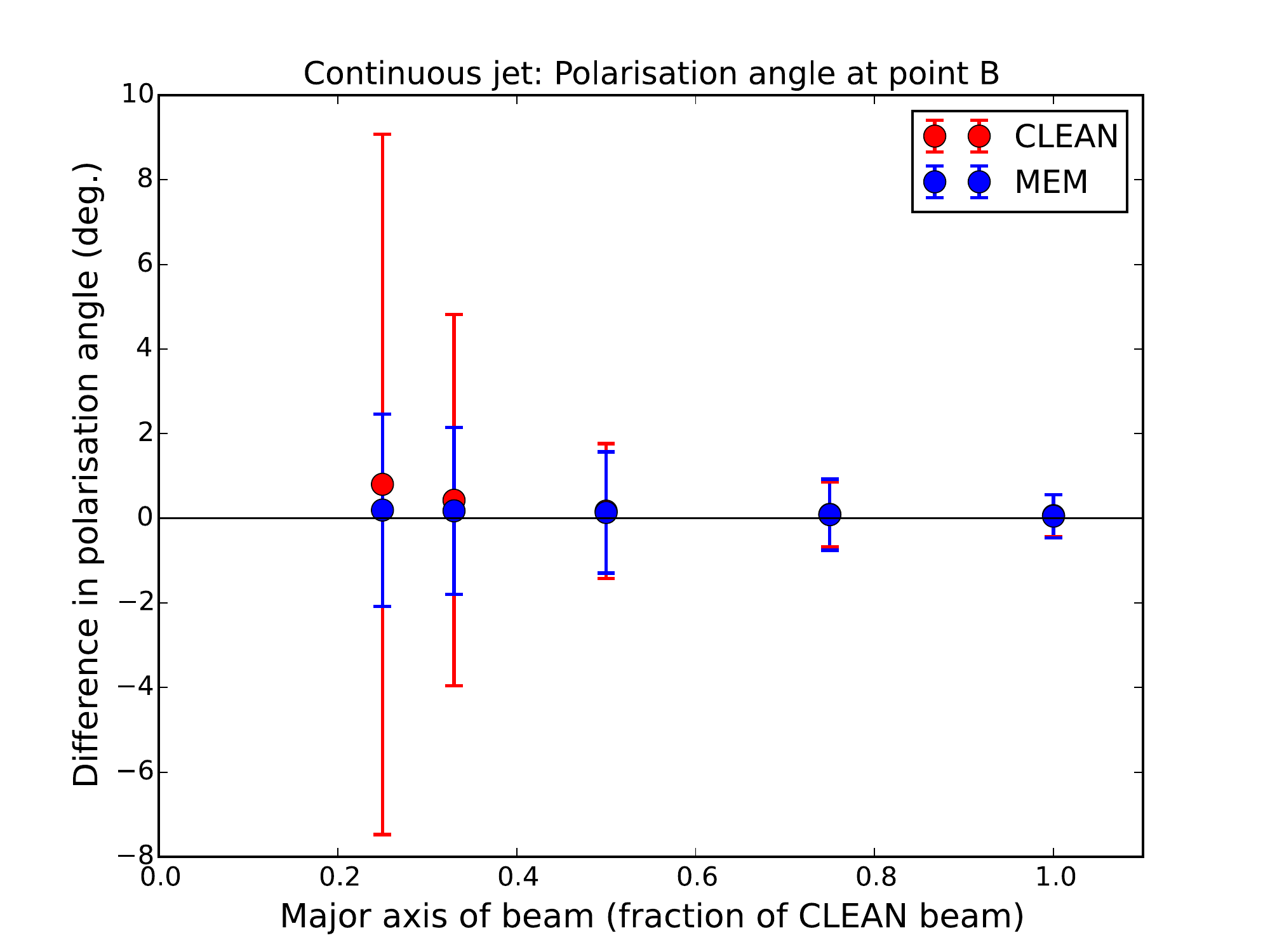}
}
\caption[Sample of some of the Monte Carlo results.]{A sample of some of the results from the Monte Carlo simulations. The left-hand columns are from the Triple Gaussian (TG) source, the right-hand ones are from the continuous bent jet (CJ) model. Panels (a, b) show the polarised flux $p$, (c, d) the fractional 
polarisation $m$ and (e, f) the polarisation angle $\chi$, for component B in each model. The size of the FWHM of the convolving beam used is indicted on the x-axis as a fraction of the standard CLEAN FWHM (note that the beam area is the square of this factor).}
\label{mem-main-fig-sims}
\end{center}
\end{figure*}

Figure \ref{mem-main-fig-sims} shows a summary of some of the the Monte Carlo results shown in full in Figures \ref{mem-main-fig-tg-fluxes-total}--\ref{fig-app-cts-pmchi-BC} of Appendix \ref{app}. In all cases, the
CLEAN results are shown by hollow circles and the MEM results by
filled circles.  Results for the individual points (A, B, C etc.) were obtained by summing $I$, $Q$ and $U$ within a
0.3~$\times$~0.3~mas$^{2}$ (i.e., 3~$\times$~3~pixel$^{2}$) area centred on the point 
in question, then using the $Q$ and $U$ values to derive $p = 
\sqrt{Q^{2} + U^{2}}$, $m = \sqrt{Q^{2} + U^{2}}/I$ and the polarisation 
angle $\chi = \frac{1}{2} \arctan{U/Q}$.
In all cases, the points in these plots indicate the
mean differences obtained by comparing the 100 realizations of the 
model source map with the ``true'' map (the original model convolved with the appropriate beam), and the error bars 
indicate the standard deviation of the various difference measurements
about these mean values.

Figure \ref{mem-main-fig-sims} shows the results for point B for both the triple Gaussian and continuous bent jet models. The top panels show the results for Stokes I, the centre panels show fractional polarisation ($m$) and the bottom panels show polarisation angle ($\chi$). When the full CLEAN beam is 
used for the convolution, 
both methods perform very well and parameters close to the 
correct ones are recovered. The performance of both CLEAN and MEM 
continues to be good when images are restored with 0.75 and 0.5 of the standard CLEAN FWHM, but below this point both CLEAN and MEM start to show increased errors in most cases -- though with MEM generally showing smaller mean errors and a tighter grouping than CLEAN.

This general trend of MEM out-performing CLEAN at high resolution can also be seen in the full results presented in Appendix \ref{app}, however there are some notable exceptions. In both simulations, the CLEAN algorithm significantly out-performs MEM in measuring the total flux, however this is somewhat offset by MEM's better performance in measuring total polarised flux and the overall fractional polarisation and polarisation angle. In most cases, MEM achieved an accuracy at high resolutions that was as good as or better than CLEAN in measuring Stokes I, total and fractional polarisation and polarisation angle at the points considered, however there are multiple exceptions to this. While Figure \ref{mem-main-fig-sims} shows MEM performing better than CLEAN at measuring fractional polarisation at Point B, Figure \ref{mem-main-fig-pmchi-BC} in the Appendix shows CLEAN is actually better at recovering the total polarised flux. Similarly, while MEM achieves more accurate polarisation angle results for the triple Gaussian source, the difference between the two algorithms' polarisation angle results is much smaller for the continuous bent jet model, with CLEAN achieving the better result in the case of point A.

\subsection{Discussion}

The collected results of our Monte Carlo simulations show that, in 
general, both deconvolution methods show decreasing accuracy and 
increased sensitivity to noise (increasing error bars) as the 
restoring beam is decreased. CLEAN outperformed MEM 
in reproducing the total and polarised flux in some regions, while
MEM performed better than CLEAN in others. The performance of both 
CLEAN and MEM is excellent for restoring beams with a FWHM of 0.75 times the standard CLEAN FWHM (this corresponds to almost half the standard beam area), and even when a beam with 0.5 of the CLEAN beam FWHM is used the accuracy of both MEM 
and CLEAN usually remains
comparable to the accuracy when the images are restored with the full 
CLEAN beam. These trends are in agreement with the general observation that, given sufficient signal to noise, features and positional accuracies can be measured on scales significantly smaller than the convolving beam 
\citep{Condon1997}. They are also in agreement with the Monte Carlo 
studies of 
\citet{Murphy2012} and \citet{Mahmud2013}, where the CLEAN algorithm was 
found capable of reliably detecting variations in polarisation features 
on scales substantially smaller than the full CLEAN beam.

Our images restored with $\frac{1}{3}$ and $\frac{1}{4}$ of the standard CLEAN beam FWHM show significantly
greater inaccuracies in both total and local flux than when larger
convolving beams are used.
While MEM often out-performs CLEAN 
at these higher resolutions, there are many cases where CLEAN is equally reliable, and some where CLEAN is arguably better than MEM (see full discussion in \ref{app}). One
might expect the latter to be the case in more compact regions where 
CLEAN's $\delta$ function model of the source is most accurate, but
there is no obvious correlation between smaller component size and
improved performance of CLEAN in these simulations.

The significant variation in errors across the points and sources considered makes it is difficult to predict the uncertainty that could reasonably be expected in measuring similar points in real observations. It is similarly difficult to make a quantitative recommendation as to which imaging algorithm to use. However, a few general points can be taken from the simulations: both algorithms perform very well with mild to moderate super-resolution, and while MEM often performs better than CLEAN at higher resolutions, CLEAN is capable of out-performing MEM in some structures. In all cases super-resolution to below a quarter of the standard CLEAN beam area becomes increasingly risky and even though MEM is often more accurate than CLEAN this accuracy varies greatly with source structure and is likely related to the similarity of the true structure to the ``pointy'' or smooth structure assumed implicitly by CLEAN and MEM, respectively.

Bearing these results in mind, Section \ref{sec-obs} now describes the application of both CLEAN and MEM to a real VLBI observation and the use of both algorithms to achieve a high resolution picture of its complex polarisation structure.

\section{Application to 0716+714}
\label{sec-obs}

0716+714 is a BL Lac object, whose distance is not known due to the lack of a redshift \citep{Lister2009a}. \citet{Mahmud2013} reported the detection of a transverse Faraday
rotation-measure (RM) gradient that reversed its direction between the
core region and inner jet. It is noteworthy that \citet{Mahmud2013} used a restoring beam equal to about 60\% of the full CLEAN beam for the lowest frequency observed. The conclusions from the Monte Carlo simulations of \citet{Mahmud2013} and the new simulations presented in Section \ref{section-montecarlo} both indicate that reliable results should be obtained with such modestly ``super-resolved'' CLEAN maps.

The rotation in the polarisation angle due to Faraday rotation is given 
by $\chi - \chi_\mathrm{o} = RM \lambda^2$, where $\chi_\mathrm{o}$ is the unrotated 
polarisation angle and $\lambda$ is the observing wavelength. The 
coefficient RM is proportional to the integral along
the line of sight of the product of the density of charge carriers (presumed to be electrons) and the line of sight magnetic field.
Therefore, gradients in the Faraday rotation measure reflect spatial 
variations of the electron density and/or the component of the magnetic field along the line of sight. One reasonable interpretation of a 
monotonic gradient in the RM across an AGN jet is that it is due to 
a toroidal magnetic field or the toroidal component of a helical jet 
magnetic field. In the case of a helical field, depending on the 
helical pitch angle and the angle at which the jet is viewed, the
resulting transverse RM gradient may include RM values of a single sign
or of both signs; monotonic RM gradients that encompass both positive 
and negative RM values can only be explained by a change in the sign 
of the line of sight magnetic field, providing strong evidence that the 
gradient is associated with a toroidal magnetic-field component, rather
than, say, a gradient in the electron density.

The detection of a Faraday rotation measure gradient in 0716+714 by \citet{Mahmud2013} provided firm evidence for the presence of a toroidal component of the magnetic field in the jet, however the detection of another oppositely directed gradient further along the jet suggests the presence of a complex magnetic field geometry. \citet{Mahmud2013} proposed a number of possible explanations for this reversal, including torsional oscillations in the jet, the reversal of the pole of the super-massive black hole facing Earth, and the presence of a nested helical magnetic field structure in the jet. The authors favoured the nested helical magnetic field structure interpretation due to its relative simplicity and suggested that the opposing gradients in Faraday RM arise from a jet which has both an inner and an outer helical magnetic field with different parts of the field dominating the observed Faraday rotation at different distances from the base of the jet. This is an intriguing scenario and motivated us to use the PMEM code
to re-image the data and test the robustness of the mildly super-resolved CLEAN images in the original paper.

\begin{figure*}
\begin{center}

\subfloat{
	\begin{overpic}[width=0.75 \columnwidth,trim={0 2.3cm 0 2.5cm}]{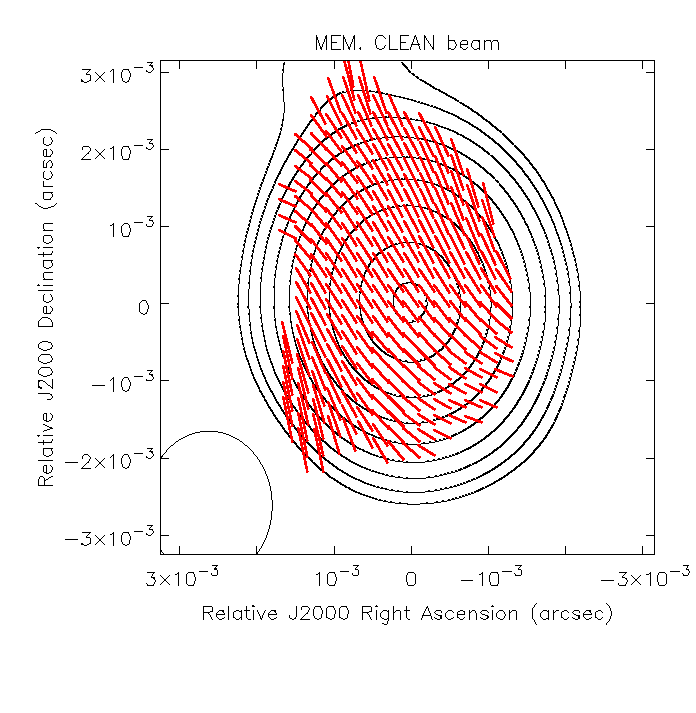}
		\put(25,78){(a)}
\end{overpic}
}
\subfloat{
	\begin{overpic}[width=0.75 \columnwidth,trim={0 2.3cm 0 2.5cm}]{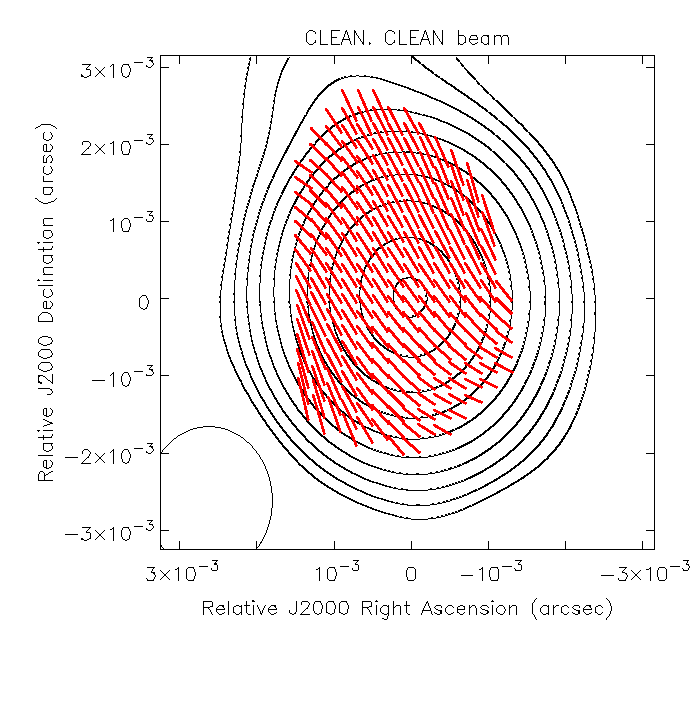}
		\put(25,78){(b)}
\end{overpic}
}

\subfloat{
	\begin{overpic}[width=0.75 \columnwidth,trim={0 2.3cm 0 2.2cm}]{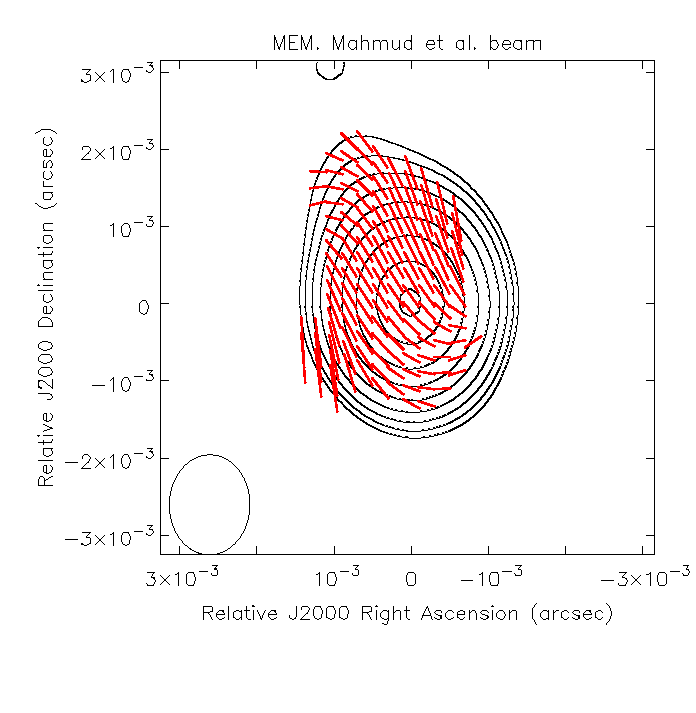}
			\put(25,78){(c)}
\end{overpic}
}
\subfloat{
	\begin{overpic}[width=0.75 \columnwidth,trim={0 2.3cm 0 2.2cm}]{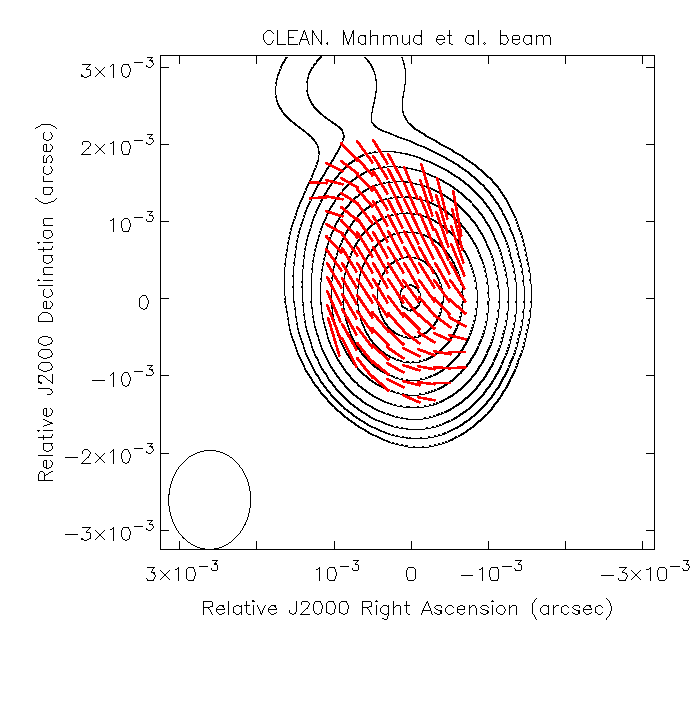}
			\put(25,78){(d)}
\end{overpic}
}

\subfloat{
	\begin{overpic}[width=0.75 \columnwidth,trim={0 2.3cm 0 2.2cm}]{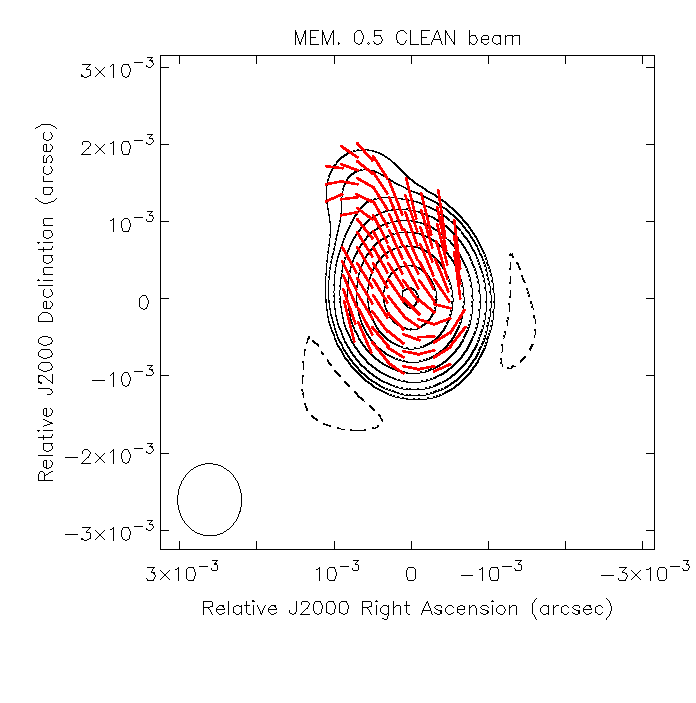}
			\put(25,78){(e)}
\end{overpic}
}
\subfloat{
	\begin{overpic}[width=0.75 \columnwidth,trim={0 2.3cm 0 2.2cm}]{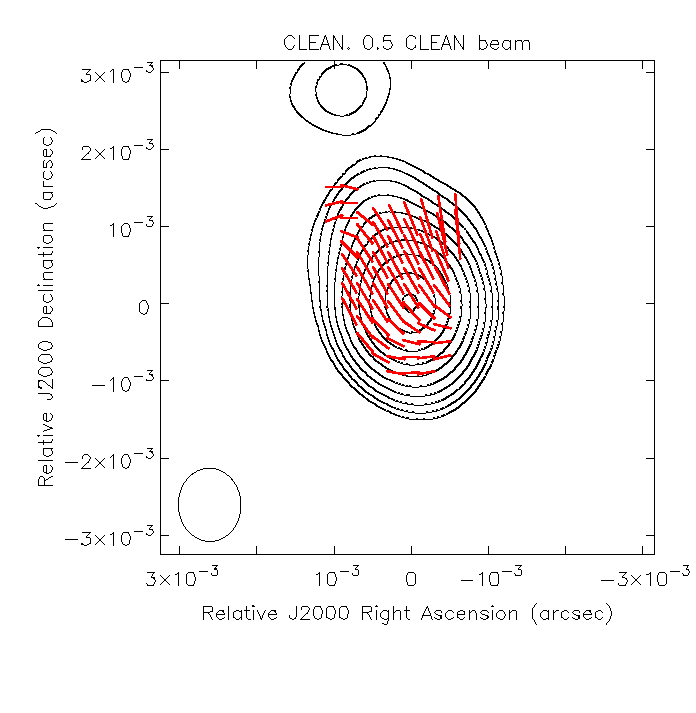}
			\put(25,78){(f)}
\end{overpic}
}

\subfloat{
	\begin{overpic}[width=0.75 \columnwidth,trim={0 2.3cm 0 2.2cm}]{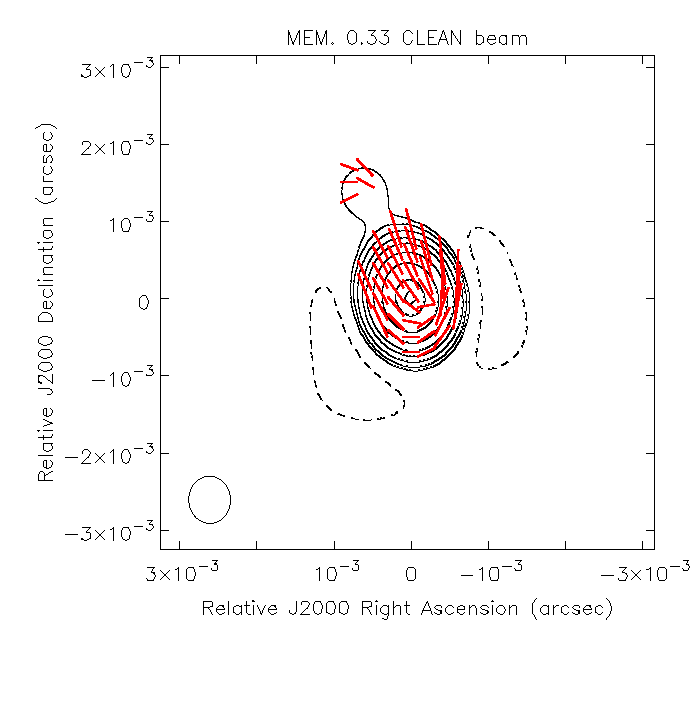}
			\put(25,78){(g)}
\end{overpic}
}
\subfloat{
	\begin{overpic}[width=0.75 \columnwidth,trim={0 2.3cm 0 2.2cm}]{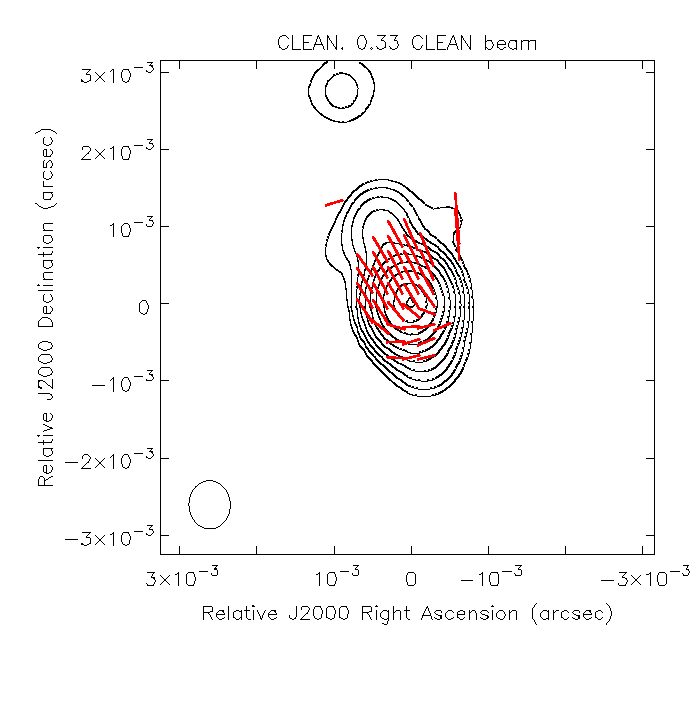}
			\put(25,78){(h)}
\end{overpic}
}
\vspace{-0.3cm}
\caption[Polarisation angle maps.]{Polarisation angle maps of 0716+714 at 4.6 GHz made with PMEM (left) and with the CLEAN task in {\sc CASA}  (right). 
The contours show Stokes $I$ and the sticks the polarisation angles. The
FWHM of the restoring beams used are: (a, b) those of the full CLEAN beam (c, d) those of the Mahmud et al. (2013) beam, (e, f) 0.50 times the full CLEAN FWHM, 
(g, h) 0.33 times the full CLEAN FWHM. Contours increase in powers of two with a final contour at 95\% of the peak. The bottom contours are (top to bottom) CLEAN: 2.64~mJy, 2.59~mJy, 2.54~mJy, 2.48~mJy; MEM: 2.64~mJy, 2.52~mJy, 4.75~mJy, 5.64~mJy. Negative contours are dashed.}
\label{fig-0716-pang}
\end{center}
\end{figure*}

\begin{figure*}

\begin{center}
\begin{minipage}[][][b]{.95\columnwidth}

\hspace{-2.5cm}
\subfloat{
	\begin{overpic}[width=0.8 \columnwidth,trim={0 2.3cm 0 2.5cm}]{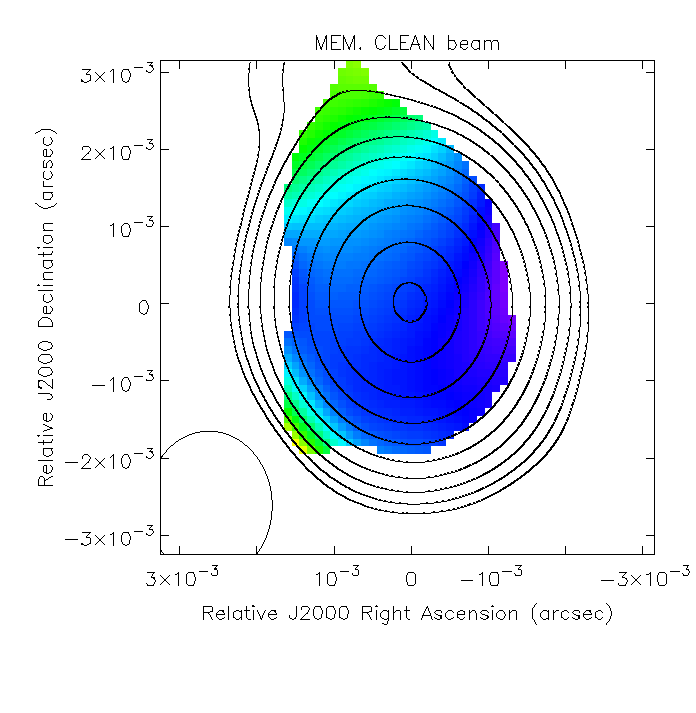}
	\put(25,78){(a)}
\end{overpic}
}
\subfloat{
	\begin{overpic}[width=0.8 \columnwidth,trim={0 2.3cm 0 2.5cm}]{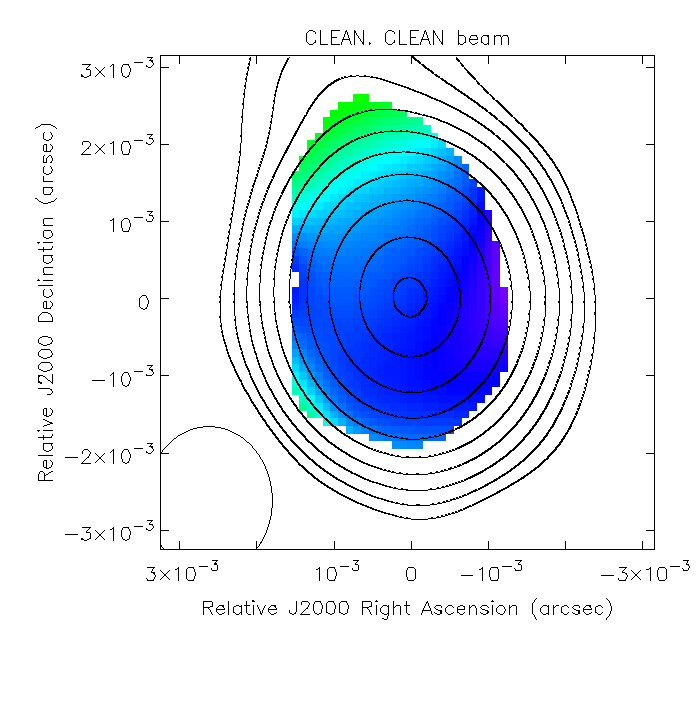}
		\put(25,78){(b)}
\end{overpic}
}

\hspace{-2.5cm}
\subfloat{
	\begin{overpic}[width=0.8 \columnwidth,trim={0 2.3cm 0 2.2cm}]{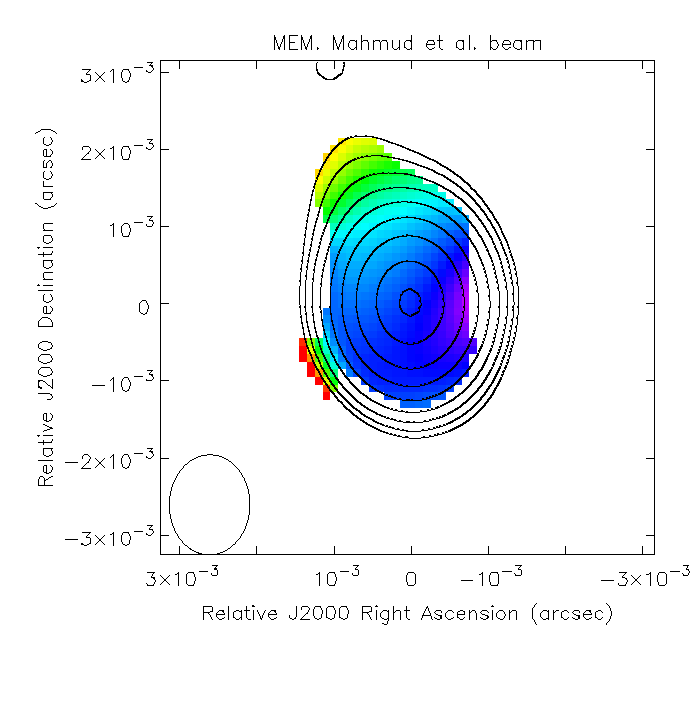}
			\put(25,78){(c)}
\end{overpic}
}
\subfloat{
	\begin{overpic}[width=0.8 \columnwidth,trim={0 2.3cm 0 2.2cm}]{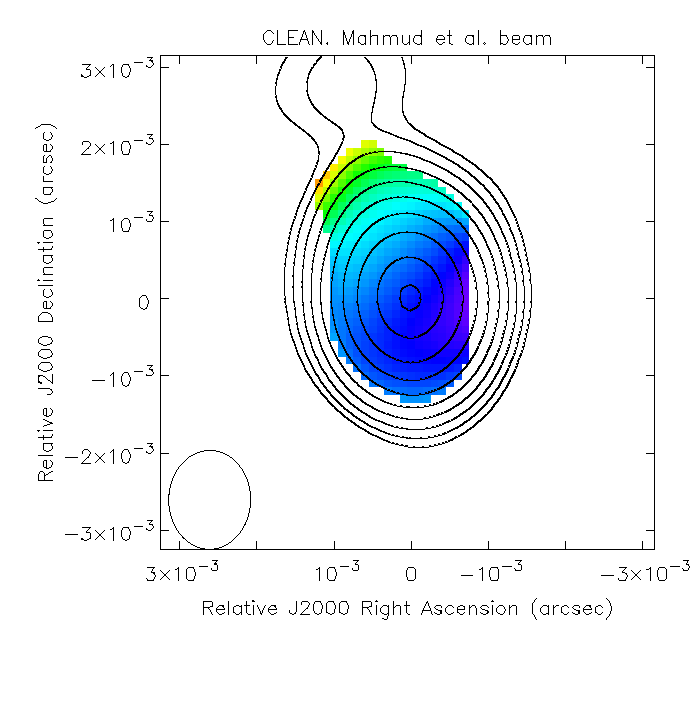}
			\put(25,78){(d)}
\end{overpic}
}

\hspace{-2.5cm}
\subfloat{
	\begin{overpic}[width=0.8 \columnwidth,trim={0 2.3cm 0 2.2cm}]{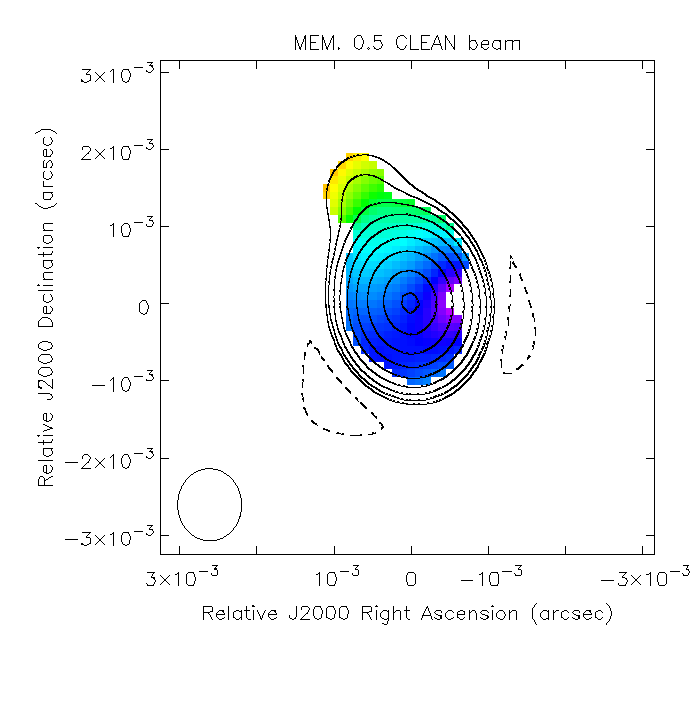}
			\put(25,78){(e)}
\end{overpic}
}
\subfloat{
	\begin{overpic}[width=0.8 \columnwidth,trim={0 2.3cm 0 2.2cm}]{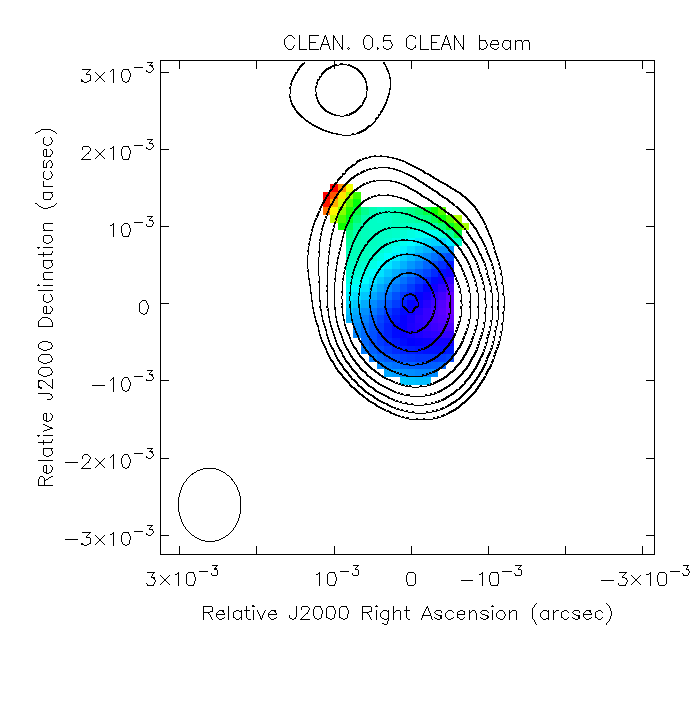}
			\put(25,78){(f)}
\end{overpic}
}

\hspace{-2.5cm}
\subfloat{
	\begin{overpic}[width=0.8 \columnwidth,trim={0 2.3cm 0 2.2cm}]{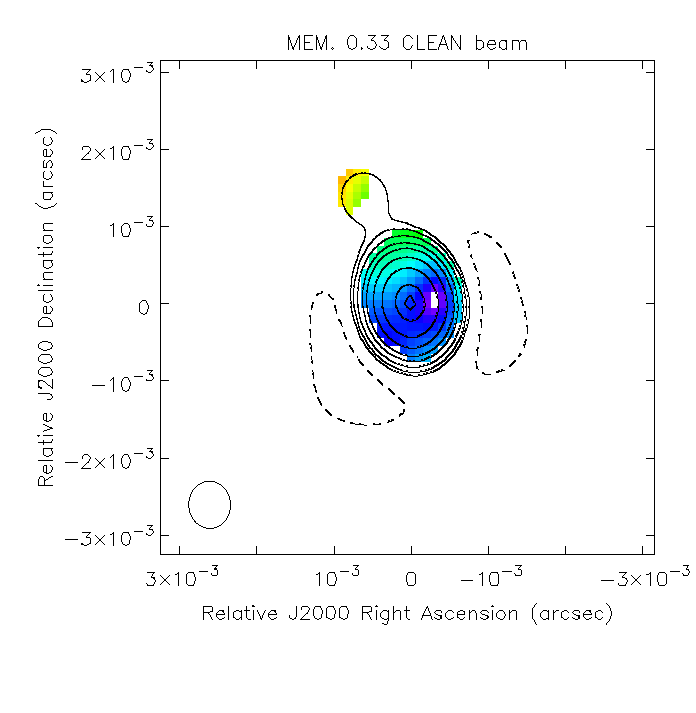}
			\put(25,78){(g)}
\end{overpic}
}
\subfloat{
	\begin{overpic}[width=0.8 \columnwidth,trim={0 2.3cm 0 2.2cm}]{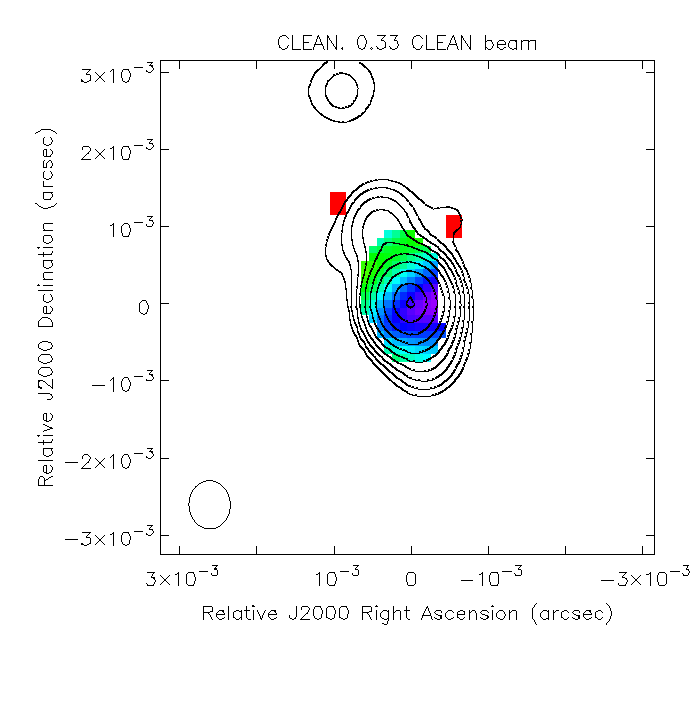}
			\put(25,78){(h)}
\end{overpic}
}
\end{minipage}
\begin{minipage}{.2\columnwidth}
\subfloat{
	\hspace{2.5cm}
	\includegraphics[width= \columnwidth]{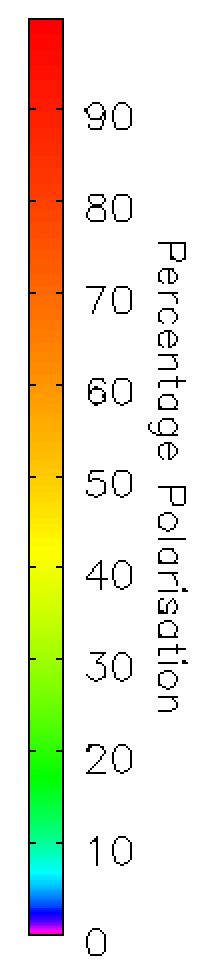}
}
\end{minipage}
\vspace{-0.3cm}
\caption[Percentage polarisation maps.]{Percentage polarisation maps of 0716+714 at 4.6 GHz made with the CLEAN task in {\sc CASA}  and the MEM as implemented in PMEM. The $I$ contours are the same as those in Figure \ref{fig-0716-pang}, while the colour scale indicates the percentage polarisation. Percentage polarisation was clipped at $3~\sigma$ in the MEM maps, and at $3.5~\sigma$ in the CLEAN maps. The restoring beams are the same as those indicated in \ref{fig-0716-pang}. }
\label{fig-0716-fpol}

\end{center}
\end{figure*}

\begin{figure*}
\begin{center}
\begin{minipage}[][][b]{.95\columnwidth}

\hspace{-2.5cm}
\subfloat{
	\begin{overpic}[width=0.8 \columnwidth,trim={0 3cm 0 1cm}]{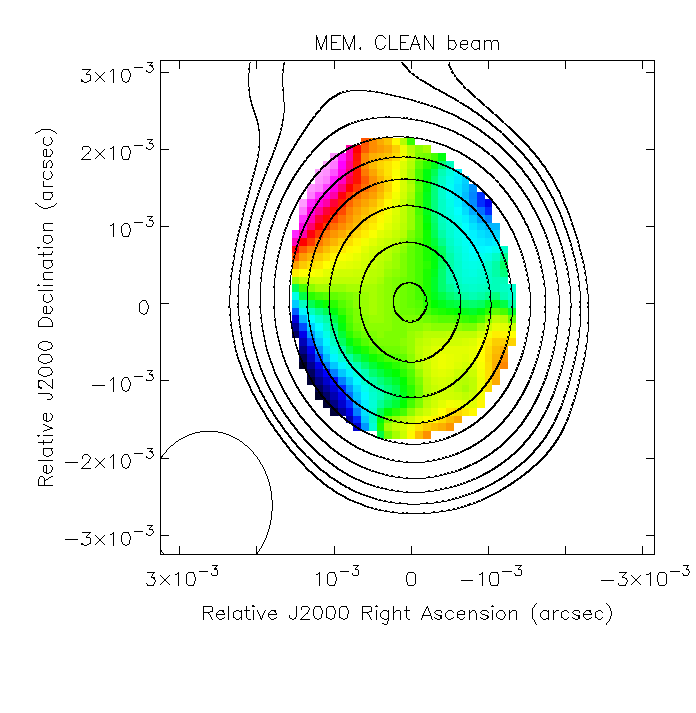}
				\put(25,75){(a)}
	\end{overpic}
}
\subfloat{
	\begin{overpic}[width=0.8 \columnwidth,trim={0 3cm 0 1cm}]{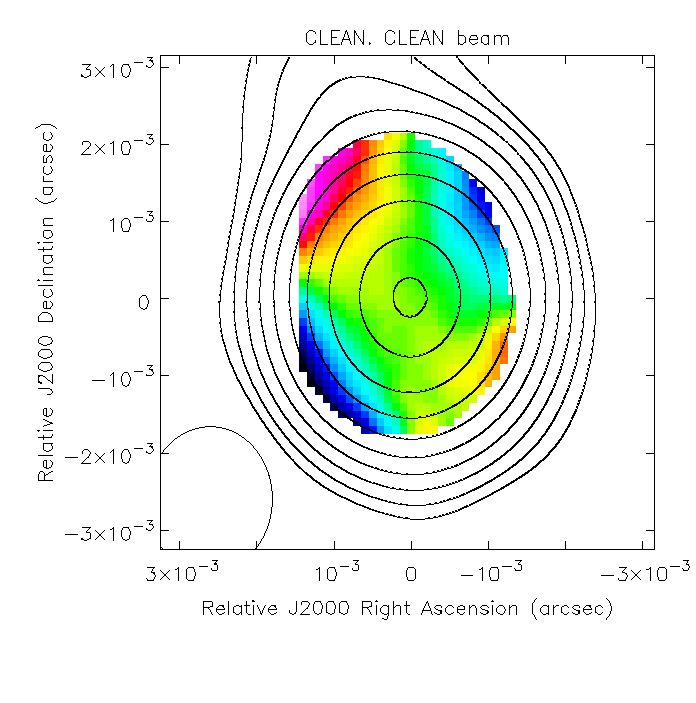}
					\put(25,75){(b)}
	\end{overpic}
}

\hspace{-2.5cm}
\subfloat{
	\begin{overpic}[width=0.8 \columnwidth,trim={0 3cm 0 1cm}]{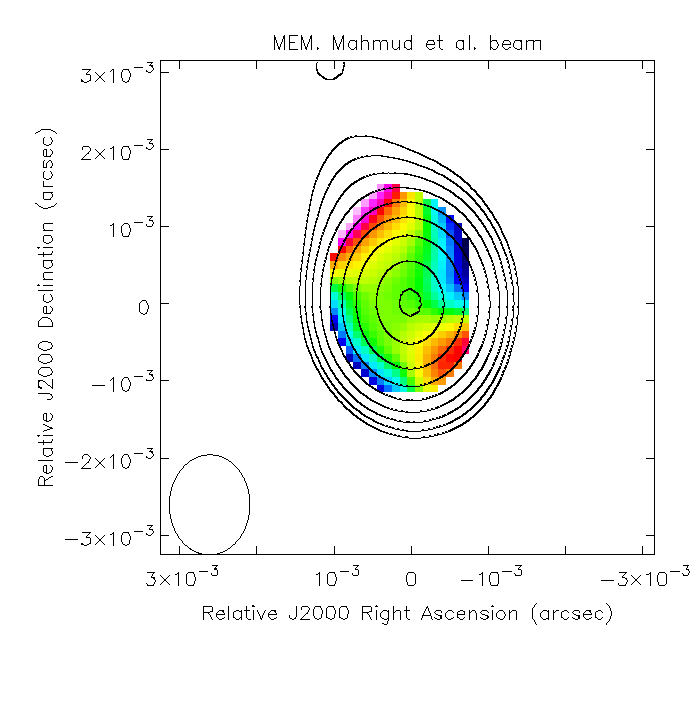}
					\put(25,75){(c)}
	\end{overpic}
}
\subfloat{
	\begin{overpic}[width=0.8 \columnwidth,trim={0 3cm 0 1cm}]{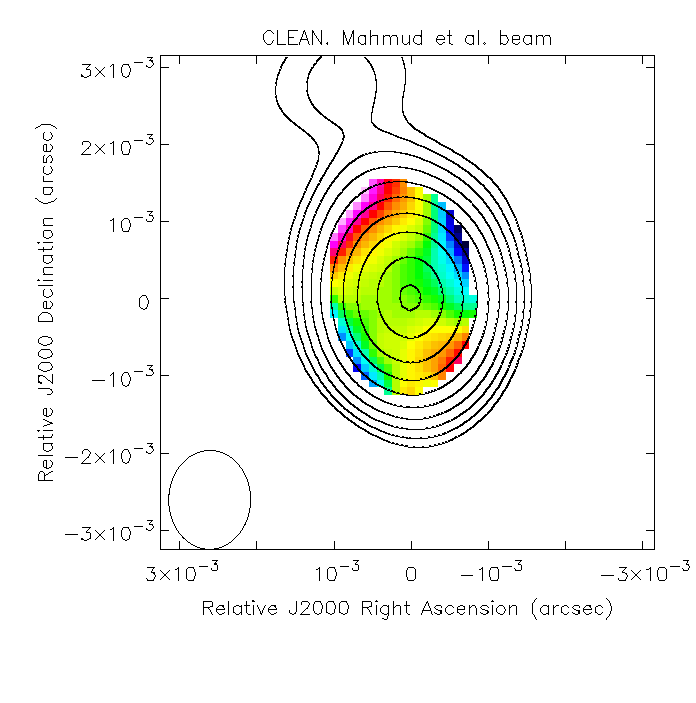}
					\put(25,75){(d)}
	\end{overpic}
}

\hspace{-2.5cm}
\subfloat{
	\begin{overpic}[width=0.8 \columnwidth,trim={0 3cm 0 1cm}]{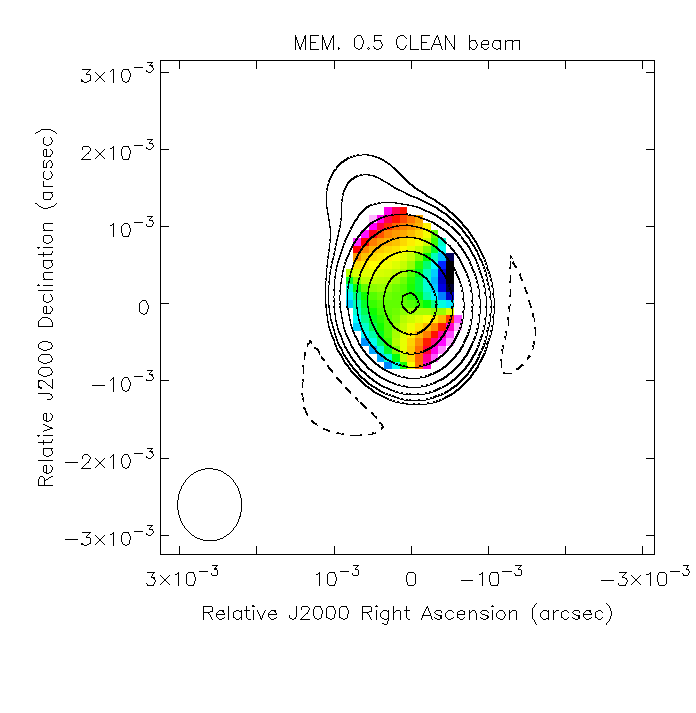}
					\put(25,75){(e)}
	\end{overpic}
}
\subfloat{
	\begin{overpic}[width=0.8 \columnwidth,trim={0 3cm 0 1cm}]{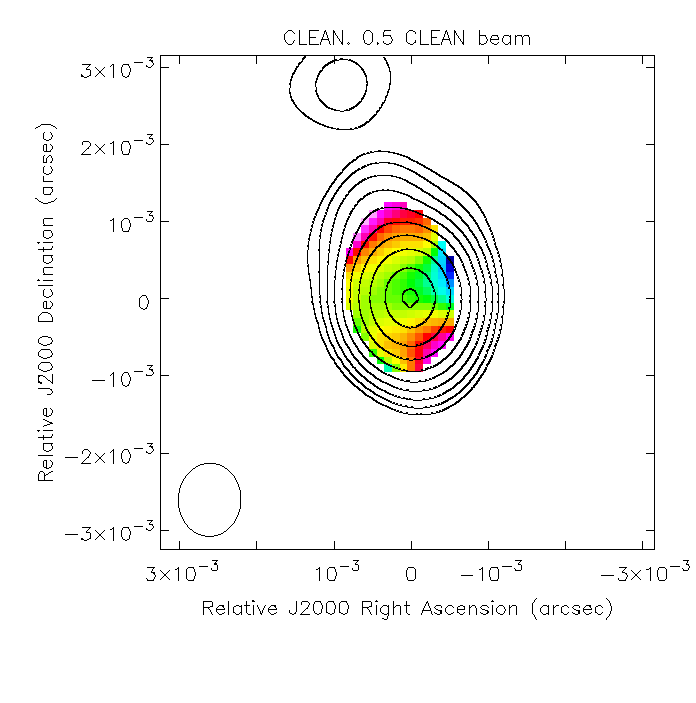}
					\put(25,75){(f)}
	\end{overpic}
}

\hspace{-2.5cm}
\subfloat{
	\begin{overpic}[width=0.8 \columnwidth,trim={0 3cm 0 1cm}]{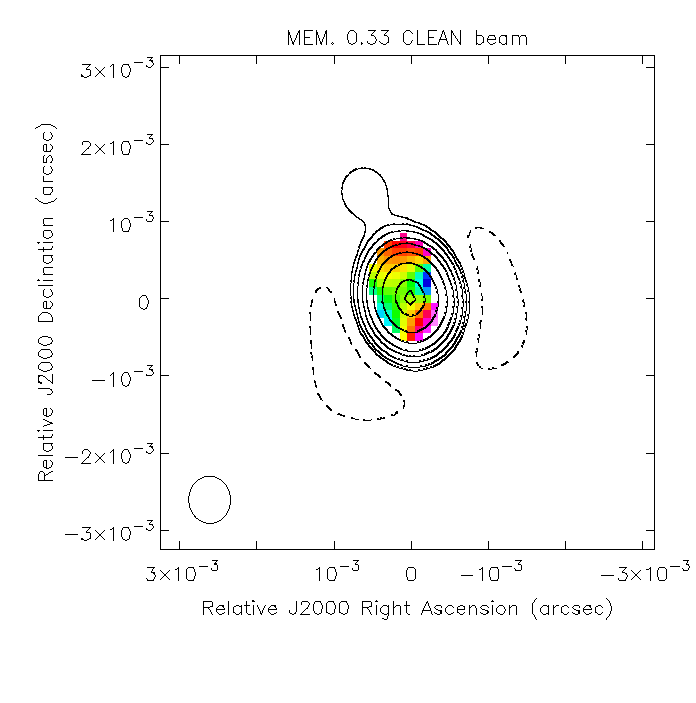}
					\put(25,75){(g)}
	\end{overpic}
}
\subfloat{
	\begin{overpic}[width=0.8 \columnwidth,trim={0 3cm 0 1cm}]{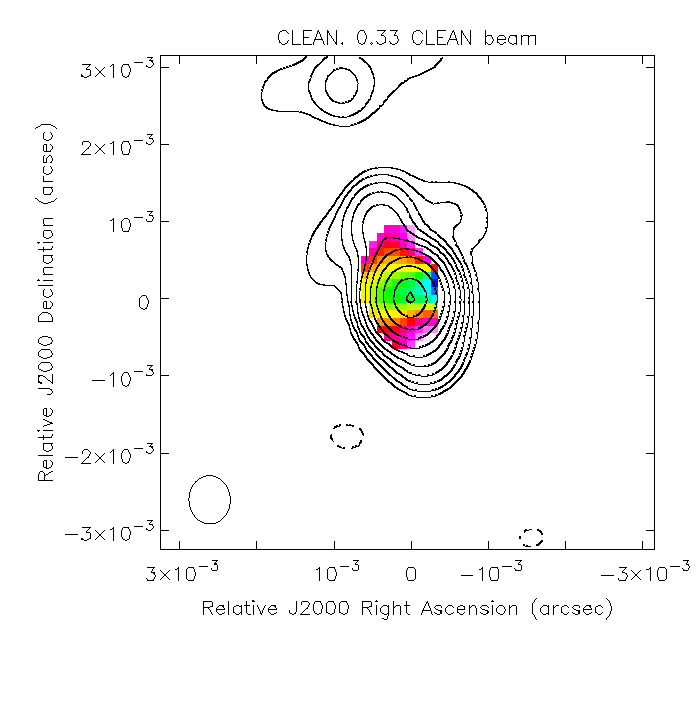}
					\put(25,75){(h)}
	\end{overpic}
}

\end{minipage}
\begin{minipage}{.2\columnwidth}
\subfloat{
	\hspace{2.5cm}
	\includegraphics[width= \columnwidth]{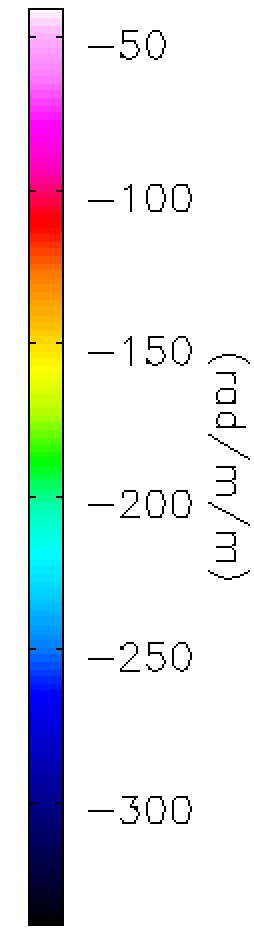}
}
\end{minipage}

\caption[RM maps]{Faraday rotation measure maps of 0716+714. The 4.6~GHz
$I$ contours are the same as those in Figure \ref{fig-0716-pang}, while the colour scale represents the Faraday rotation measure in rad/m$^2$. The restoring beams are the same as those indicated in \ref{fig-0716-pang}.}
\label{mem-obs-0716-RM}
\end{center}
\end{figure*}

\begin{figure}
\begin{center}
\subfloat{
	\includegraphics[width= \columnwidth]{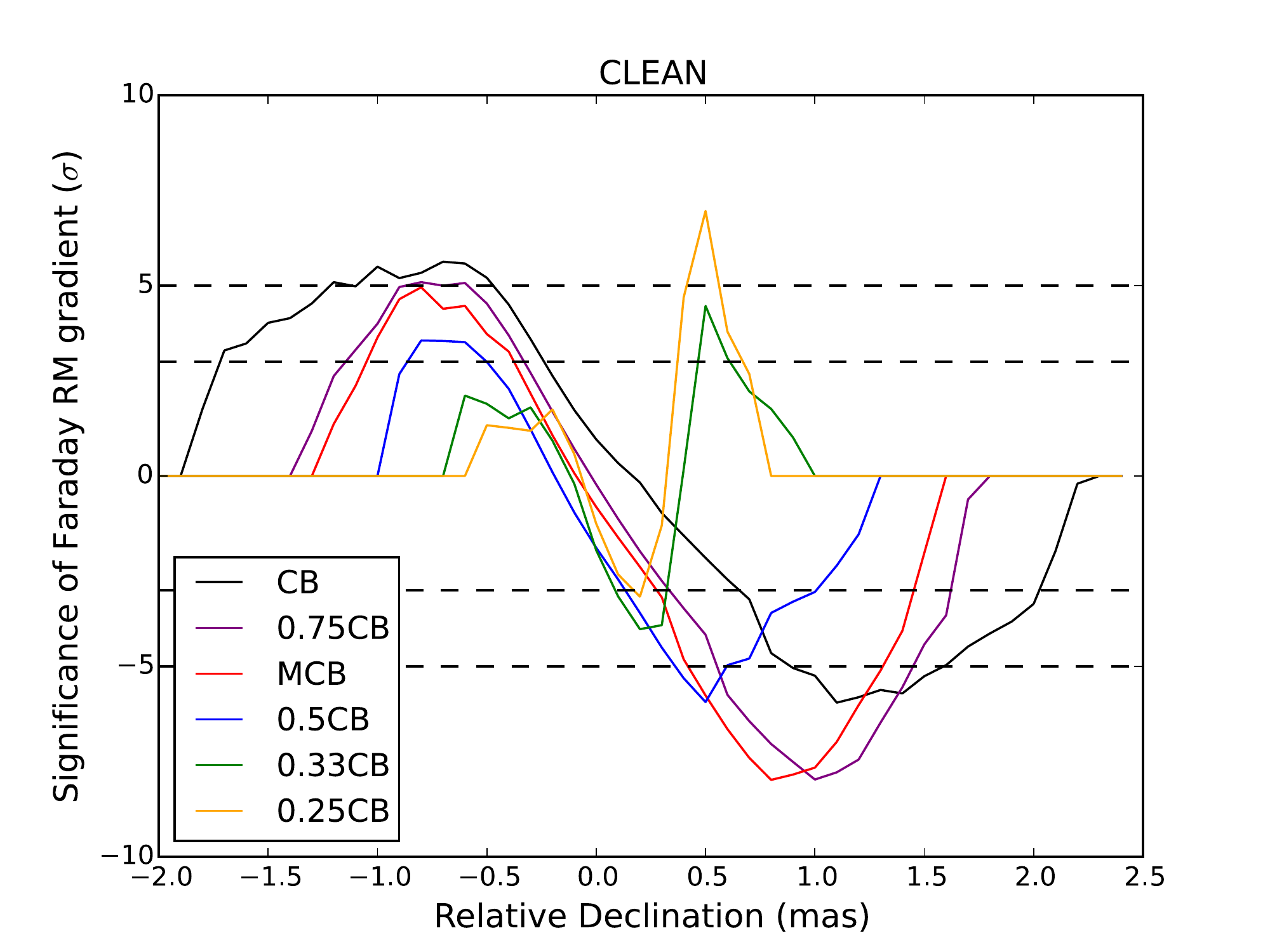}
}

\subfloat{
	\includegraphics[width= \columnwidth]{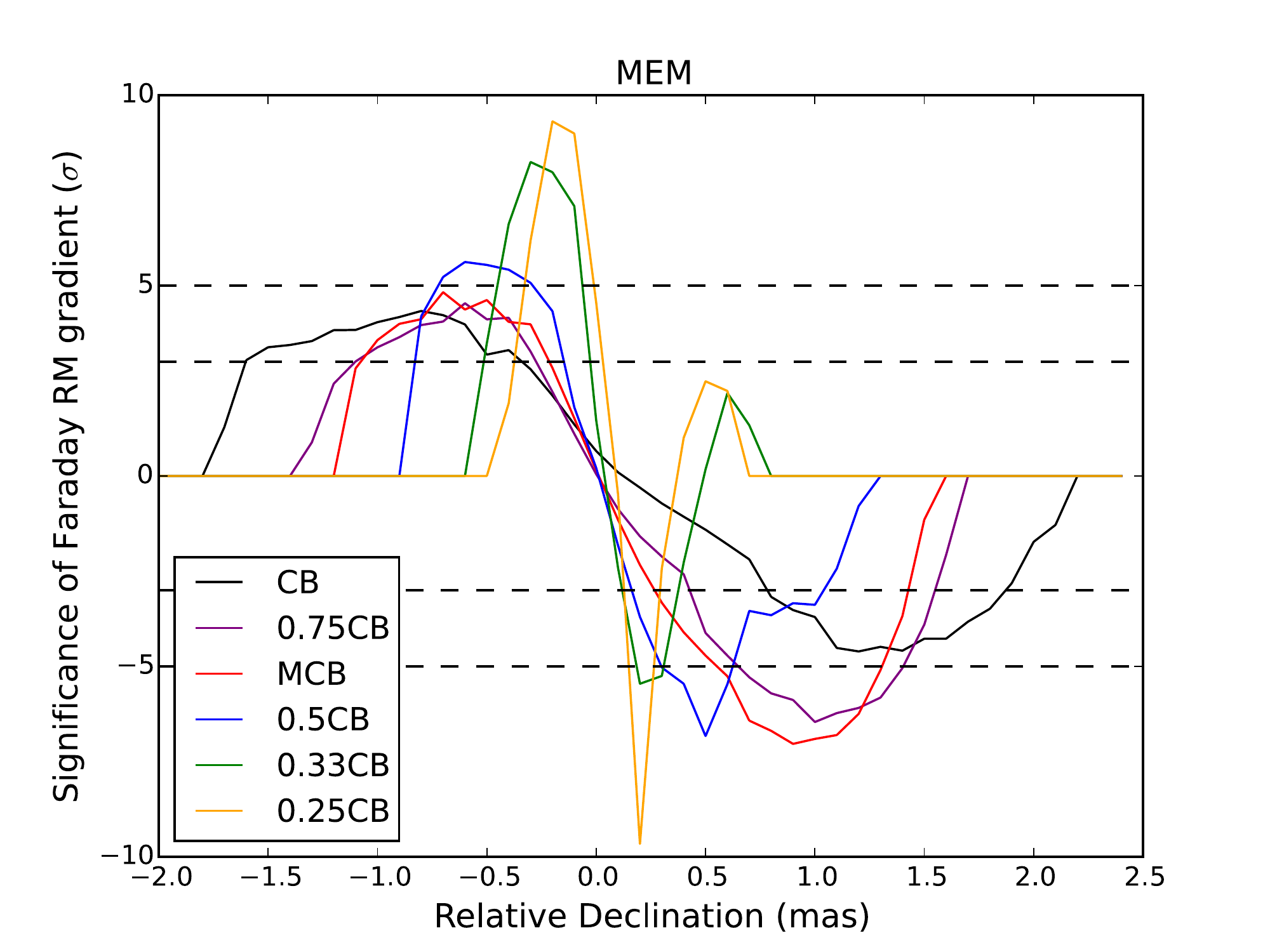}
}
\caption[Significance plot]{The significance of the gradients in Faraday rotation measure transverse to the jet direction in 0716+714. A positive significance value indicates a left-to-right increasing gradient in Figure \ref{mem-obs-0716-RM}, while a negative significance indicates a gradient running in the opposite direction. The x-axis plots the distance along the jet measured in mas from the centre of the map. The colours indicate the major axis of the restoring beam used.}
\label{fig-obs-rm-sig}
\end{center}
\end{figure}

PMEM was used to deconvolve the same 4.6~GHz, 5.1~GHz, 7.9~GHz, 8.9~GHz, 12.9~GHz and 15.4~GHz VLBA data as were used by \citet{Mahmud2013}. A standard CLEAN beam of $1.896 \times 1.624$ mas$^{2}$ with a position angle of 2.08$^{\circ}$ was found for the 4.6~GHz data. Restoring beams with both the major and minor axes scaled by a factor of 1.0, 0.60, 0.50 and 0.33 were then used to make maps. The exact beam used by Mahmud et al., $1.28 \times 1.06$ mas$^{2}$, position angle -0.8$^{\circ}$, was also examined. In all cases with a Briggs robustness parameter of $0$. 

In preparation for the PMEM imaging of 0716+714, images of the dirty maps and beams were made for all six frequencies using {\sc CASA}'s clean task. PMEM was then run on each frequency with the fluxes achieved by earlier CLEAN imaging used as the best guesses for the total flux and final noise parameters. Flux conservation was turned off, a suitable number of edge pixels were ignored and a mask was used. All other parameters were left at their default settings.

Corresponding PMEM and CLEAN $Q$ and $U$ images were made at each of the 6 frequencies, and polarisation angle maps were generated in {\sc CASA}. Following \citet{Mahmud2013}, each polarisation angle map was corrected for local galactic Faraday rotation using the data from \citet{Pushkarev2001}. The same electric vector position angle corrections used by \citet{Mahmud2013} were also applied.

Figure \ref{fig-0716-pang} shows the resulting 4.6-GHz PMEM (left) and CLEAN (right) Stokes $I$ images with polarisation angle sticks superposed. Maps are presented with restoring beam FWHM equal to: (a, b) those of the full CLEAN beam (c, d) those of the Mahmud et al. (2013) beam, (e, f) 0.50 times the full CLEAN FWHM and
(g, h) 0.33 times the full CLEAN FWHM. The PMEM and CLEAN maps made with the full CLEAN beam and the Mahmud et al. (2013) beam are essentially quite similar, while the two maps are slightly
different with a restoring beam of 0.5 CLEAN FWHM, with MEM showing more polarised emission further along the jet. At both 0.5 and 0.33 CLEAN resolution significant negative Stokes I emission can be seen in the off-source region of the MEM maps, a result of MEM's inability to describe negative flux (see Appendix \ref{app} for details). It is difficult to quantitatively determine which algorithm is performing better at high resolutions, though MEM at least shows more polarisation information and appears to have less suspicious-looking ``patches'' of polarised emission and can thus be regarded as more plausible than CLEAN. This is in keeping with the conclusion of \citep{Gull1984} that while MEM maps may not quantitatively be more accurate than alternative methods, they might hope to be ``preferrable'' via some other metric.

Figure \ref{fig-0716-fpol} shows the 4.6-GHz PMEM (left) and
CLEAN (right) Stokes $I$ images (contours) with fractional polarisation superposed as a colour scale. Maps restored using the same beams as in
Fig.~\ref{fig-0716-pang} are shown. Again, the PMEM and CLEAN maps made with beams corresponding to the full CLEAN beam and the Mahmud et al. (2013) beam are essentially 
identical. The PMEM map made with 0.50 CLEAN resolution
looks slightly more reliable than the corresponding CLEAN map (the CLEAN
map has started to show a small region with implausibly high fractional
polarisation). The two maps made with 0.33 CLEAN resolution
beams are appreciably different: the PMEM map continues to yield physically
plausible degrees of polarisation for both the core and the jet, whereas
the CLEAN map shows reliable fractional polarisation only in the core,
with a spurious increase in fractional polarisation to the south of the
map peak. These differences, and similar differences in the corresponding PMEM and CLEAN maps obtained at the other 5 frequencies, illustrate some of the advantages in simultaneously deconvolving all Stokes parameters using MEM. It is particularly notable that these advantages are available ``for free'' -- no additional observations are needed, and little additional processing is required.

Polarisation-angle maps at the six frequencies made with a given restoring beam were then used to create a Faraday rotation measure (RM) map by making linear fits to the polarisation angles plotted as functions of the wavelength squared for each pixel. This fitting was conducted using {\sc CASA}'s ``rmfit'' routine.

Figure \ref{mem-obs-0716-RM} shows the 4.6-GHz PMEM (left) and
CLEAN (right) Stokes $I$ images (contours) with the RM superposed as a colour scale. Maps restored using the same beams as in
Fig.~\ref{fig-0716-pang} are shown. Note that the region just south of the map peak corresponds to emission from the core region, and the region north of the map peak to emission from the inner jet. The two sets of RM maps are very
similar down to restoring beams of with 0.5 times the CLEAN FWHM. The RM maps created
using restoring beams of 0.33 CLEAN FWHM show significant differences,
with the PMEM RM map looking more similar to those obtained with larger
restoring beams than is the case for the CLEAN RM map. This smooth transition between different resolutions is a feature of the continuous multi-stokes modelling applied by MEM which appears to  persist to multi-frequency data products such as Faraday RM maps. The PMEM RM map obtained with \citet{Mahmud2013} beam is very similar
to the CLEAN map presented in their paper, with transverse gradients in opposite directions visible to the North and South of the map peak. If anything, this structure appears to be slightly more prominent in the PMEM map. It is notable that the RM values themselves have varied significantly with the changes in the restoring beam, but the direction of the gradient has been preserved (see also the appendix of \citet{Mahmud2013}).

The results of the Monte Carlo simulations discussed in Section \ref{section-montecarlo} suggest that mild super-resolution can be reliable in both CLEAN and MEM, and that MEM in particular may give accurate polarisation details at high resolution. Figure \ref{fig-obs-rm-sig} compares the significance of the transverse gradient in Faraday RM across the jet of 0716+714 for a variety of resolutions and for both algorithms. The value of the transverse gradient was determined by taking the difference between the first and last RM values in a straight horizontal line transverse to the jet direction such that $RM_{grad} = RM_{2} - RM{1}$ (making a horizontal cut of the RM values shown in each plot of Figure \ref{mem-obs-0716-RM} and taking the first and last pixels to be $RM_{1}$ and $RM_{2}$ respectively). The uncertainty in the gradient was then calculated using standard propagation of errors; i.e. $\Delta RM_{grad}^{2} = \Delta RM_{1}^{2} + \Delta RM_{1}^{2}$. The significance of the gradient was then found by dividing the gradient value by its uncertainty. Where this significance is positive it indicates a gradient that increase from left to right as indicated in Figure \ref{mem-obs-0716-RM}, while a negative significance indicates a gradient in the opposite direction. The significance of each gradient was plotted for each horizontal row of pixels, beginning just before the start of non-masked RM values in the CLEAN beam resolution core and ending shortly after the termination of non-masked RM values in the CLEAN beam resolution jet.

The information presented in Figure \ref{fig-obs-rm-sig} is in agreement with the general conclusions of the Monte Carlo simulations in Section \ref{section-montecarlo}. Both algorithms show broadly similar performance up to a resolution of about 0.5 the standard CLEAN FWHM, after which the difference between the two algorithms becomes much greater. The gradient in the core (the southern region in Fig. \ref{mem-obs-0716-RM}, negative declinations in Fig. \ref{fig-obs-rm-sig}) is detected with significances of approximately 5.6, 5.1 and 5.0~$\sigma$ by CLEAN at 1.0, 0.75 and 0.60 times the CLEAN FWHM, and with a significance in excess about 3.6~$\sigma$ at 0.5 CLEAN FWHM. CLEAN does not detect a significant gradient in the lower part of the jet after this point. MEM detects the same gradient with less significance at the lowest resolutions (significances of 4.3~$\sigma$, 4.5~$\sigma$ and 4.8~$\sigma$ for 1.0, 0.75 and 0.6 times the full CLEAN beam), though goes on to detect it with increasing significance at the highest resolutions - reaching 5.6, 8.2 and 9.3~$\sigma$ at 0.5, 0.33 and 0.25 CLEAN FWHM, respectively.

A similar pattern is seen for the stronger, northern gradient across the inner jet. This gradient is in the opposite direction to the southern (core region) gradient and is detected by CLEAN with a significance of 6.0~$\sigma$ at CLEAN beam resolution. The significance of the gradient then increases with resolution, reaching 8~$\sigma$ at 0.75 and 0.6 CLEAN FWHM resolution, before declining again to 5.9~$\sigma$, 4.0~$\sigma$ and 3.2~$\sigma$ as the resolution goes to 0.5, 0.33 and 0.25 CLEAN FWHM respectively.

MEM's performance relative to CLEAN is similar to the first gradient - confirming the gradient, but with slightly lower significances at the lowest three resolutions, before rising in significance at the highest resolutions while CLEAN falls off. The significances are 4.6, 6.5, 7.0, 6.8 and 5.5~$\sigma$ for resolutions of 1.0, 0.75, 0.60, 0.5, and 0.33 CLEAN FWHM respectively. Again, the peak significance detected by MEM is at 0.25 CLEAN beam resolution with a value of 9.7~$\sigma$.

At the highest two resolutions an apparently significant third gradient is detected at approximately 4.5~$\sigma$ and 7.0~$\sigma$ for 0.33 and 0.25 CLEAN FWHM, respectively. The Monte Carlo results suggest a high degree of caution be applied to this result however. Notably, though apparently highly significant when imaged with CLEAN at the highest resolutions, this dubious third gradient is evident in the MEM maps, but with significances of less than 3~$\sigma$. Thus, with the understanding gained from the Monte Carlo simulations, this third gradient can be rejected as an artefact.

High resolution imaging with both CLEAN and MEM suggests that the gradient reversal seen in 0716+714 is significant and that an erroneous gradient due to systematic algorithmic error can be ruled out in this case. It does not exclude the possibility that at least one of the gradients could arise from an external Faraday screen rather than the nested helical magnetic field structure proposed by \citet{Mahmud2013} and, without sufficient bandwidth to perform a full Faraday rotation measure synthesis analysis \citep{Brentjens2005}, the Faraday depth of the screen responsible for the rotation cannot be estimated. There is however significant evidence that helical magnetic fields are detectable across the jets of many different AGN \citep{Asada2002,Gabuzda2004,Zavala2004,Gabuzda2007,Asada2008,Asada2008a,Gabuzda2014,Gabuzda2015a} - consistent with the interpretation of the observed reversal as arising from a nested helix structure.

\section{Conclusions}

This paper has described the production, testing and first application of a new Maximum Entropy Method deconvolution code, PMEM, suitable for VLBI polarisation observations. Section \ref{section-montecarlo} outlined the results of Monte-Carlo simulations of PMEM's ability to deconvolve both total intensity and polarisation maps of a realistic VLBI observation of two model sources. Two major conclusions can be drawn from these simulations. Firstly, both CLEAN and MEM yielded reliable results when resolving beams with a FWHM down to half of the standard CLEAN FHWM were used, even in regions of relatively low signal to noise ratio (this corresponds to a beam one quarter the size of the standard CLEAN beam). Thus, the use of modest super-resolution does not yield untrustworthy results, especially if the CLEAN and MEM maps produced at such a resolution agree with each other. Secondly, PMEM can outperform CLEAN when restored with beams significantly smaller than the full CLEAN beam, down to $\frac{1}{3}$ to $\frac{1}{4}$  of the CLEAN FWHM in some cases - though such performance is difficult to predict and quantify.

This paper has also presented PMEM maps made using real VLBA data for the AGN 0716+714. The fractional polarisation maps made using PMEM and CLEAN indicate the additional small-scale polarisation information that can be obtained by using PMEM alongside the CLEAN algorithm. The PMEM Faraday rotation measure map of 0716+714 is the first MEM RM map that has been made, to our knowledge. This MEM RM map confirms the validity of the results presented in \citet{Mahmud2013}, and demonstrates that those results were not been affected by any systematic bias due to the CLEAN algorithm. The plots in Figure \ref{fig-obs-rm-sig} also show the worth of investigating polarised VLBI sources with multiple algorithms and resolutions. PMEM will be applied to additional VLBA data for other AGN and the results will be presented in further papers.

\section*{Acknowledgements}

The authors would like to thank the anonymous referee for their thoughtful report, which has lead to numerous improvements in this paper. This research was funded by an EMBARK scholarship from the Irish Research Council (IRC) and Science Foundation Ireland grant 13/ERC/I2907. The authors wish to acknowledge the DJEI/DES/SFI/HEA Irish Centre for High-End Computing (ICHEC) for the provision of computational facilities and support. The authors would like to thank Quentin Michelas for developing the CASA interface for PMEM.

\bibliographystyle{mnras}
\bibliography{MEM_Paper.bib}
\nocite{*}

\appendix

\section{PMEM details}
\label{app-pmem}

\subsection{Aliasing}

As repeated convolutions are required for every iteration of the MEM, it is computationally much more efficient to use the FFT method, which has a complexity of $O(n \log n)$, than perform a direct Discrete Fourier Transform (DFT) with a complexity of $O(n^2)$ \citep{Brault1971}.

This increase in speed can come at a price -- the FFT is susceptible to aliasing. When FFTs are used to convolve MEM models with the dirty beam, aliasing artefacts can appear as as artificially enhanced fluxes in outer pixels of the resulting convolved map. This effect can perpetuate itself when this false flux creates a local distortion in the residual maps calculated from the convolved map and the MEM actually begins to include the flux in its model of the source, leading to a runaway effect.

One way in which it is possible to reduce aliasing related to the use of the FFT is to ``zero-pad'' the images being transformed, i.e. to add extra pixels at the edge of the map to contain the higher visibilities that would otherwise go unsampled. However, in addition to the extra computation time needed in the FFT of a padded map, the potentially sharp features in such zero-padded maps can cause further aliasing (the optimisation of such a process is currently a field of active study). As the majority of sources do not suffer greatly from aliasing effects, it was found easier to clip the offending aliased pixels rather than zero pad the entire image. The amount of clipping required to remove aliasing effects can vary from a few pixels, to up to 5 or 10\% of the image and can be specified using the $N_{\mathrm{exclude}}$ parameter. In cases where the clipping needed becomes large, the dirty maps may need to be made with a larger size. This effectively zero pads the image, and gives more room to clip unneeded pixels from the edge.  Note that the clipping is performed in the model and residual maps, therefore the value of the final map at the outer $N_{\mathrm{exclude}}$ pixels is not reliable.

While the AIPS task ``VTESS'' automatically clips the outer 25\% of an image, PMEM allows the user to specify any value deemed appropriate. Such a value can be determined by examining the outer pixels of the model and residual maps for any signs of aliasing (higher or lower than expected flux) and setting the $N_{\mathrm{exclude}}$ parameter to exclude such pixels. Alternatively, making a large image and taking a similar approach to VTESS would allow a conservative number of pixels to be clipped without needing to examine the model and residual maps. PMEM also accepts a MEM mask, which operates in a similar way to a CLEAN mask and can be used to prevent sidelobes or aliasing artefacts from being included in the MEM model.

\subsection{Using PMEM}

PMEM has been tested on a variety of machines, from dual-core laptops with small amounts of RAM to 24-core servers. Typical run times for simulated and real data have been of the order of a small number of seconds, though this increases with image size and also depends on the source structure. Memory requirements are light enough to be negligible for all but the largest images ($> 10^{7}$ pixels). PMEM can be run from the command line via a parameter set file, from a simple graphical user interface, or via an interface to CASA through which it can be called like any CASA task. Default parameters are provided for the parameters discussed above, along with documentation on how the user might choose to vary them.

PMEM requires the user to provide FITS images of the dirty maps for each of the Stokes parameters to be imaged, as well as a FITS image of the dirty beam corresponding to the observation. The user is then asked to estimate the flux of the source and the final rms noise that might be achieved, as well as values for the parameters discussed in the sections above. This represents a degree of overhead compared to making a straight-forward CLEAN image -- indeed, the best way to estimate the final rms noise can be to make a CLEAN map first, however PMEM does not require any user interaction during the deconvolution process and often shows best results with no masking other than near the edge of the map to avoid aliasing.

It is suggested that the user begin by excluding the outer 5--10~\% of the image and setting $w_{p} = a_{factor} =  \Delta_{step} = 1$. If the final noise in the Stokes I maps is comparable to CLEAN and the residuals do not show strong features near the source then  $a_{factor}$ and $\Delta_{step}$ do not need to be changed. If the map does not appear to converge well, the user should first test restricting model emission by either using a mask or simply increasing the number of excluded pixels (this is particularly helpful when the residual map shows evidence of aliasing). If this also fails, the parameters $a_{factor}$ can be increased and $\Delta_{step}$ reduced to have MEM converge with smaller steps and reach a better solution. In cases where the Stokes I residuals appear low but significant residuals are still present in Stokes Q and U the parameter $w_{p}$ can be increased to achieve better polarisation maps - but it is strongly recommended to stay within the range $1<w_{p}<10$ to avoid making maps where with a poor Stokes I model.

The code, while numerically intensive compared to CLEAN and slower than non-polarisation implementations of MEM, runs in the order of seconds for maps of $10^{6}$ pixels. The results are written out into multiple FITS files along with a log of the deconvolution process. The following FITS files are generated for each of the Stokes parameters:

\begin{itemize}
\item The MEM model map.
\item The MEM model map convolved with the restoring beam.
\item The residual map (the difference between the MEM model convolved with the dirty beam and the dirty map).
\item The final MEM map. This is the MEM map convolved with the restoring beam with the appropriately scaled residual map added (by default PMEM and CLEAN do not apply any scaling).
\end{itemize}

\section{Monte Carlo simulations}
\label{app}

\subsection{Methodology details}

The model sources for the Monte Carlo simulations were constructed using the Python module ``Astropy'' and a typical Very Long Baseline Array (VLBA) ``snapshot'' $UV$ coverage for a relatively high-declination source observed at 4.6~GHz, shown in Figure \ref{mem-main-fig-uvcore}. A new C++ program (``simuv'') was written and used to generate simulated observations of the model sources using the $UV$ coverage in Figure \ref{mem-main-fig-uvcore}. Thermal noise was added to the visibilities in such a way as to create realistic noise levels in the final CLEAN maps. This was done by examining the root-mean-square (rms) deviation of the flux density in regions far from the source in real maps made using very similar $UV$ coverage and adding thermal noise to the model visibilities such that the resulting CLEAN maps had similar noise levels. The random thermal noise added to the visibilities was generated using the GNU Scientific Library random Gaussian function with zero mean and a user specified standard deviation, seeded with the current time multiplied by the process ID of the current CPU thread running simuv.

One hundred UV FITS datasets were generated for each model with different thermal noise added in each dataset. The data were loaded into {\sc CASA}  and the ``clean'' task used to generate a dirty map corresponding to each image and Stokes parameter as a FITS file. (Note that no cleaning was performed on these maps -- the {\sc CASA}  ``clean'' task was just used to make a dirty image.) All imaging was performed without any $UV$ tapering and with a Brigg's robustness level of 0, resulting in an even compromise between natural and uniform $UV$ weighting. A cell size of $0.1$ mas was used in all cases, providing a compromise between the issues discussed in Section \ref{sec-diag-hess} and having enough pixels across the full width at half maximum (FWHM) of the CLEAN beam ($1.84 \times 1.64$ mas, in $-78^{\circ}$) to adequately test the ability of MEM and CLEAN to reconstruct structures on the scales probed. The MEM imaging was performed using estimates of the final rms noise from CLEAN images made in the usual way and the known fluxes of the model source. It was found that better images were achieved with flux conservation turned off ($\gamma = 0$) for the triple Gaussian source, while conserving flux gave better results for the continuous bent jet model. The estimated flux from a CLEAN image at full resolution was used to initialise the MEM flux in both cases.

The final MEM and CLEAN maps were convolved with the full fitted CLEAN beam, and with scaled beams corresponding to $\frac{3}{4}$, $\frac{1}{2}$, $\frac{1}{3}$ and $\frac{1}{4}$  of the CLEAN beam.
Results from the standard (non-multiscale) Cotton-Schwab CLEAN algorithm as implemented by the ``clean'' task in {\sc CASA}  were generated using a {\sc CASA}  script and standard CLEAN imaging techniques. One realization of the model map was imaged manually and used to set the threshold parameter -- the lowest CLEAN component flux allowed, for the imaging script to have three times the background noise observed in the manual image. The maximum number of iterations of the CLEAN algorithm was set high enough that the threshold parameter was the limiting factor in the automated CLEAN performed by the script. A gain value of $0.1$ and a CLEAN mask encompassing the source region in the image was used in each case. 

Maps corresponding to each Stokes parameter and beam size were compared with the model maps convolved with the corresponding beam. Distributions of the difference between the model and imaged fluxes in both total and
polarised flux in regions of interest on the source were made and the performance of the two algorithms compared. Distributions of the fractional polarisation $m = \sqrt{Q^{2} + U^{2}}/I$ and the polarisation angle $\chi = \frac{1}{2} \arctan{U/Q}$ were also generated.

\subsection{Results}

\begin{figure*}
\begin{center}
\subfloat[Total $I$]{
	\includegraphics[width=  1.0\columnwidth]{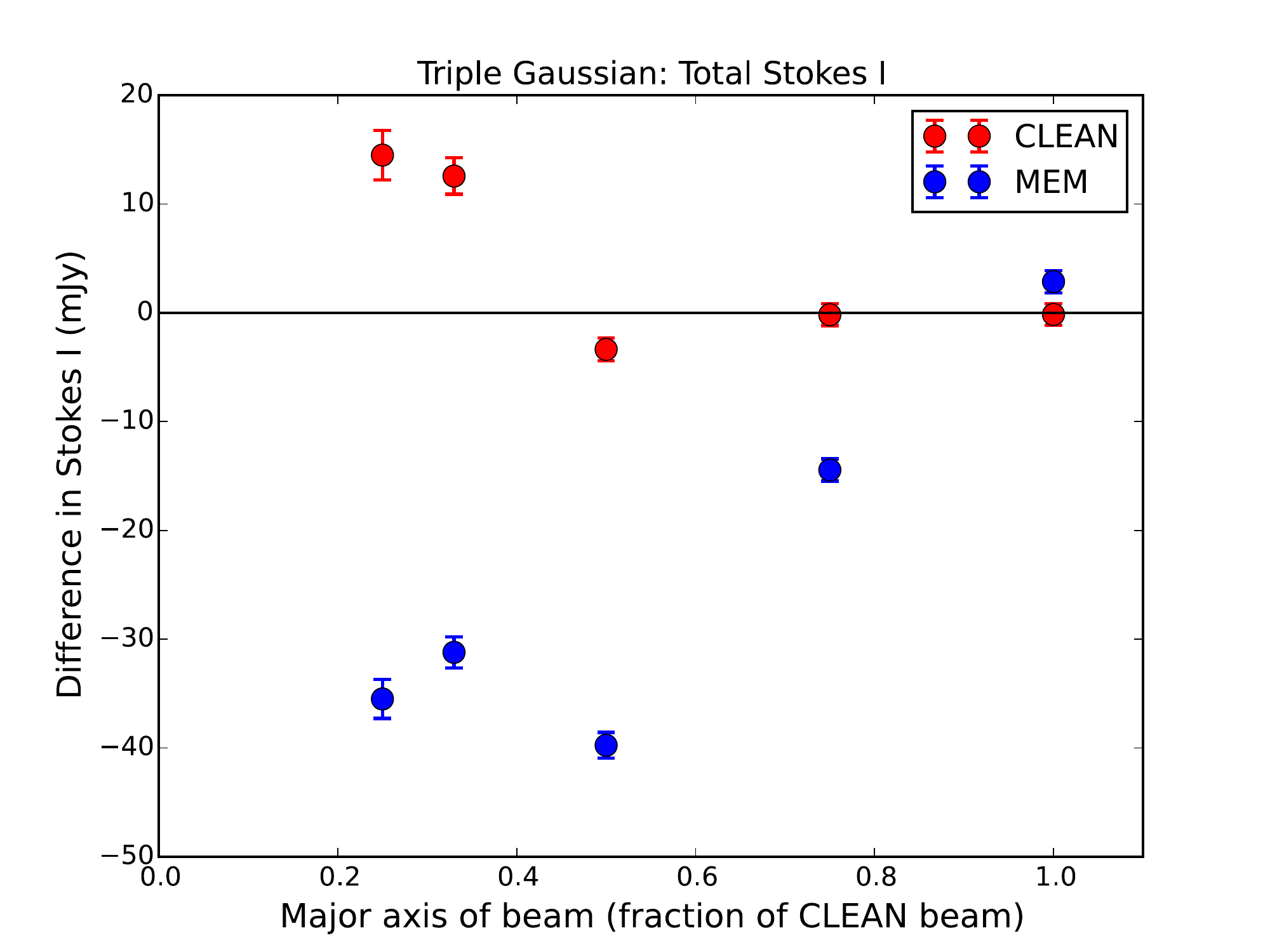}
}
\subfloat[Point A $I$]{
	\includegraphics[width=  1.0\columnwidth]{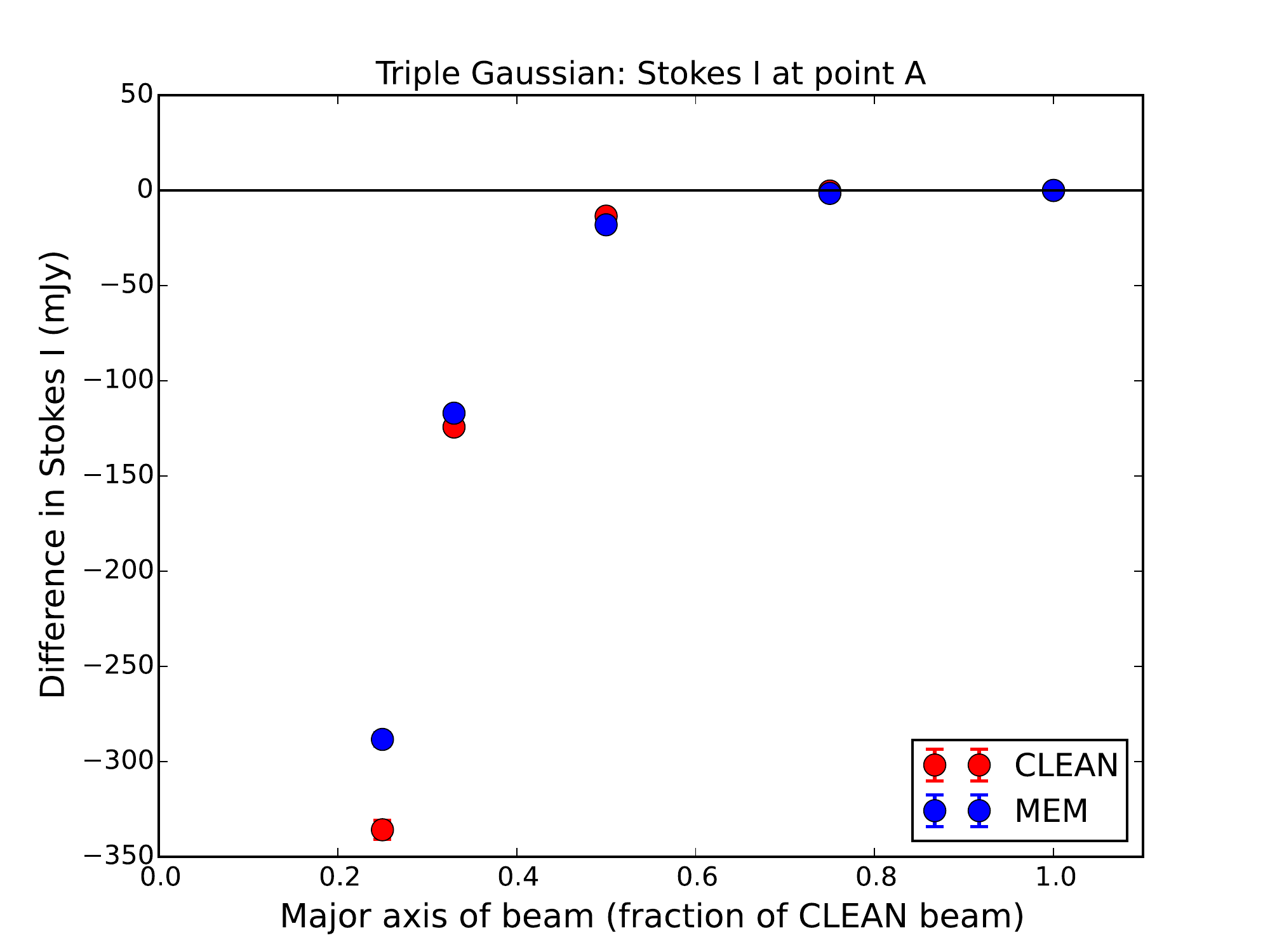}
}

\subfloat[Point B $I$]{
	\includegraphics[width=  1.0\columnwidth]{mem-sims-images/TGP2/TGP2_SI_B-eps-converted-to.pdf}
}
\subfloat[Point C $I$]{
	\includegraphics[width=  1.0\columnwidth]{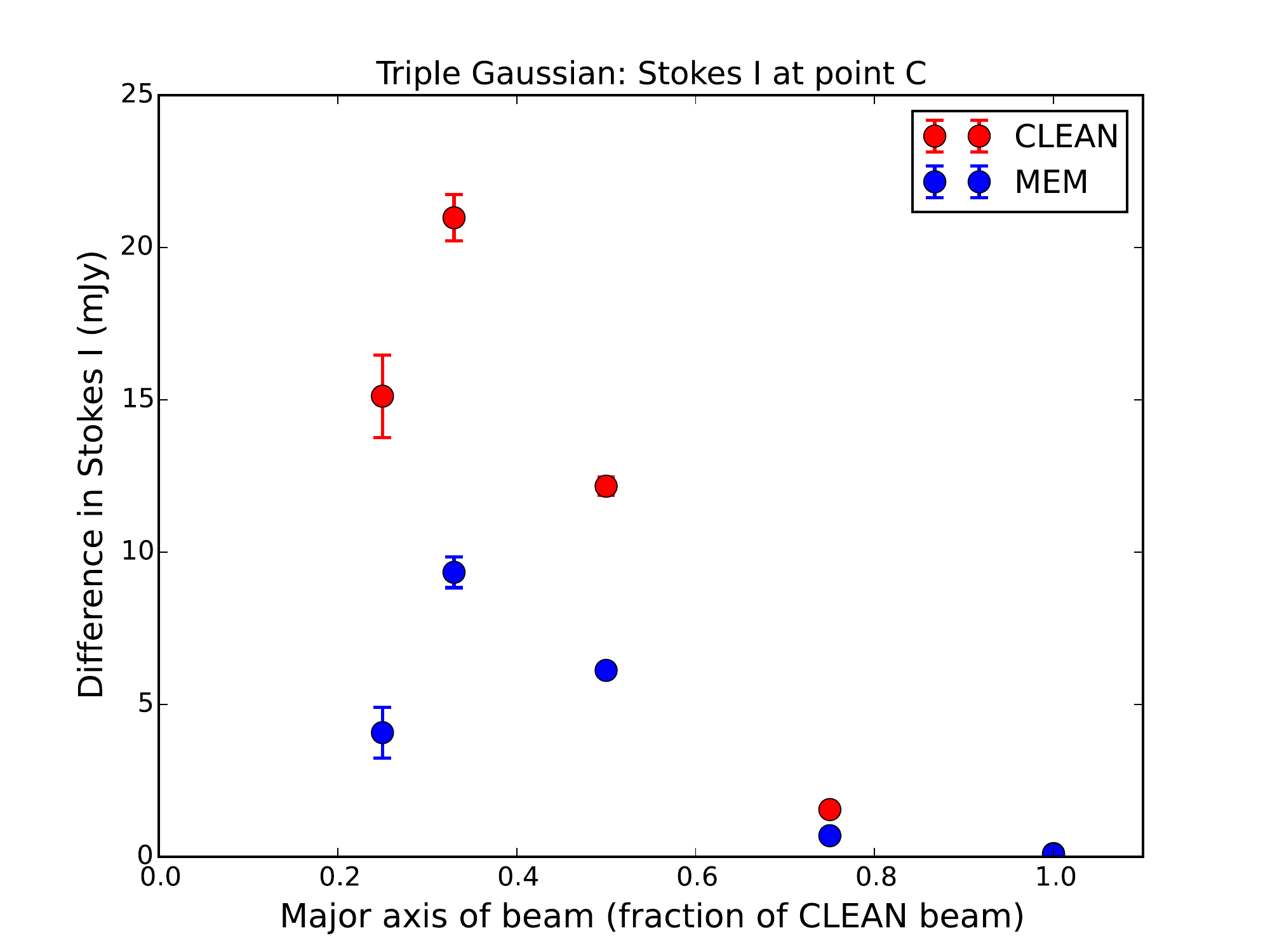}
}
\caption[Error distributions for the total flux.]{Distribution of the errors in the Stokes $I$ flux for the (a) entire source and at the positions of components (b) A, (c ) B and (d) C for the MEM and CLEAN maps of the 
triple Gaussian model source convolved with beams comprising various fractions of the full CLEAN beam. The size of the major axis of convolving beam used is indicted on the x-axis as a fraction of the standard CLEAN major axis (note that the beam area is the square of this factor).}
\label{mem-main-fig-tg-fluxes-total}
\end{center}
\end{figure*}

\begin{figure*}
\begin{center}
\subfloat[Total $I$]{
	\includegraphics[width=  1.0\columnwidth]{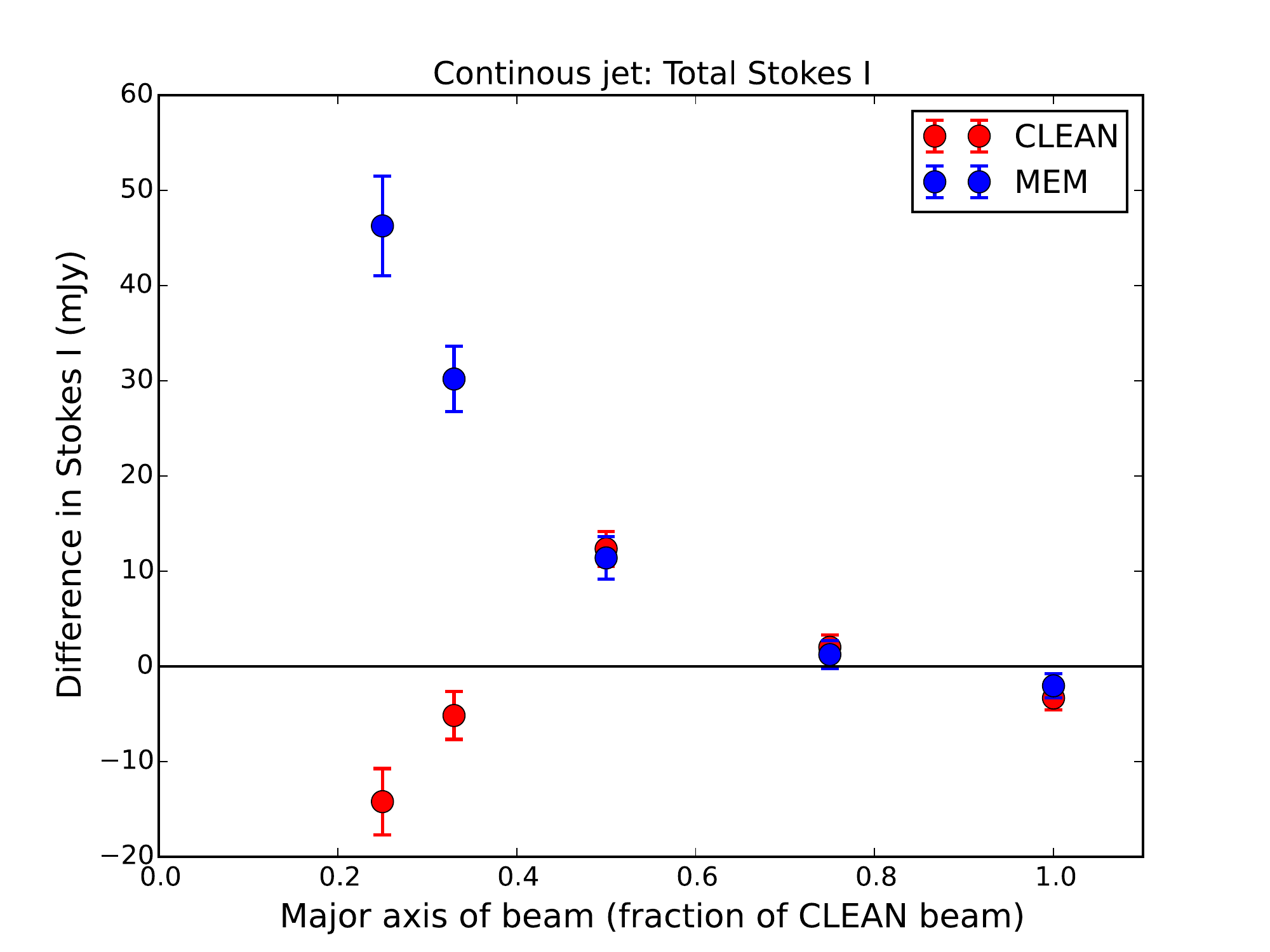}
}
\subfloat[Point A $I$]{
	\includegraphics[width=  1.0\columnwidth]{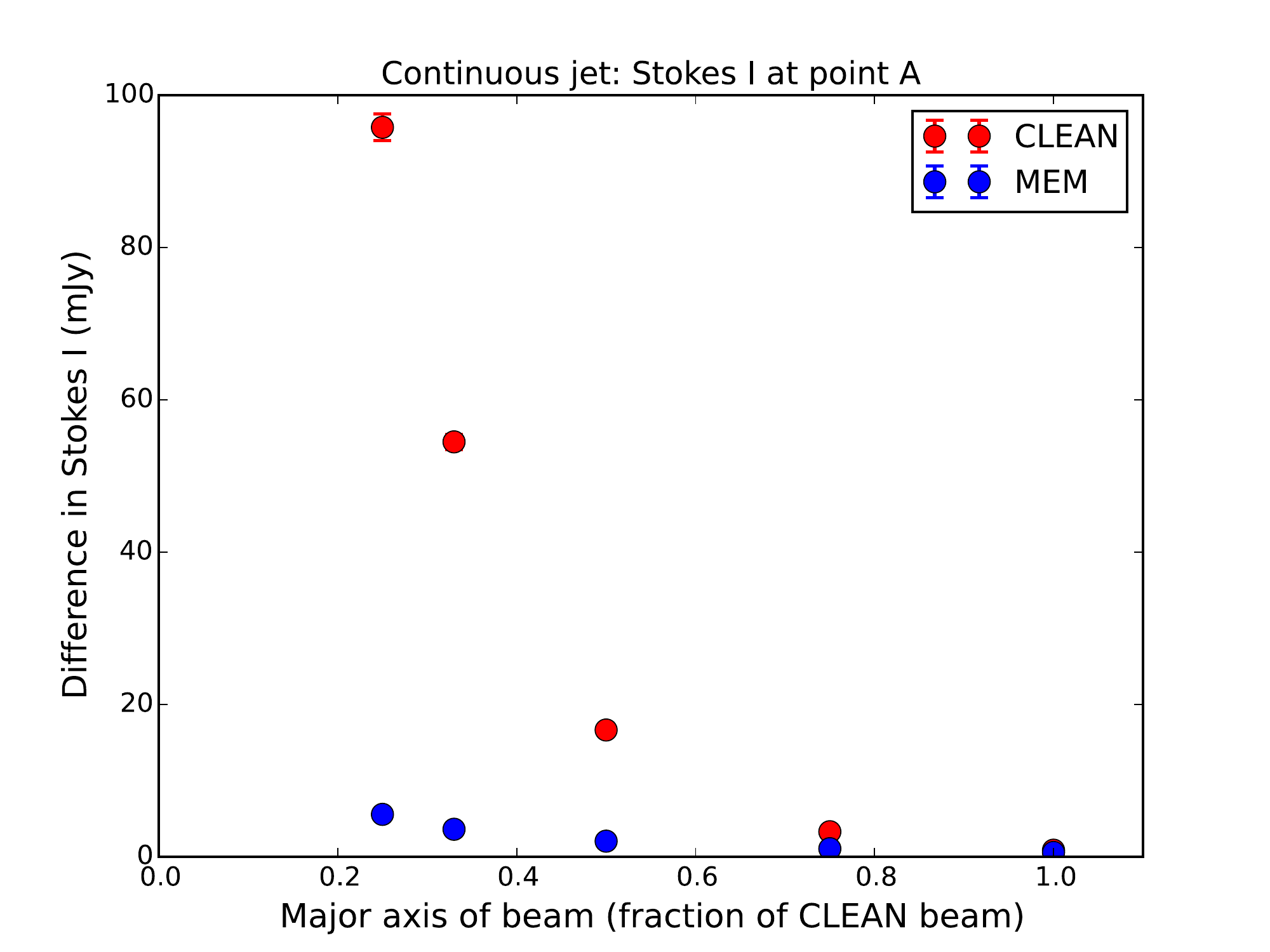}
}

\subfloat[Point B $I$]{
	\includegraphics[width=  1.0\columnwidth]{mem-sims-images/CTSJ/CTSJ_SI_B-eps-converted-to.pdf}
}
\subfloat[Point C $I$]{
	\includegraphics[width=  1.0\columnwidth]{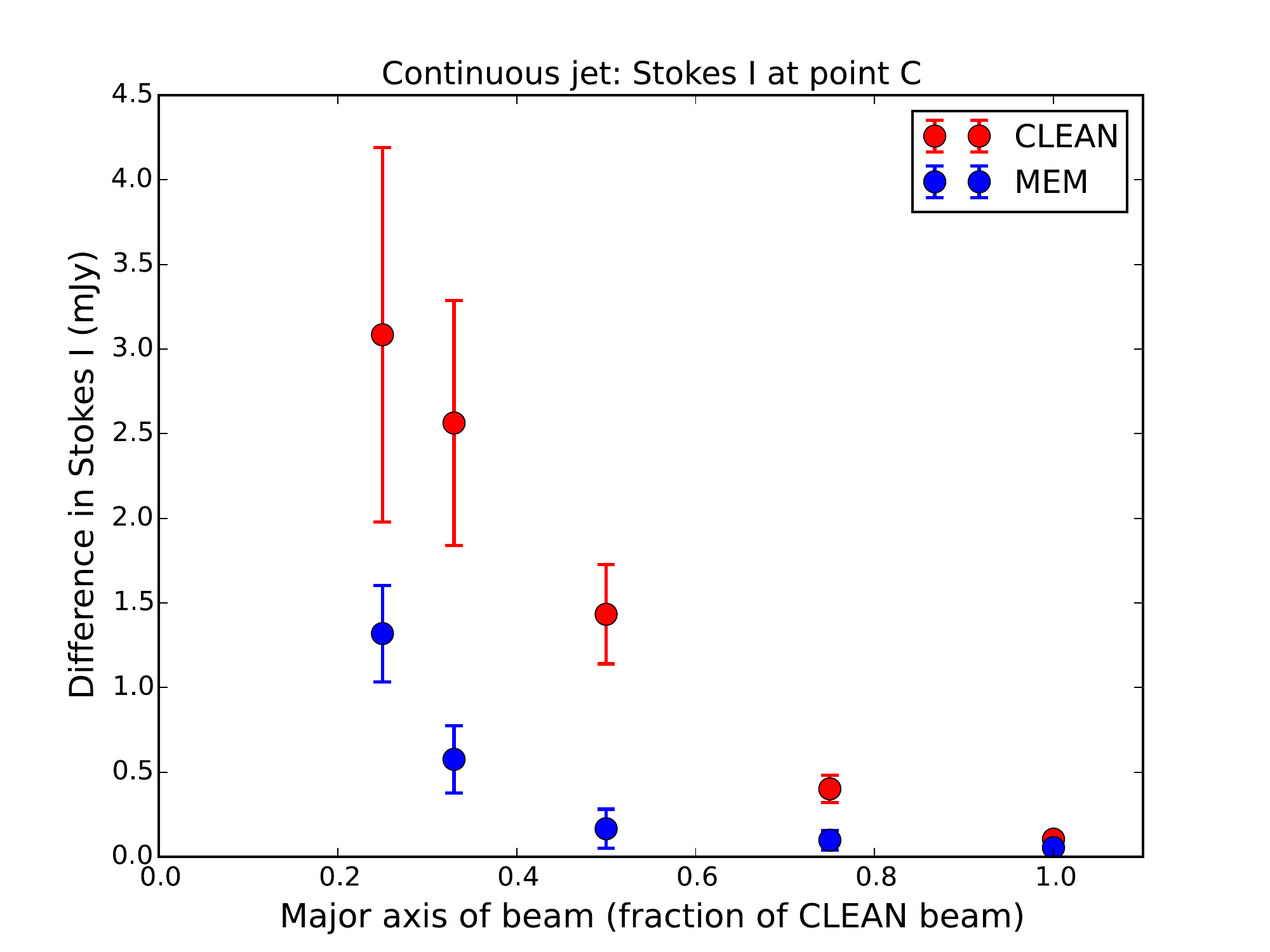}
}
\caption[Continuous, bent jet model. Error distributions for the total flux.]{Distribution of the errors in the Stokes I flux for the (a) entire source and at the positions of components (b) A, (c ) B and (d) C for the MEM and CLEAN maps of the 
continuous, bent jet model convolved with beams comprising various fractions of the full CLEAN beam. The size of the major axis of convolving beam used is indicted on the x-axis as a fraction of the standard CLEAN major axis (note that the beam area is the square of this factor).}
\label{fig-app-cts-fluxes-total}
\end{center}
\end{figure*}

Figures \ref{mem-main-fig-tg-fluxes-total} and \ref{fig-app-cts-fluxes-total} shows the performance of MEM 
and CLEAN at recovering the total flux $I$ of the model sources, as well as the Stokes I flux at points A, B and C in each source. The total fluxes were calculated by summing the flux within a mask created manually using the ``true'' convolved model images, while the fluxes at positions A, B and C were obtained by summing the flux within an area of size
$0.3~\textrm{mas}\times 0.3~\textrm{mas}$ ($~\textrm{pixel}\times 
3$-pixel) centred on these points.

In the case of the triple Gaussian source, CLEAN continues to be highly accurate down to a resolution corresponding to half of the CLEAN beam, while MEM is an order of magnitude less accurate. At higher resolutions CLEAN's accurate begins to deteriorate, with MEM continuing to perform worse. A similar trend is seen for the continuous bent jet source, though both algorithms remain accurate at slightly higher resolution. It is notable that the direction of the error in each algorithm swaps between sources, underestimate the first source while overestimating the second, and vice versa. Note also that the use of a mask to describe the collection area eliminated some regions in the MEM map which contained large amounts of unrealistic negative flux. These regions occur where MEM, unable to describe negative Stokes I pixels due to Equation \ref{mem-heqn}, is left with a negative region (off-source) in its residual maps when they are formed by subtracting the model convolved with the dirty beam from the data. This effect is exasperated at high resolutions when residual maps are added to the model map convolved with a high resolution beam without scaling applied (as in the CASA's CLEAN algorithm). Thus a relatively small sidelobe at CLEAN beam resolution can cause a large negative feature at the highest resolutions.

It is of interest that both sources show similar results for Stokes I, even though the continuous bent jet model was imaged with flux conservation on, while the triple Gaussian had $\gamma = 0$ at all times. It appears that flux can be well conserved using only a comparison between the convolved model and the dirty map as a guide, and that the choice of whether to conserve flux or not depends on which option gives the best maps in a qualitative sense (lower noise, less spurious modelling of sidelobes by MEM).

\begin{figure*}
\begin{center}
\subfloat[Total $p$]{
	\includegraphics[width=  1.0\columnwidth]{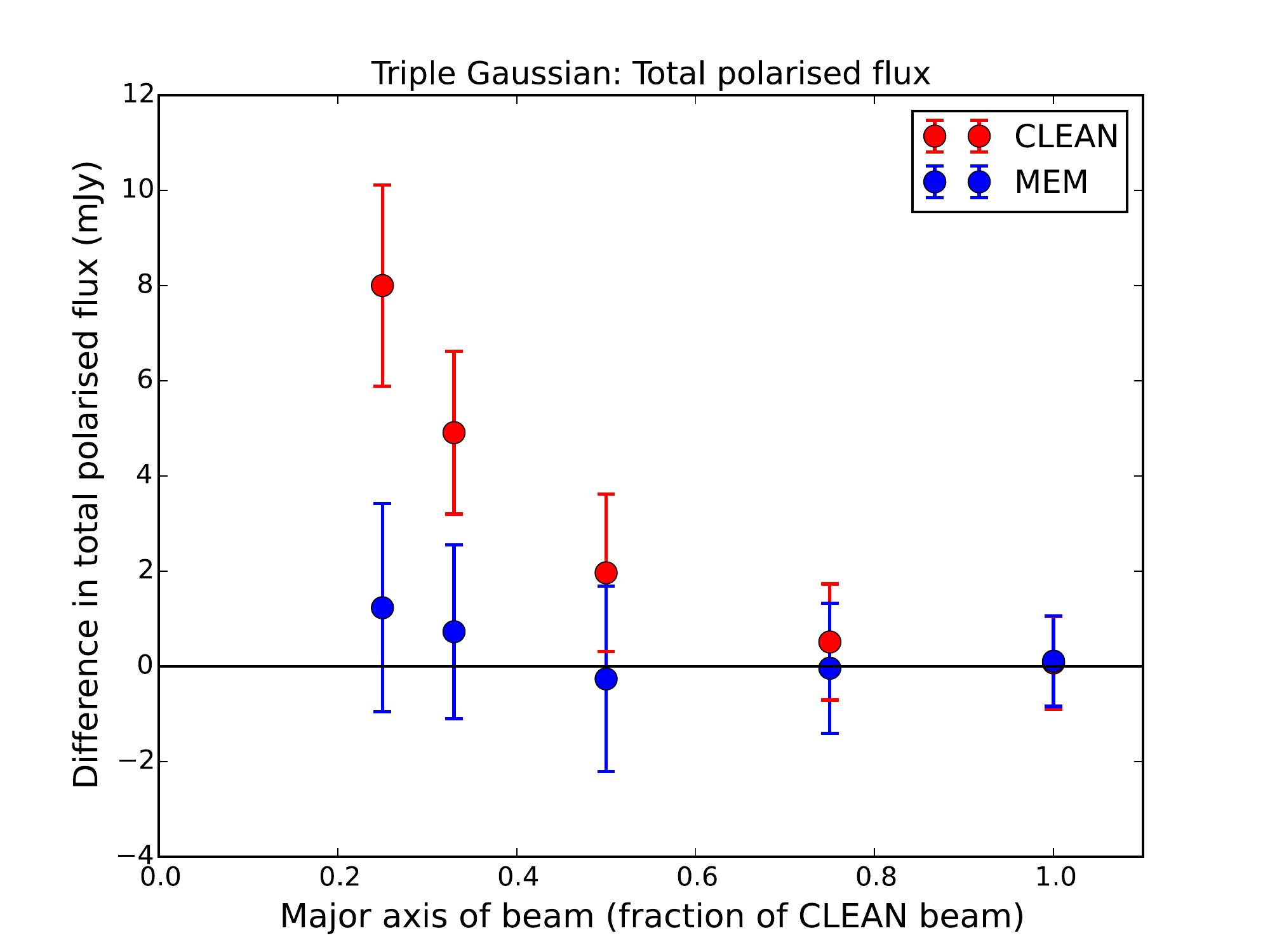}
}
\subfloat[Point A $p$]{
	\includegraphics[width=  1.0\columnwidth]{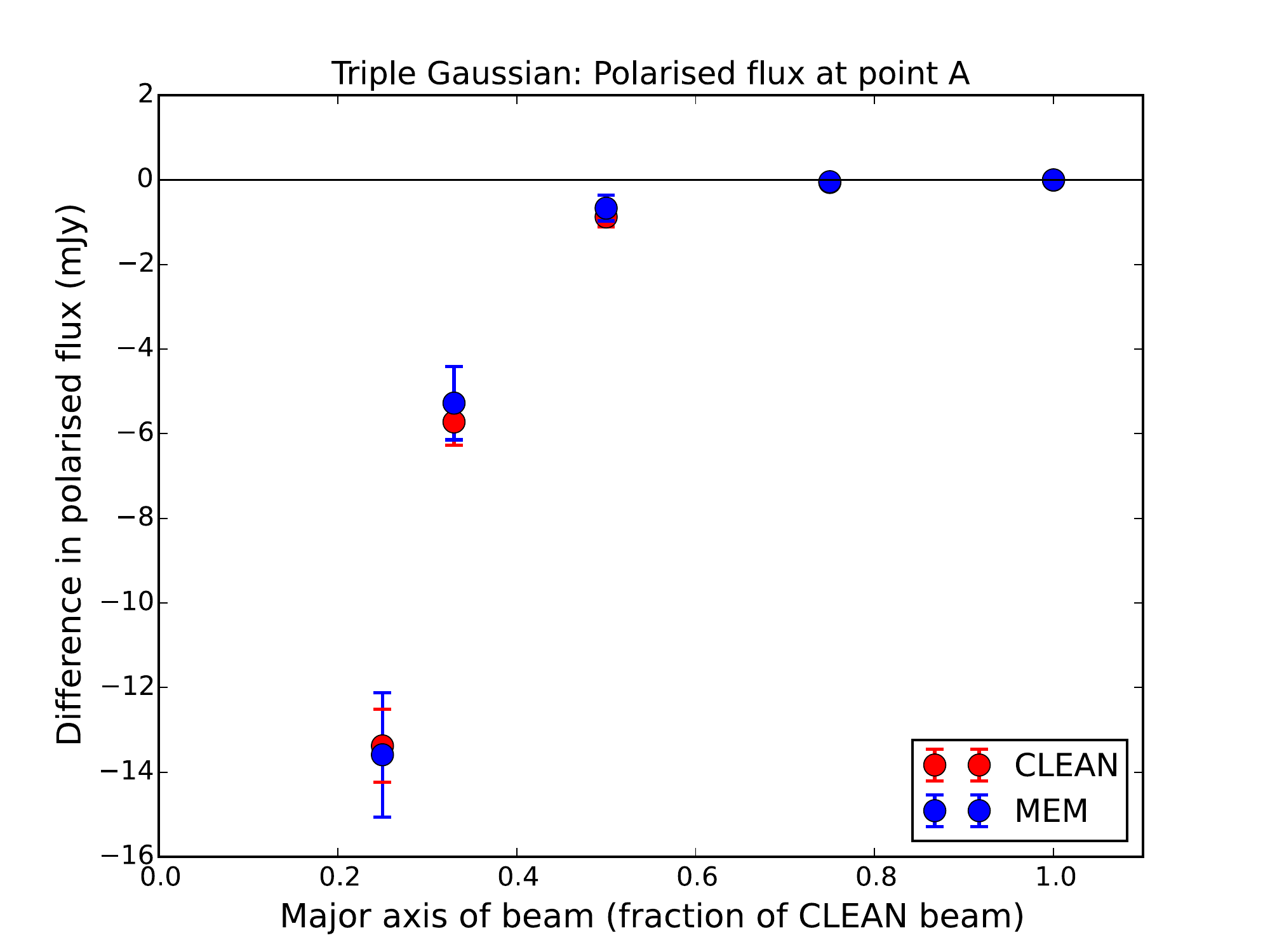}
}

\subfloat[Total $m$]{
	\includegraphics[width=  1.0\columnwidth]{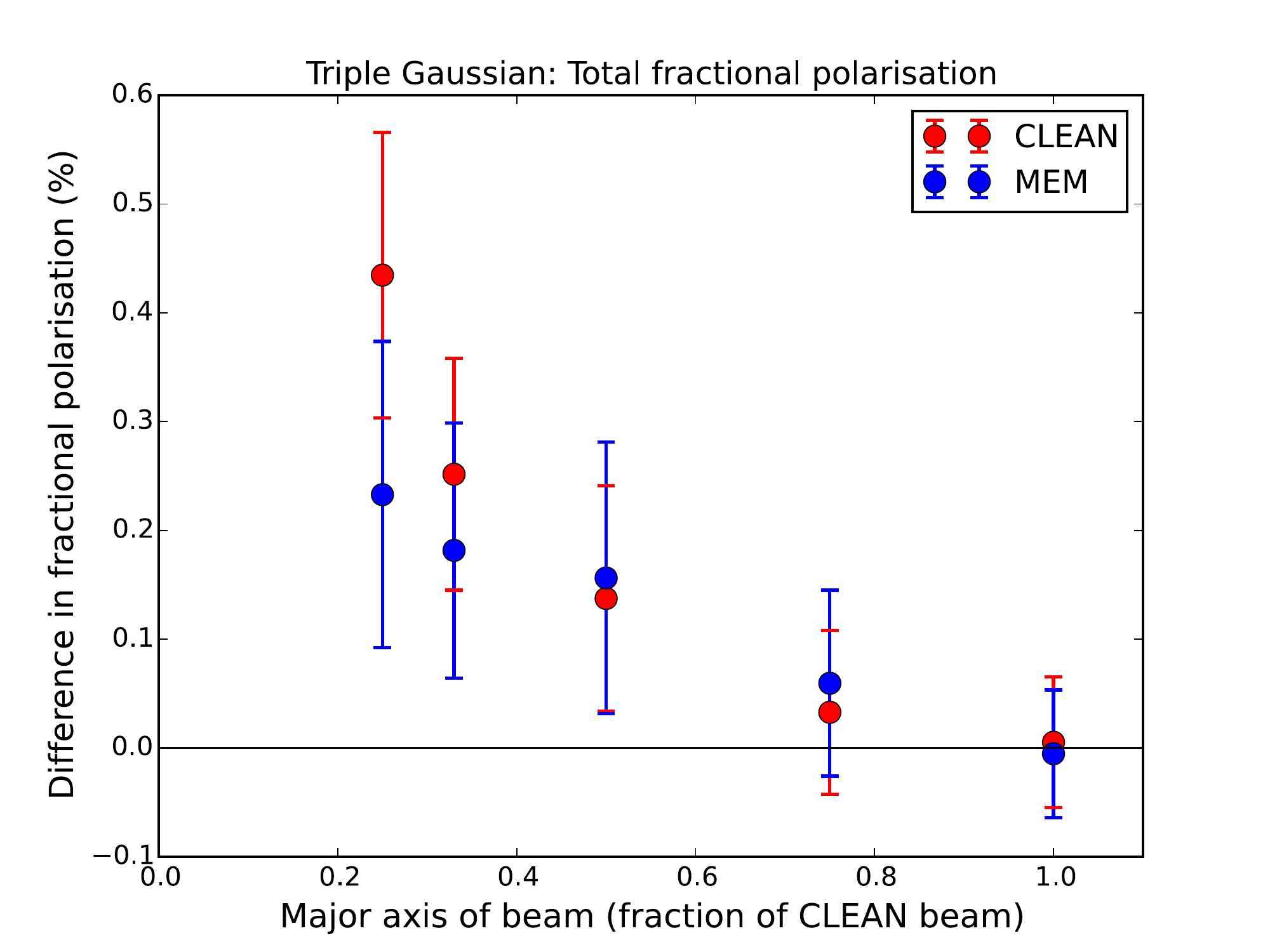}
}
\subfloat[Point A $m$]{
	\includegraphics[width=  1.0\columnwidth]{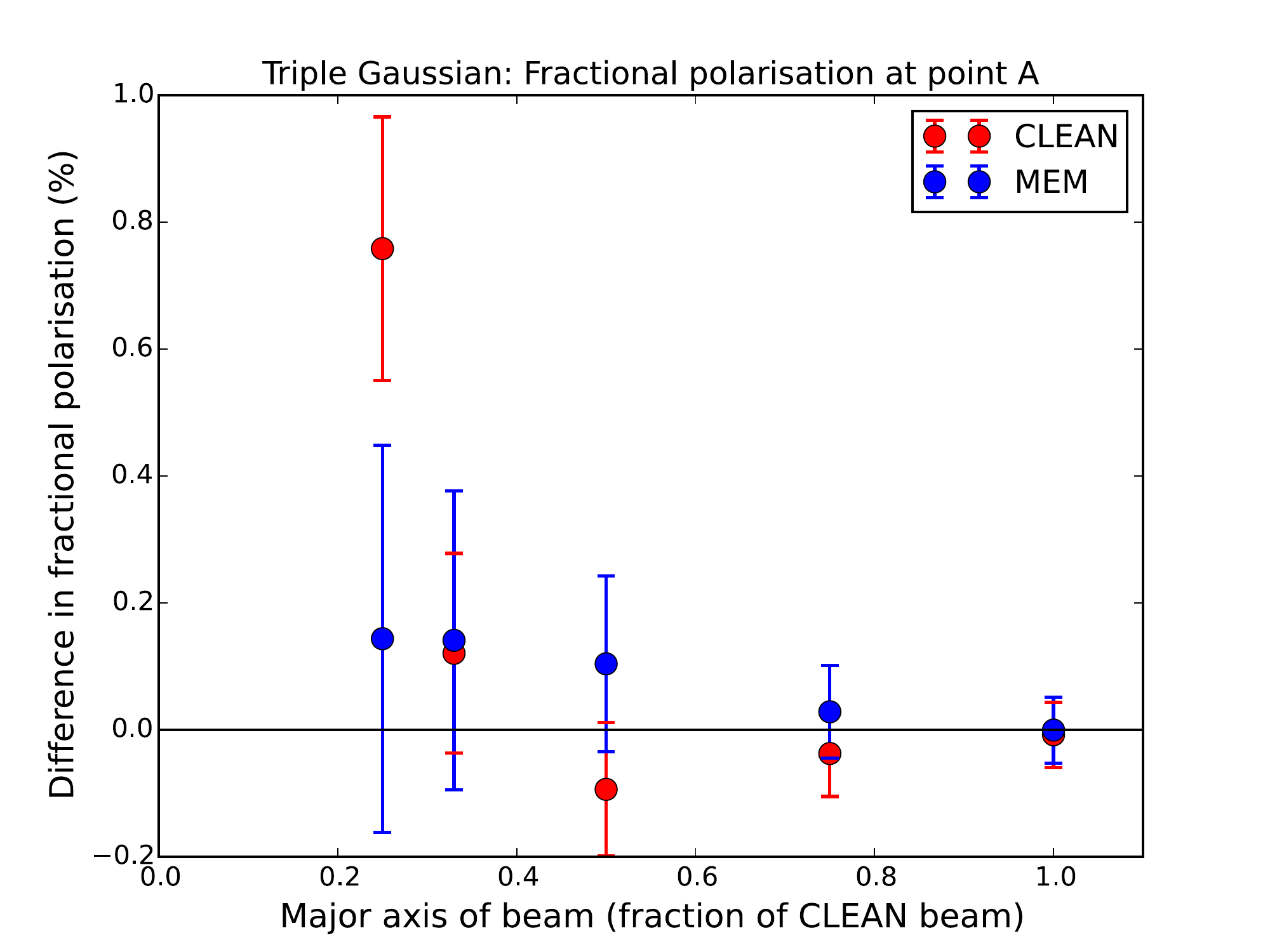}
}

\subfloat[Total $\chi$]{
	\includegraphics[width=  1.0\columnwidth]{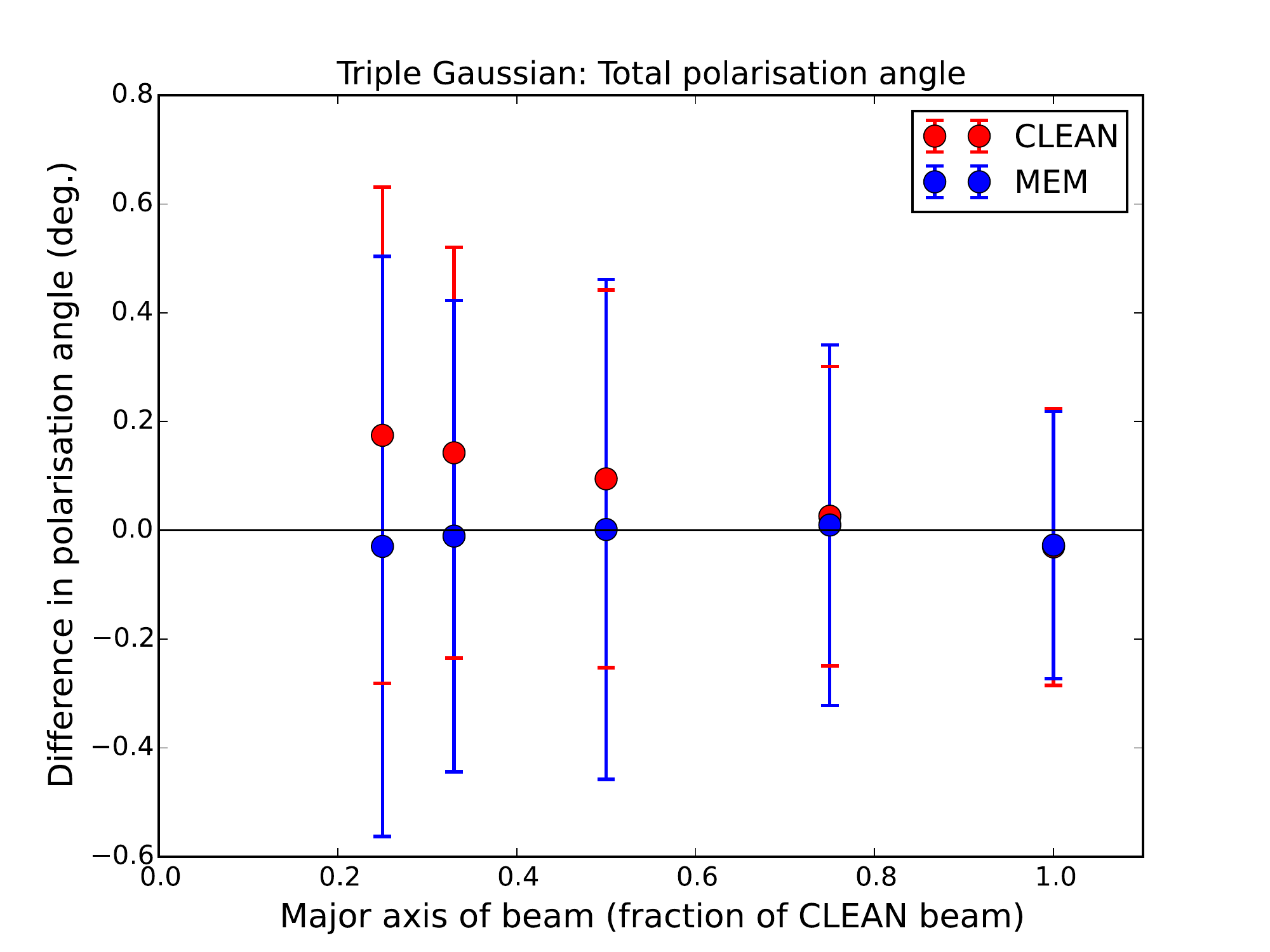}
}
\subfloat[Point A $\chi$]{
	\includegraphics[width=  1.0\columnwidth]{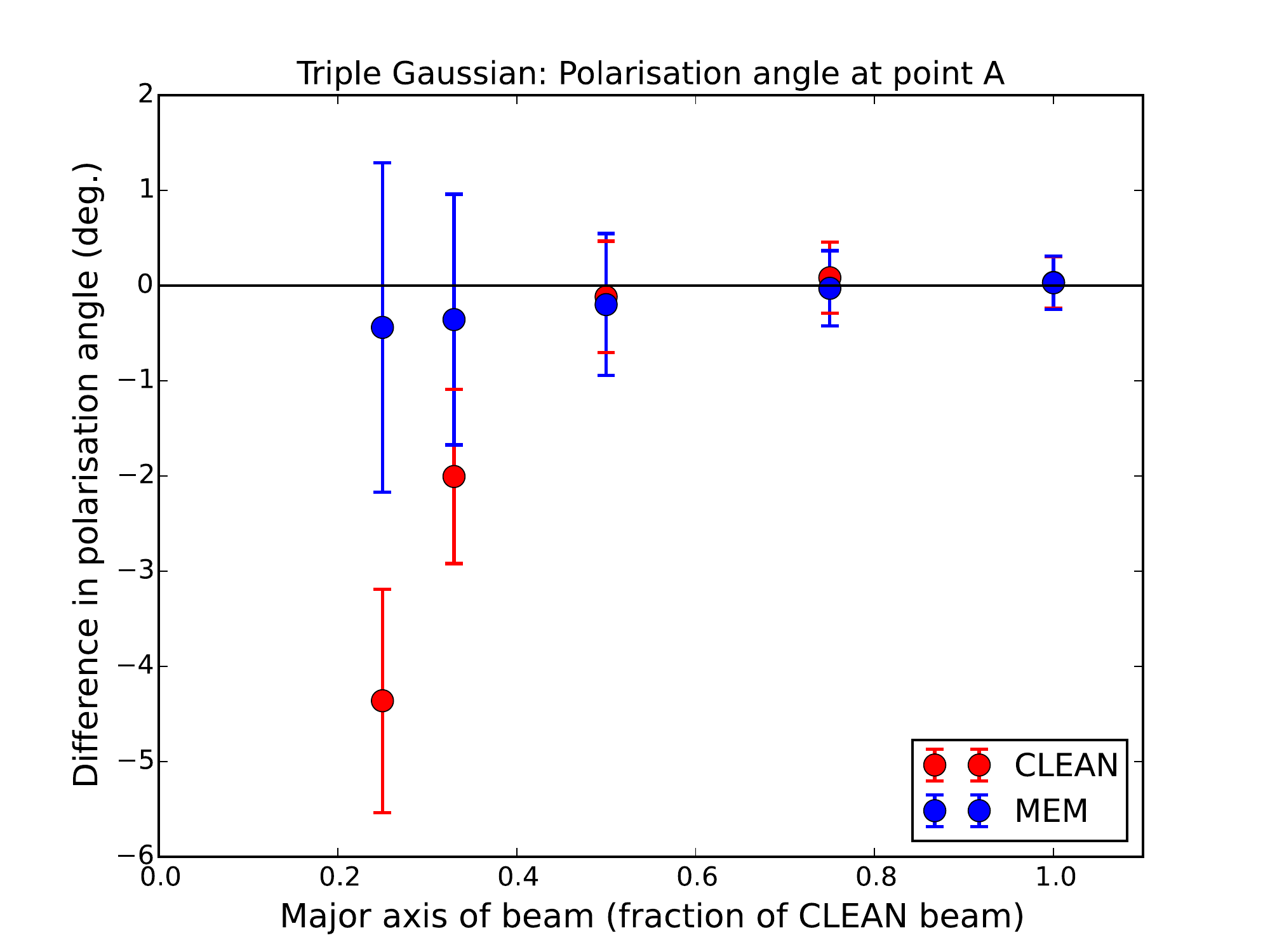}
}
\caption[Error distributions for p, m, chi for total and A.]{Distribution 
of the errors in the (a, b) polarised flux $p$, (c, d) fractional 
polarisation $m$ and (e, f) polarisation angle $\chi$, for the (a, c, e)  
entire source and (b, d, f) the position of component A, for
the MEM and CLEAN maps of the triple Gaussian model source convolved 
with beams comprising various fractions of the full CLEAN beam. The size of the major axis of convolving beam used is indicted on the x-axis as a fraction of the standard CLEAN major axis (note that the beam area is the square of this factor).}
\label{mem-main-fig-pmchi-totA}
\end{center}
\end{figure*}

\begin{figure*}
\begin{center}
\subfloat[Total $p$]{
	\includegraphics[width=  1.0\columnwidth]{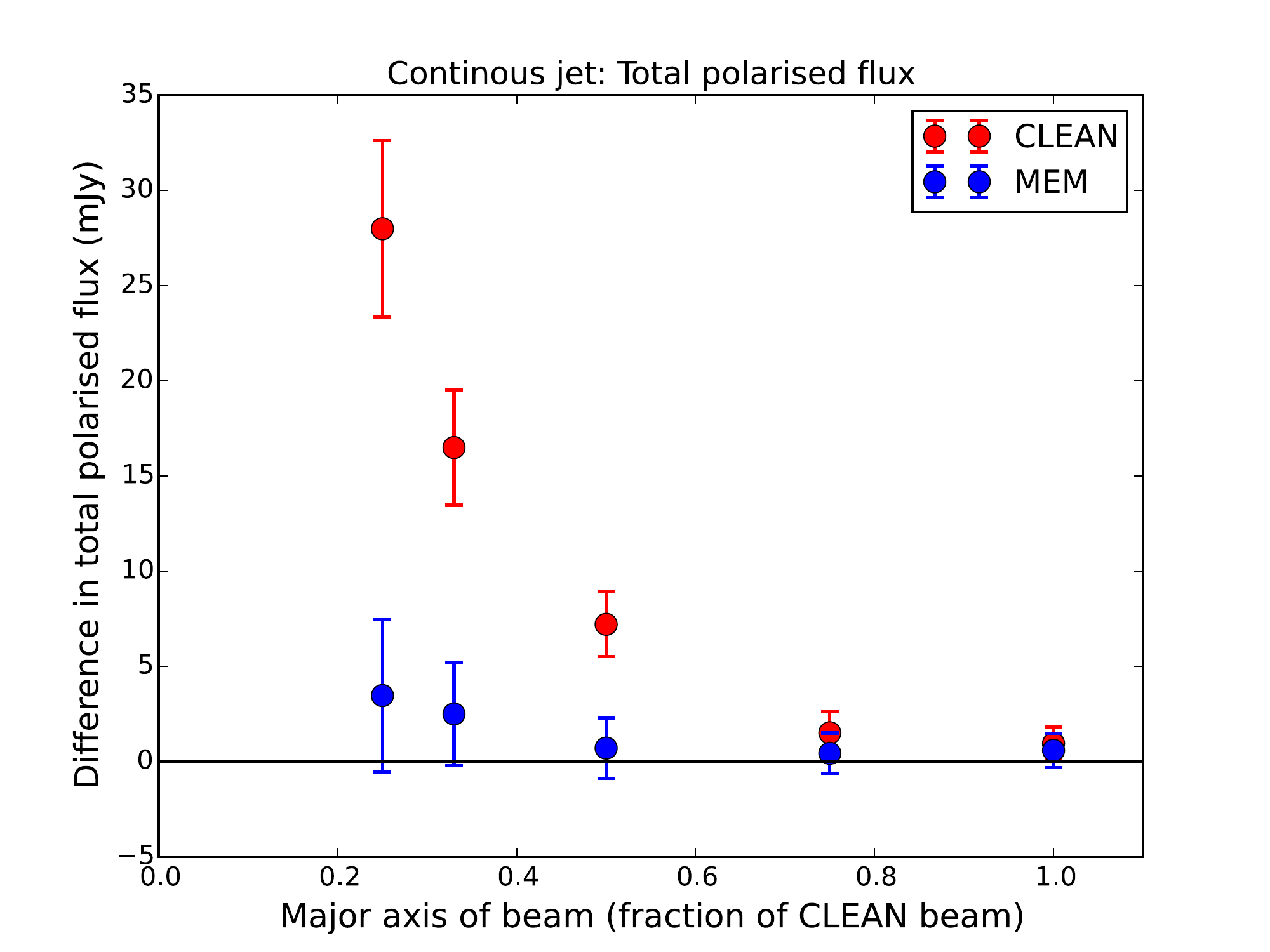}
}
\subfloat[Point A $p$]{
	\includegraphics[width=  1.0\columnwidth]{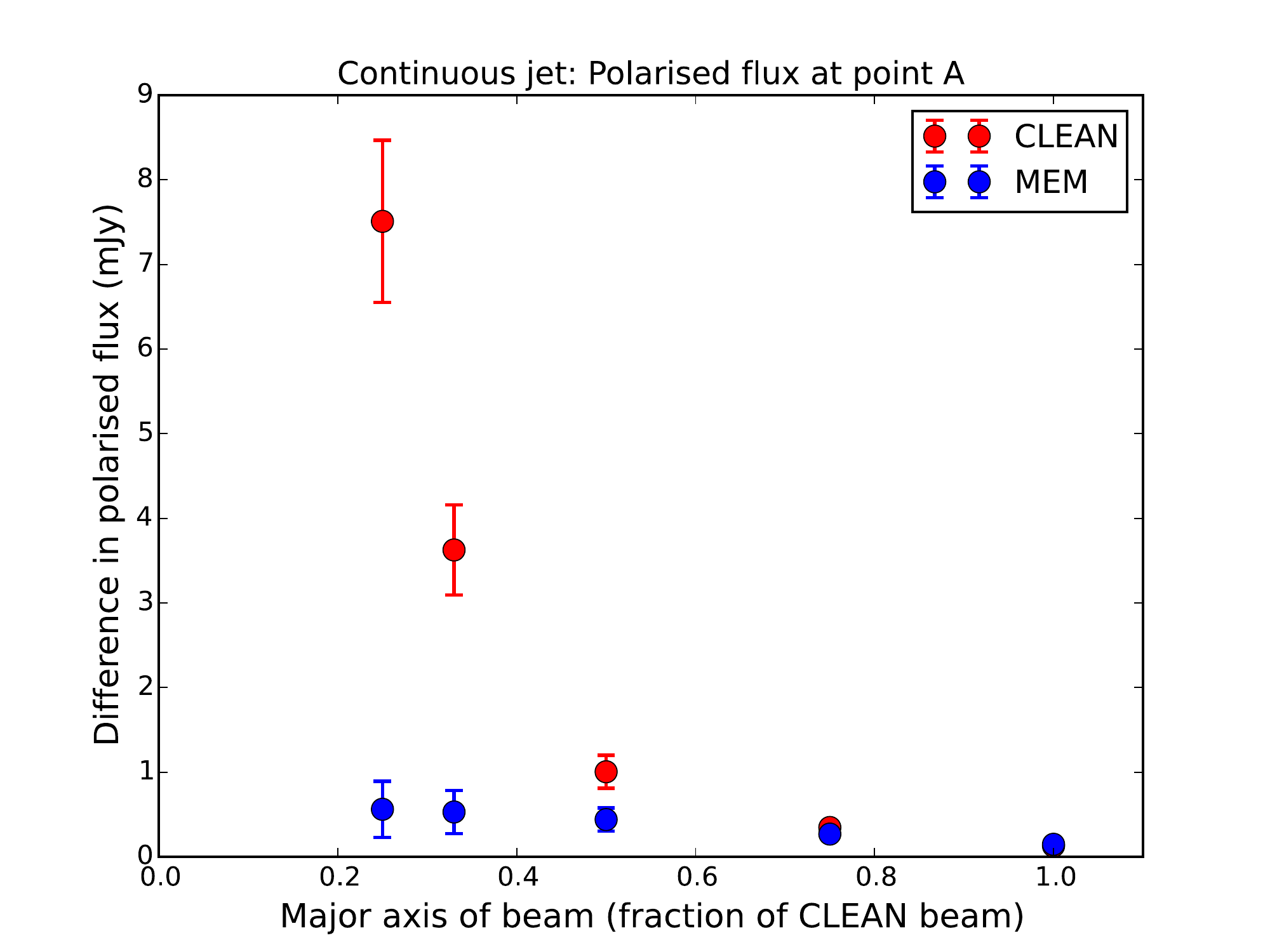}
}

\subfloat[Total $m$]{
	\includegraphics[width=  1.0\columnwidth]{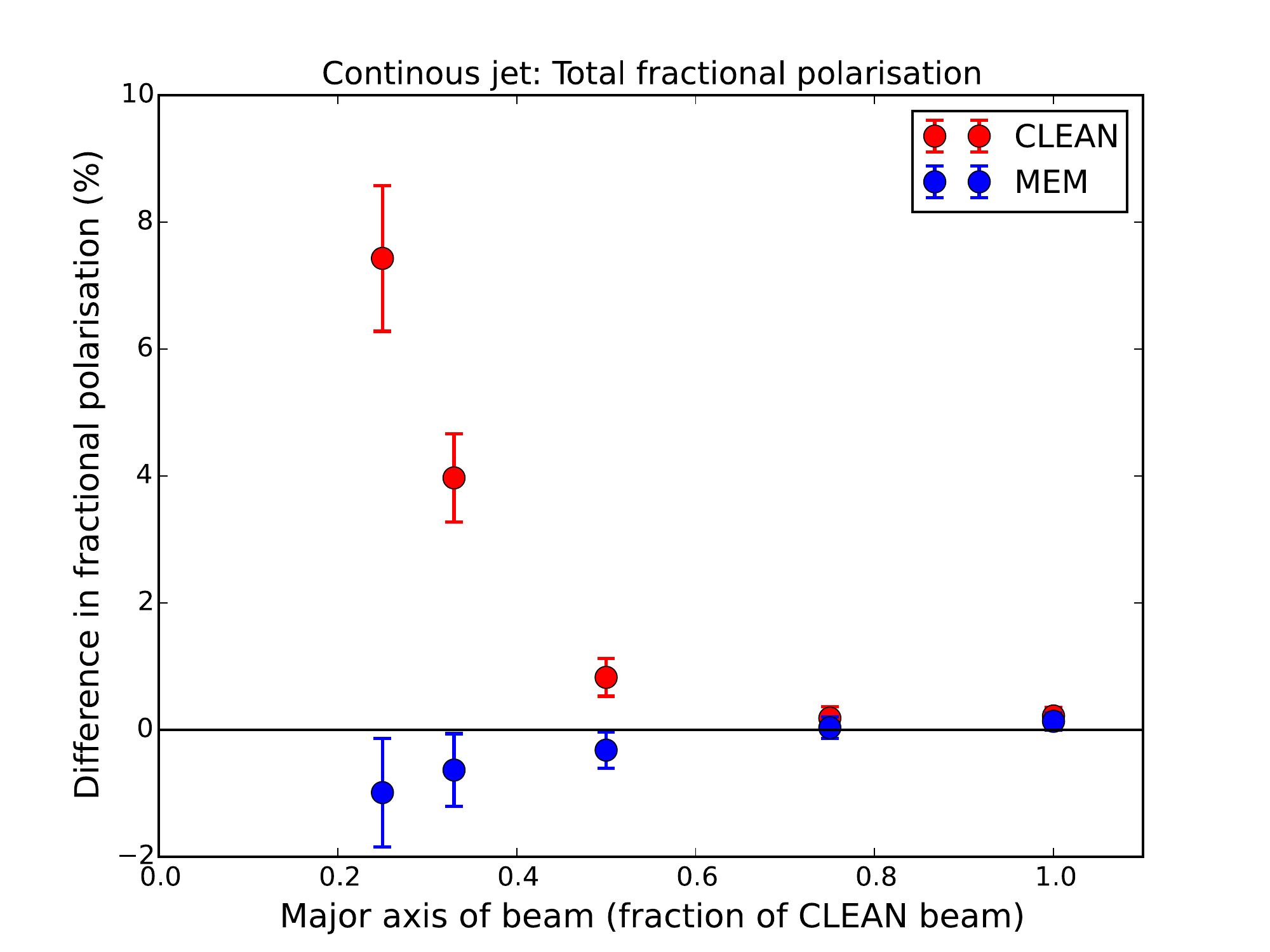}
}
\subfloat[Point A $m$]{
	\includegraphics[width=  1.0\columnwidth]{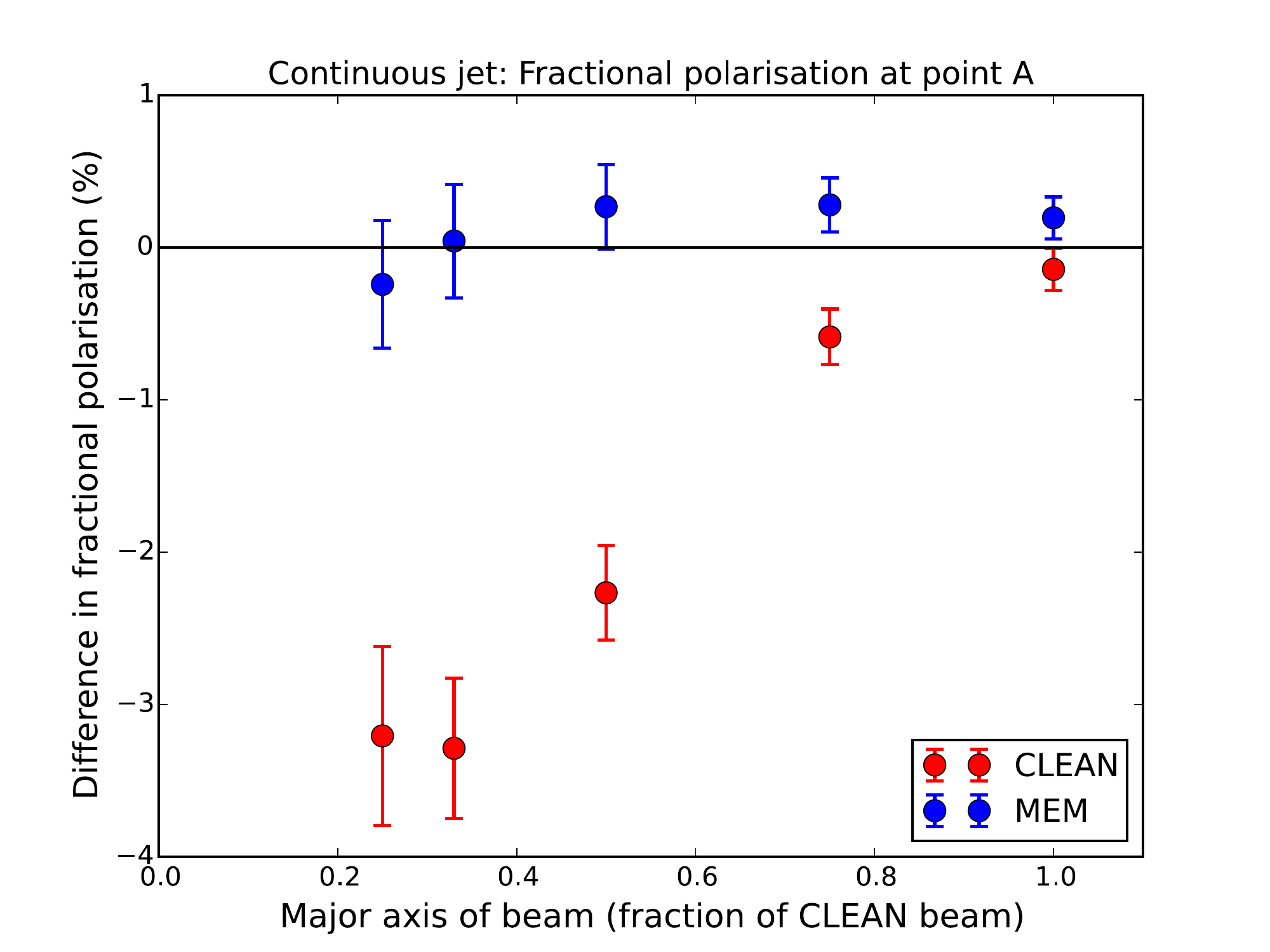}
}

\subfloat[Total $\chi$]{
	\includegraphics[width=  1.0\columnwidth]{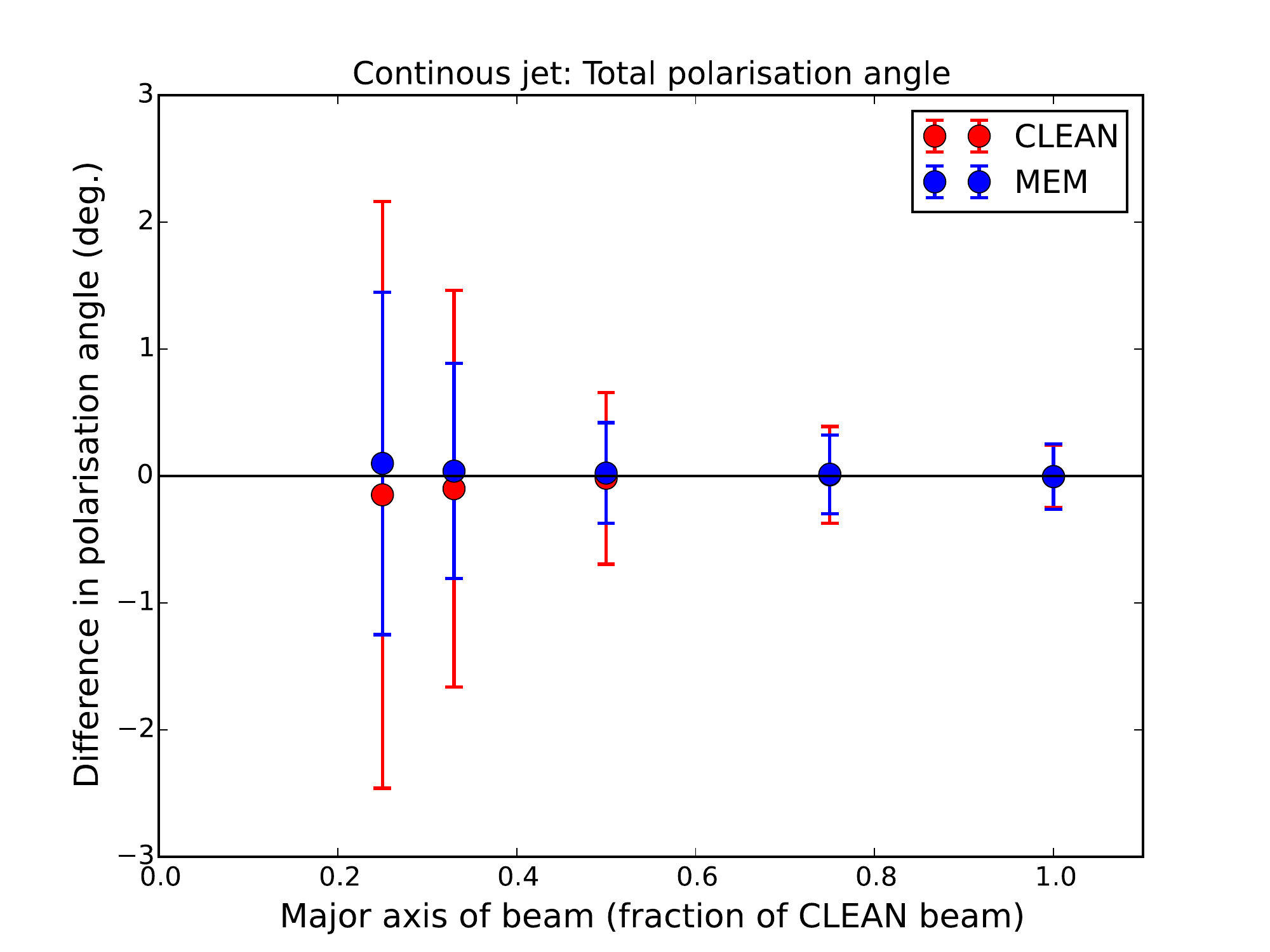}
}
\subfloat[Point A $\chi$]{
	\includegraphics[width=  1.0\columnwidth]{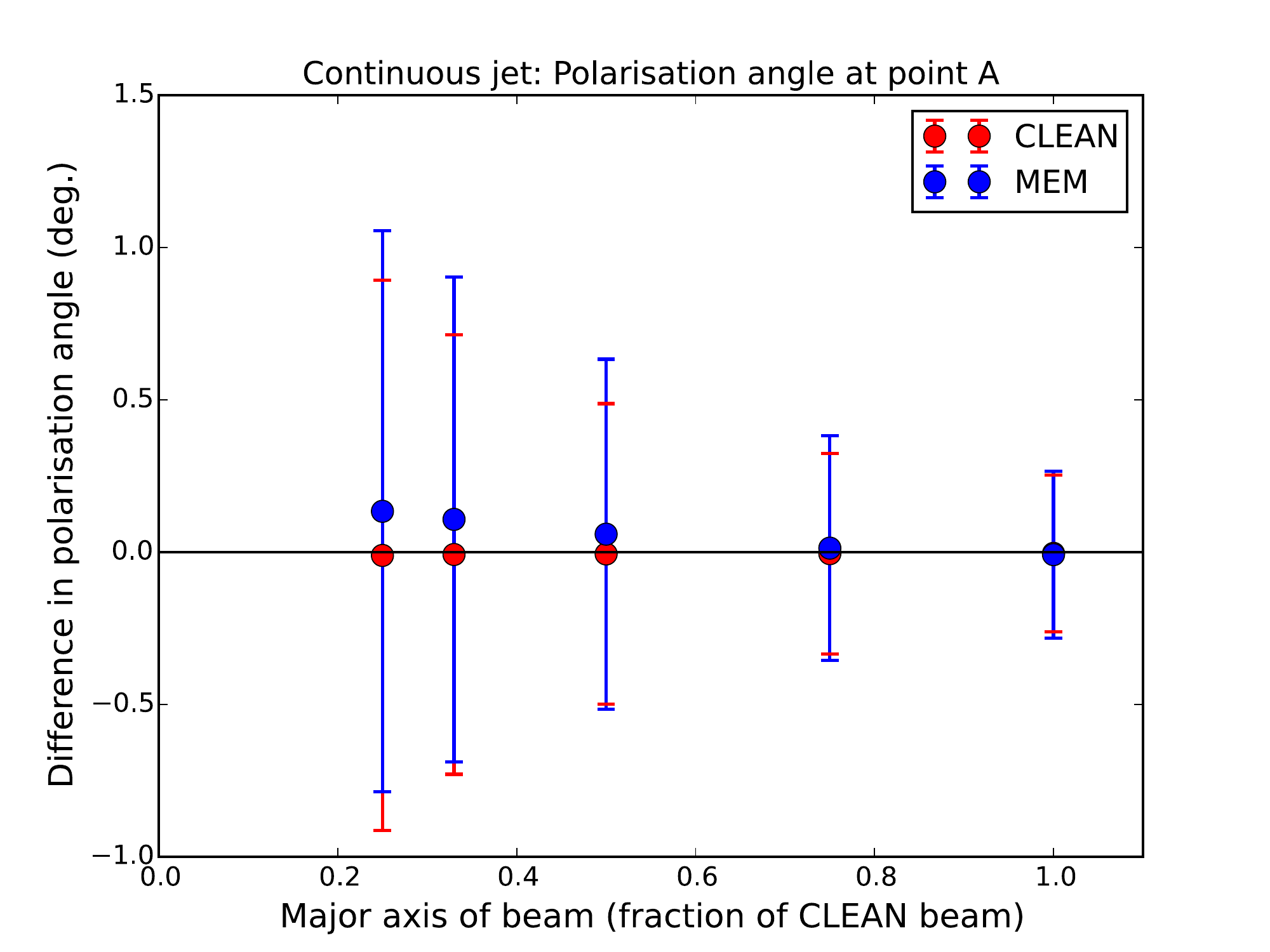}
}
\caption[Continuous bent jet model. Error distributions for p, m, chi for total and A.]{Distribution 
of the errors in the (a, b) polarised flux $p$, (c, d) fractional 
polarisation $m$ and (e, f) polarisation angle $\chi$, for the (a, c, e)  
entire source and (b, d, f) the position of component A, for
the MEM and CLEAN maps of the continuous, bent jet model convolved 
with beams comprising various fractions of the full CLEAN beam. The size of the major axis of convolving beam used is indicted on the x-axis as a fraction of the standard CLEAN major axis (note that the beam area is the square of this factor).}
\label{fig-app-cts-pmchi-totA}
\end{center}
\end{figure*}

Figures \ref{mem-main-fig-pmchi-totA} and \ref{fig-app-cts-pmchi-totA} show the performance of both algorithms at measuring the overall polarisation properties of the source. MEM performs at least as well as CLEAN in determining both total and fractional polarisation over the entire source at all resolutions, and often out-performs CLEAN by more than a factor of two at the highest resolutions. Both MEM and CLEAN measure the integrated polarisation angle accurately at all resolutions, with very little between them in performance (the mean MEM error is often smaller than CLEAN's for the triple Gaussian source, but with a larger spread of values, though it performs better in the continuous jet model).

\begin{figure*}
\begin{center}
\subfloat[Point B $p$]{
	\includegraphics[width=  1.0\columnwidth]{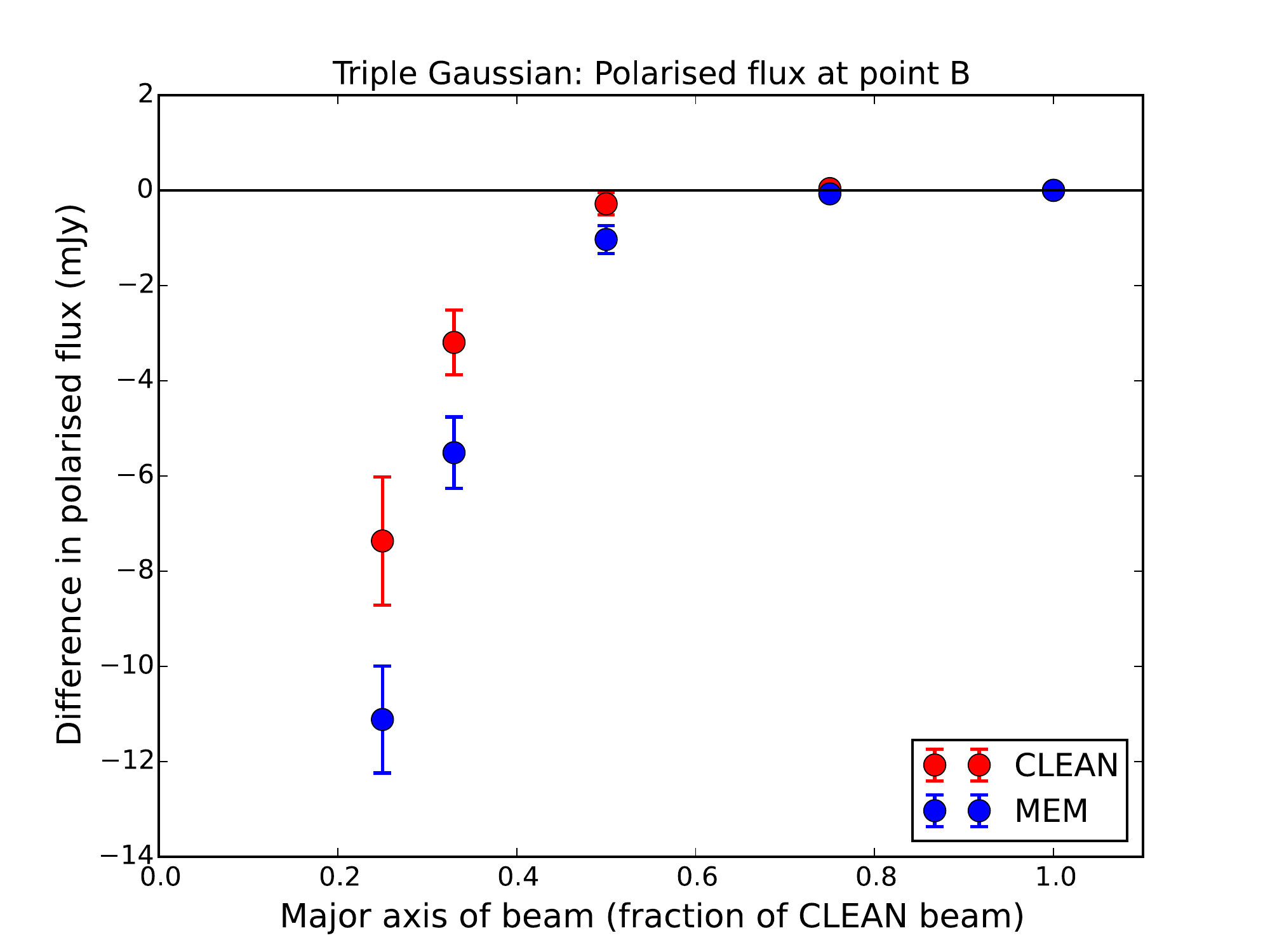}
}
\subfloat[Point C $p$]{
	\includegraphics[width=  1.0\columnwidth]{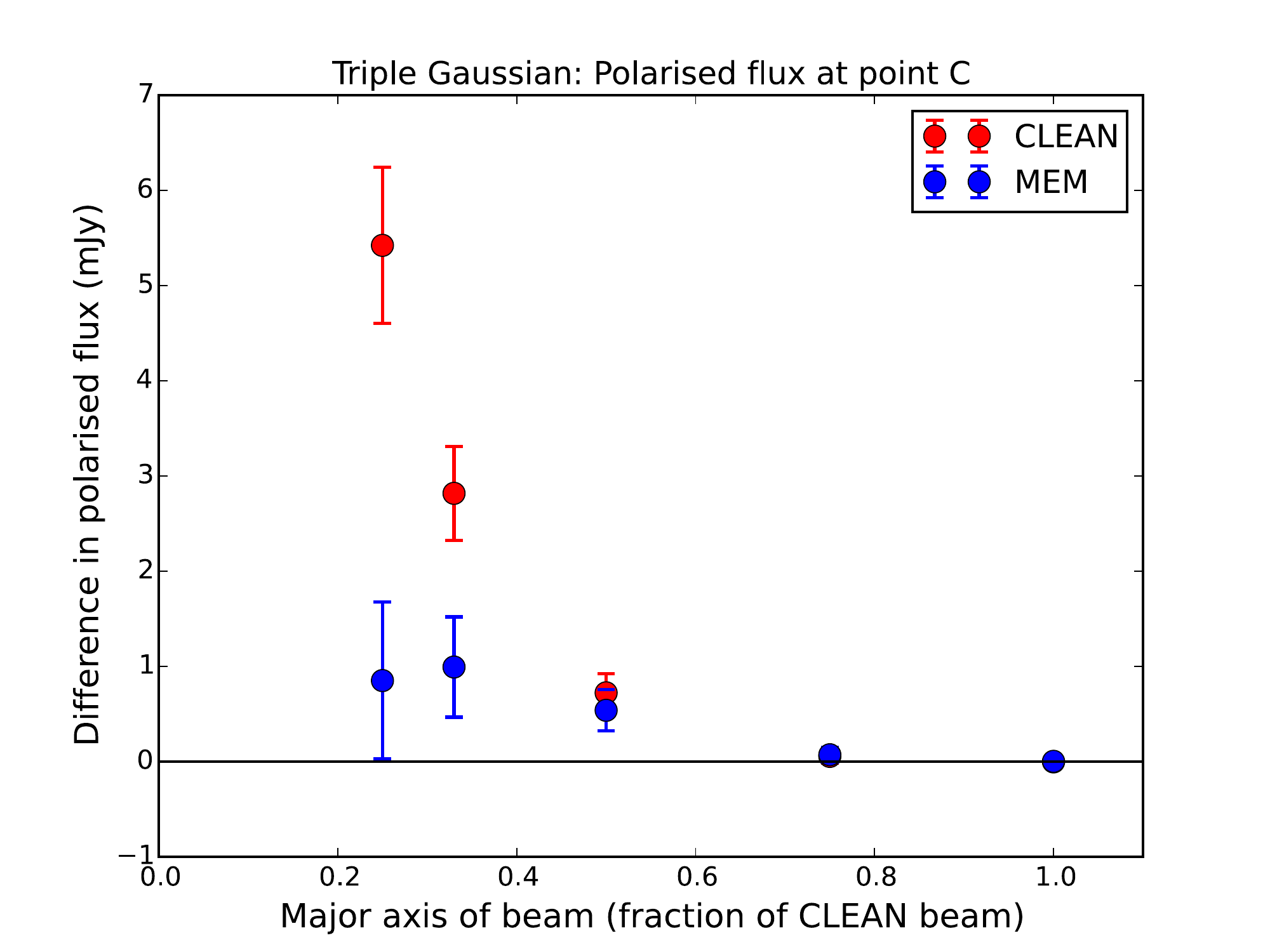}
}

\subfloat[Point B $m$]{
	\includegraphics[width=  1.0\columnwidth]{mem-sims-images/TGP2/TGP2_m_B-eps-converted-to.pdf}
}
\subfloat[Point C $m$]{
	\includegraphics[width=  1.0\columnwidth]{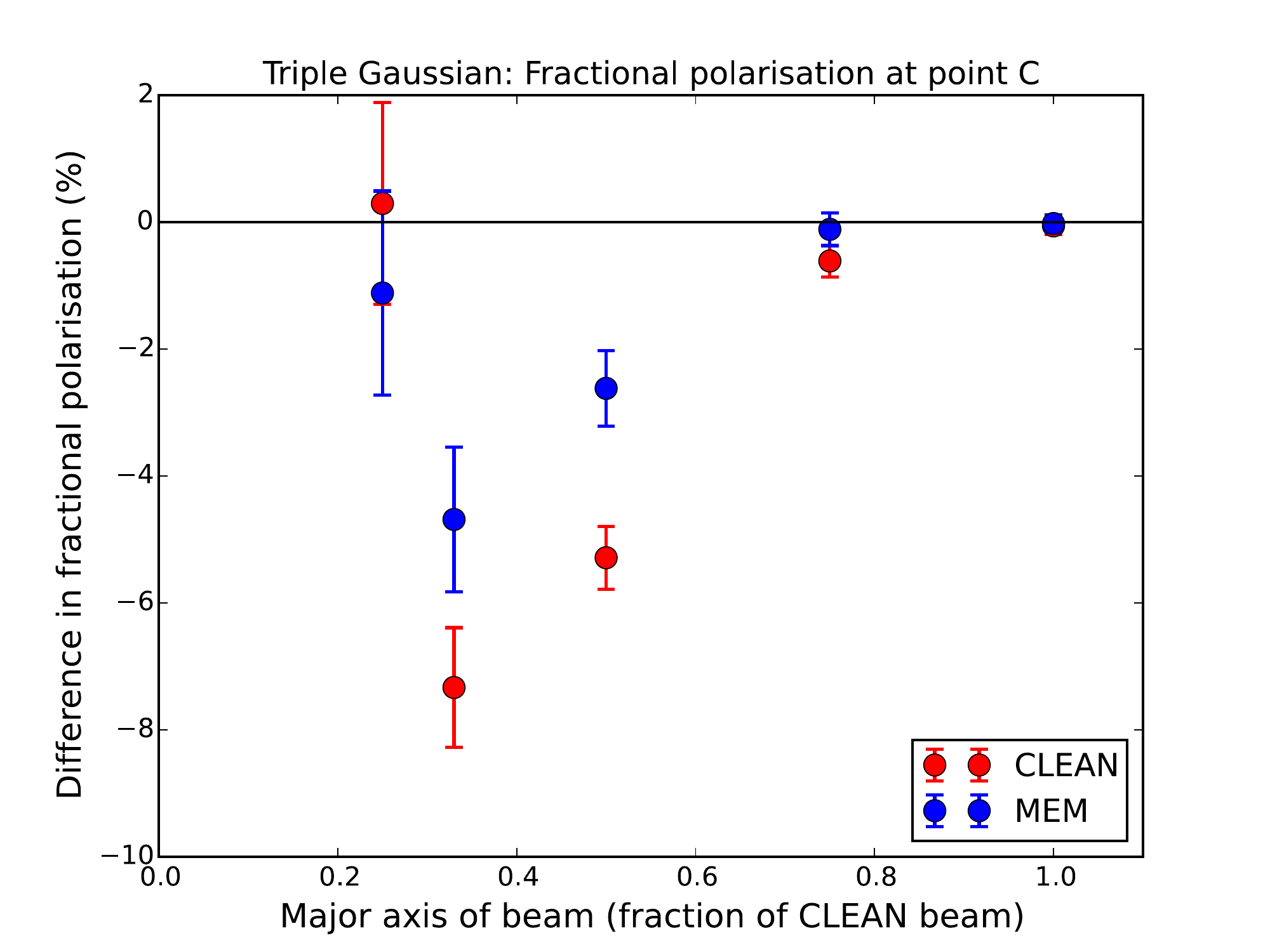}
}

\subfloat[Point B $\chi$]{
	\includegraphics[width=  1.0\columnwidth]{mem-sims-images/TGP2/TGP2_chi_B-eps-converted-to.pdf}
}
\subfloat[Point C $\chi$]{
	\includegraphics[width=  1.0\columnwidth]{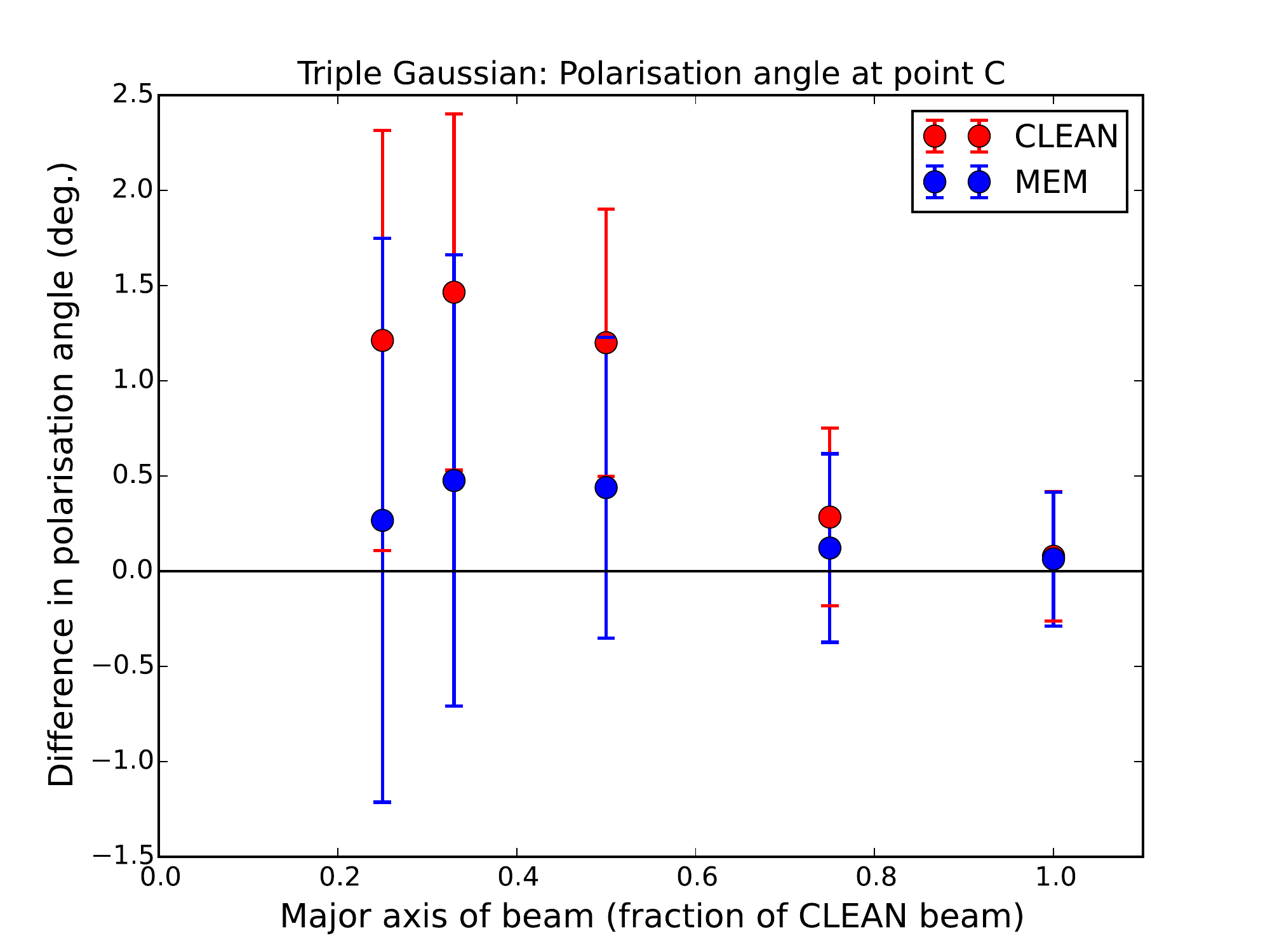}
}
\caption[Error distributions for p, m, chi for B and C.]{Distribution 
of the errors in the (a, b) polarised flux $p$, (c, d) fractional 
polarisation $m$ and (e, f) polarisation angle $\chi$ for the 
positions of components (a, c, e) B and (b, d, f) C, for
the MEM and CLEAN maps of the triple Gaussian model source convolved
with beams comprising various fractions of the full CLEAN beam. The size of the major axis of convolving beam used is indicted on the x-axis as a fraction of the standard CLEAN major axis (note that the beam area is the square of this factor).}
\label{mem-main-fig-pmchi-BC}
\end{center}
\end{figure*}

\begin{figure*}
\begin{center}
\subfloat[Point D $m$]{
	\includegraphics[width=  1.0\columnwidth]{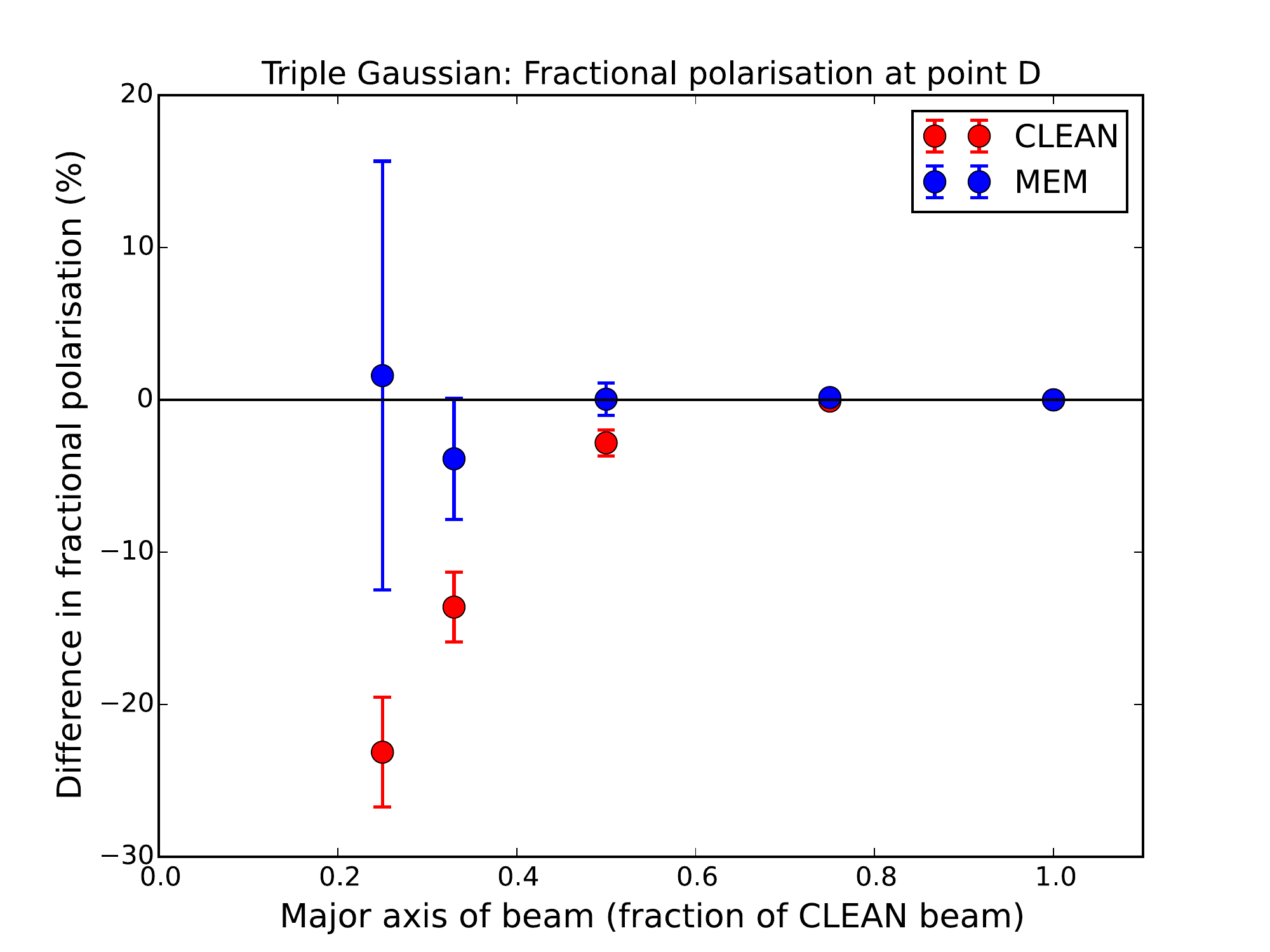}
}
\subfloat[Point D $\chi$]{
	\includegraphics[width=  1.0\columnwidth]{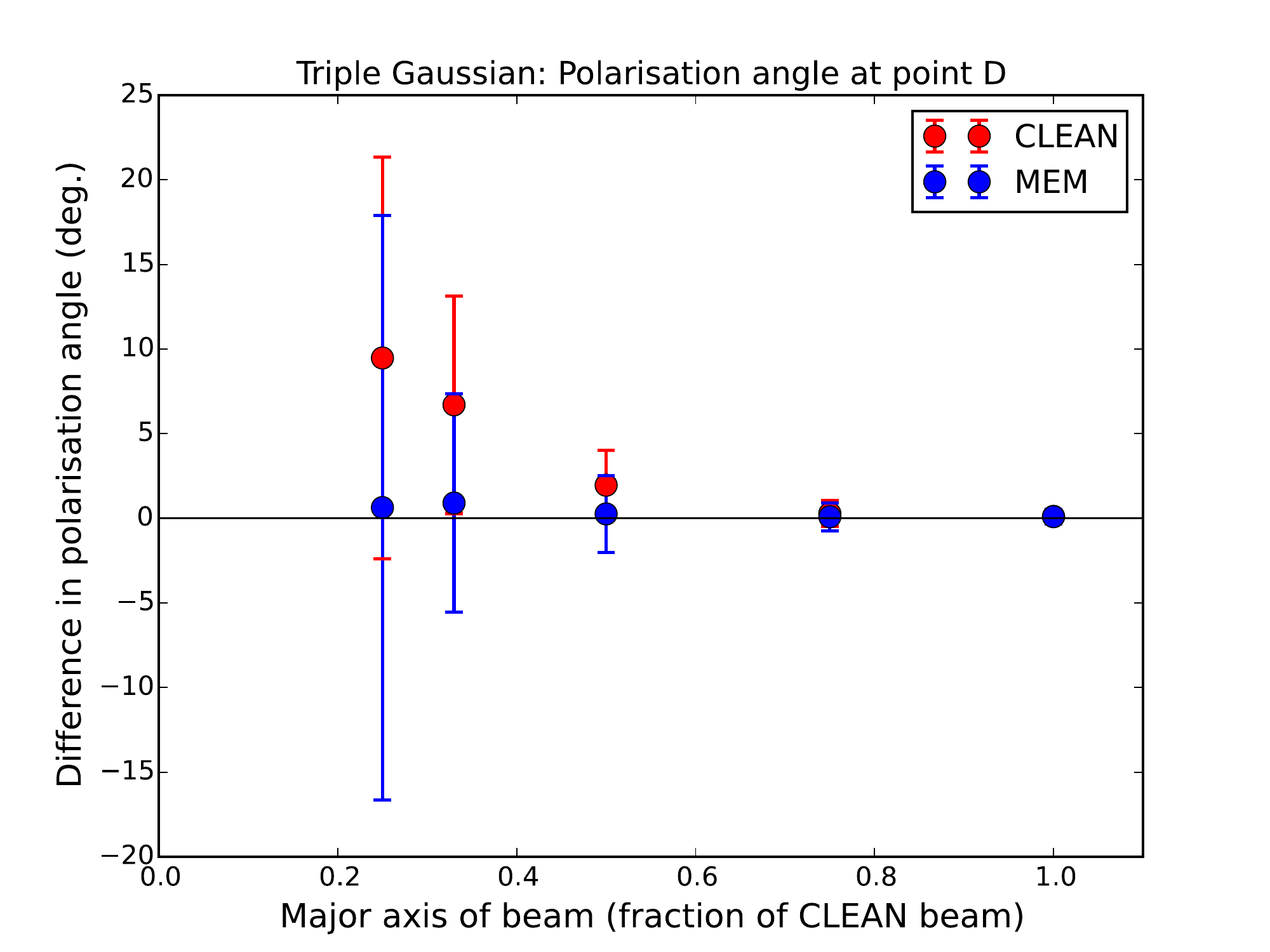}
}

\subfloat[Point E $m$]{
	\includegraphics[width=  1.0\columnwidth]{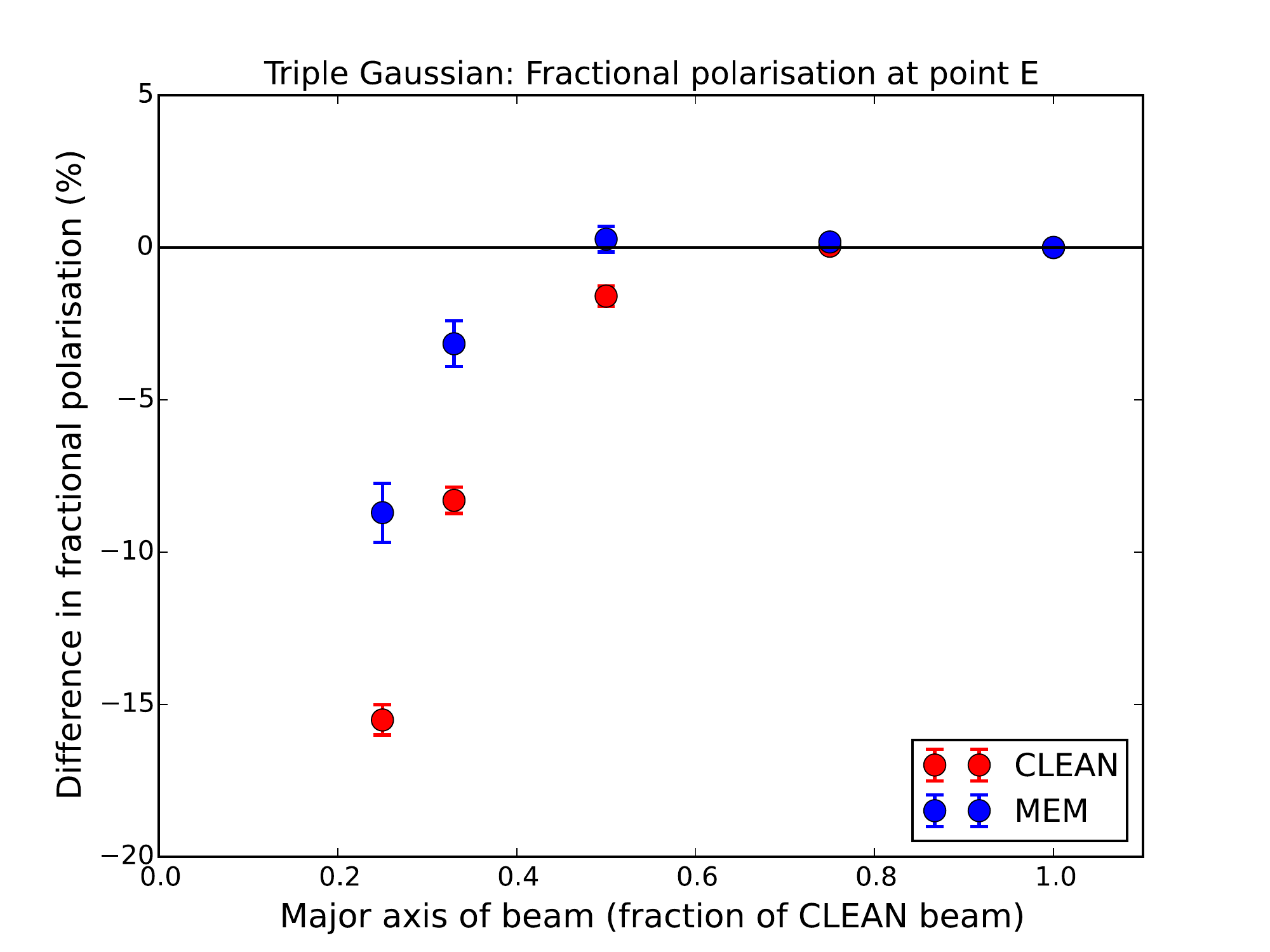}
}
\subfloat[Point E $\chi$]{
	\includegraphics[width=  1.0\columnwidth]{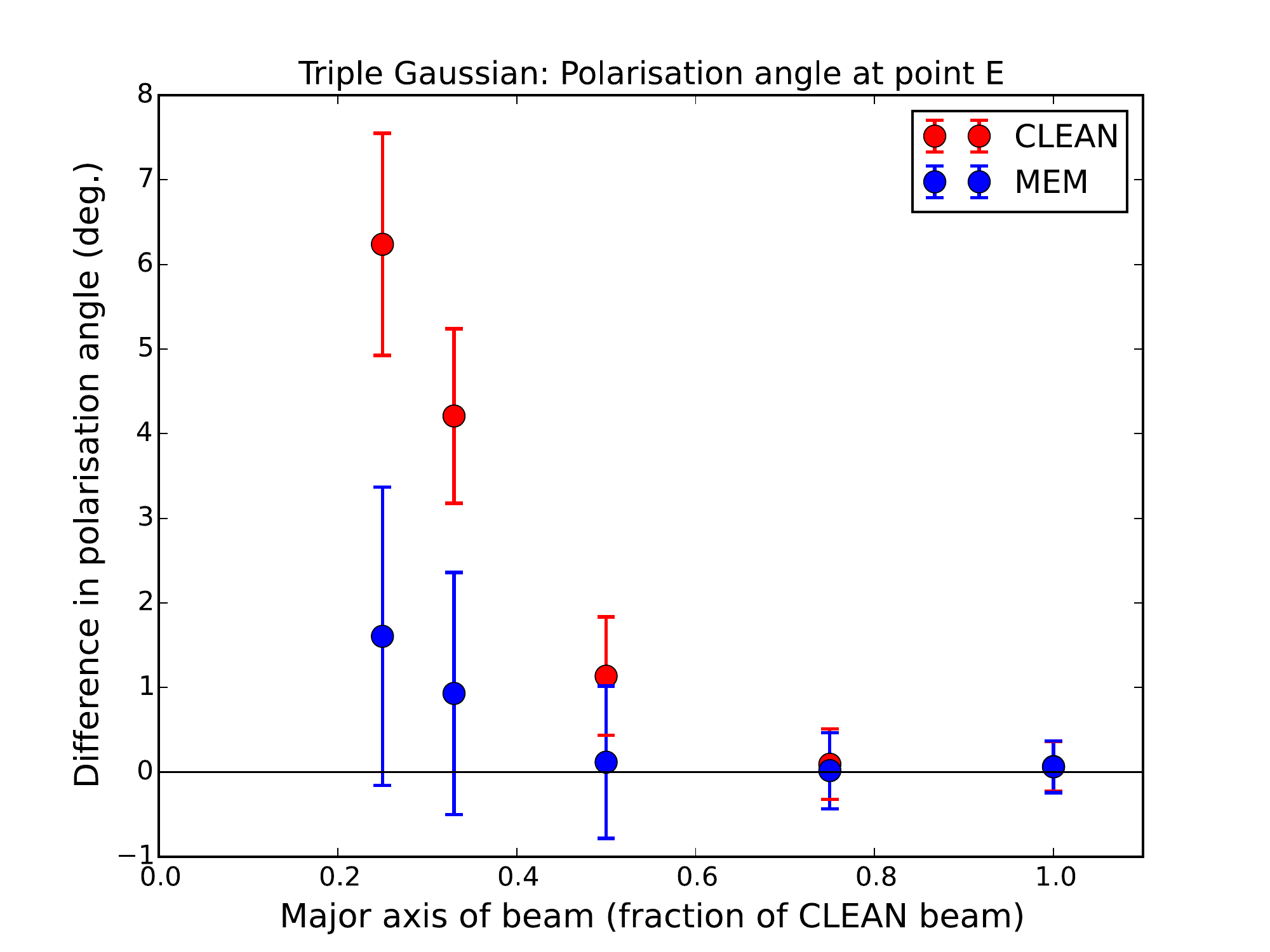}
}

\subfloat[Point F $m$]{
	\includegraphics[width=  1.0\columnwidth]{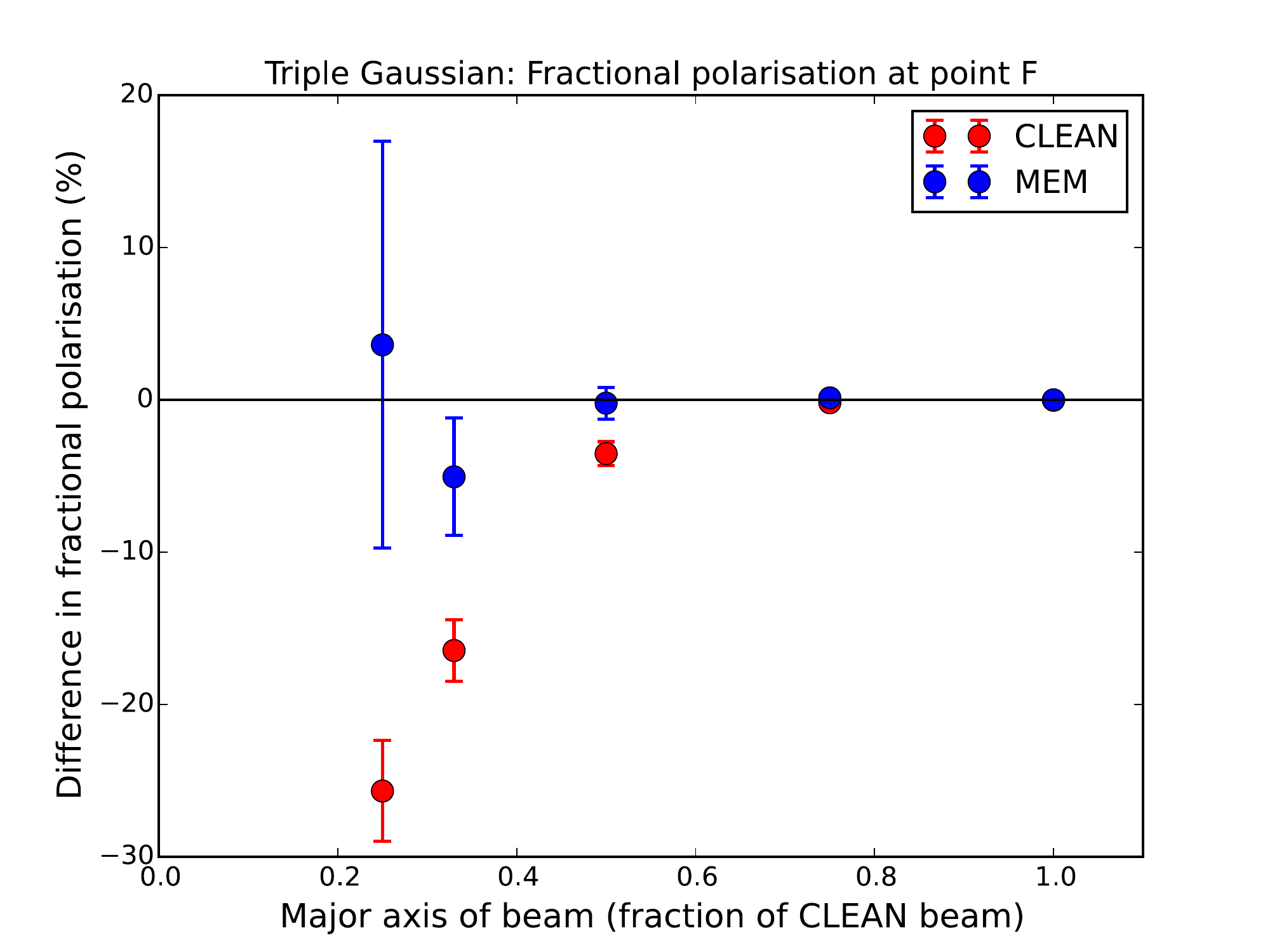}
}
\subfloat[Point F $\chi$]{
	\includegraphics[width=  1.0\columnwidth]{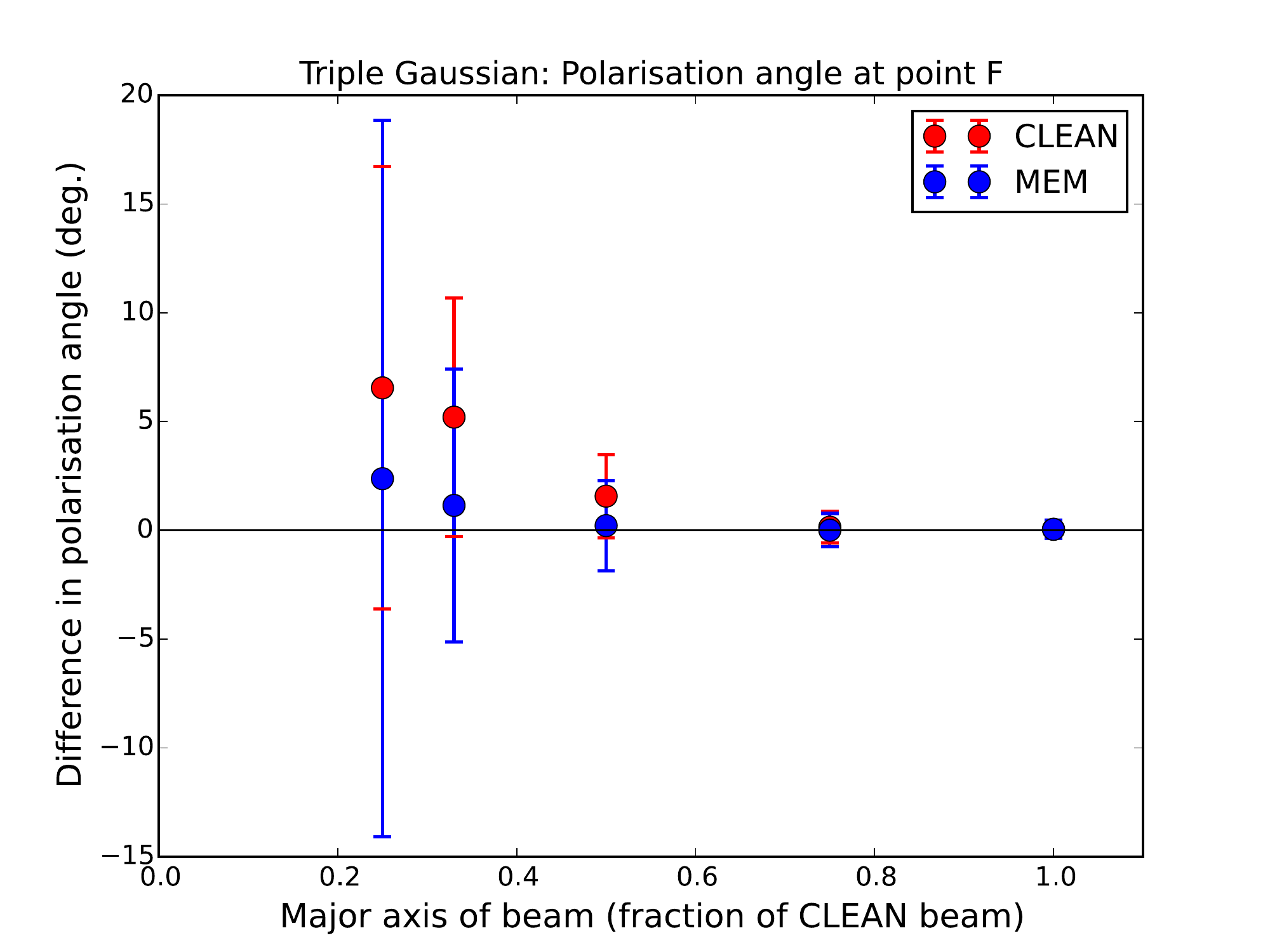}
}
\caption[Error distributions for m, chi for D, E, F.]{Distribution 
of the errors in the (a, c, e) fractional 
polarisation $m$ and (b, d, f) polarisation angle $\chi$ for 
positions (a, b) D, (c, d) E and (e, f) F, for
the MEM and CLEAN maps of the triple Gaussian model source convolved
with beams comprising various fractions of the full CLEAN beam. The size of the major axis of convolving beam used is indicted on the x-axis as a fraction of the standard CLEAN major axis (note that the beam area is the square of this factor).}
\label{mem-main-fig-mchi-DEF}
\end{center}
\end{figure*}

\begin{figure*}
\begin{center}
\subfloat[Point B $p$]{
	\includegraphics[width=  1.0\columnwidth]{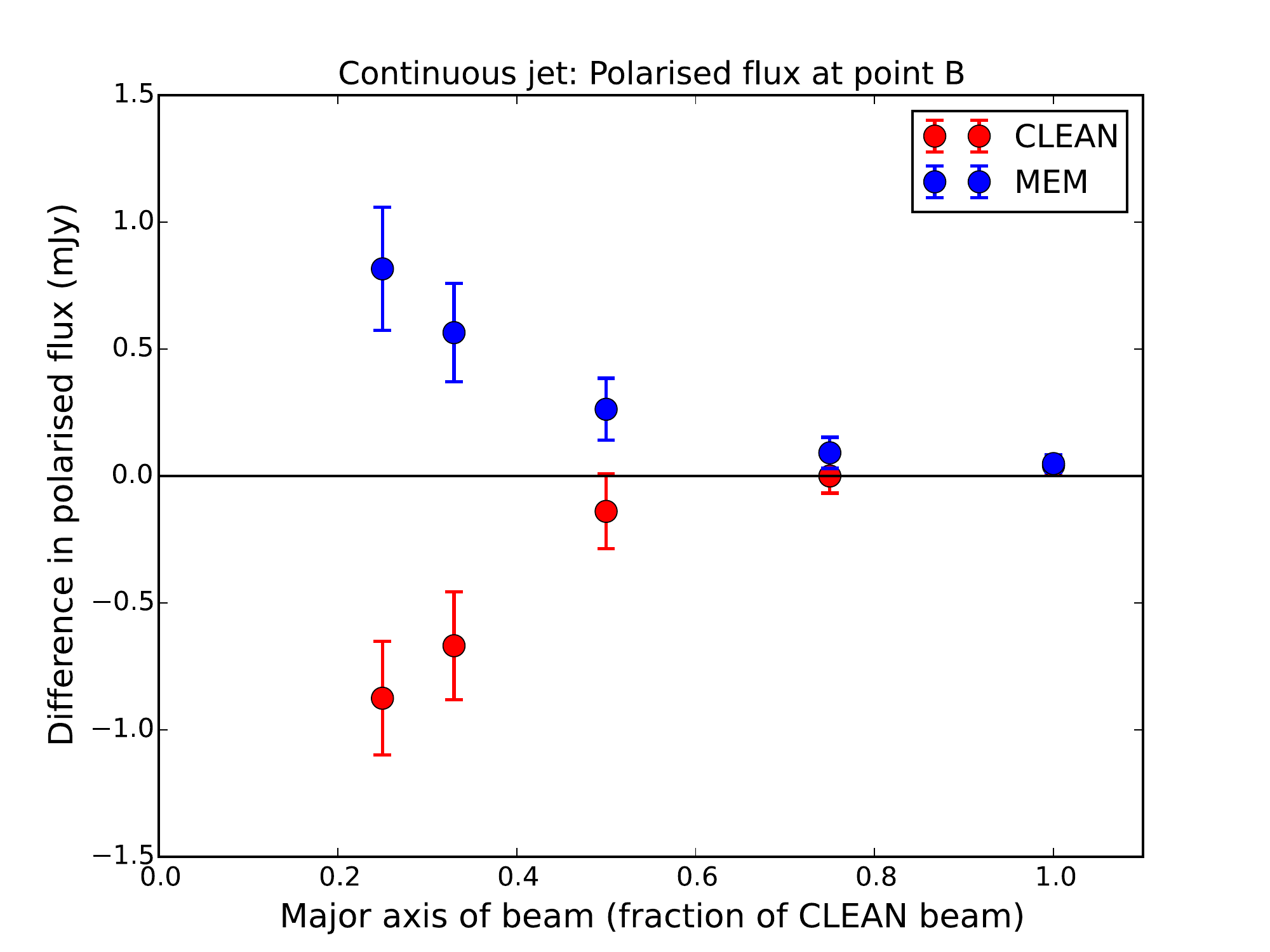}
}
\subfloat[Point C $p$]{
	\includegraphics[width=  1.0\columnwidth]{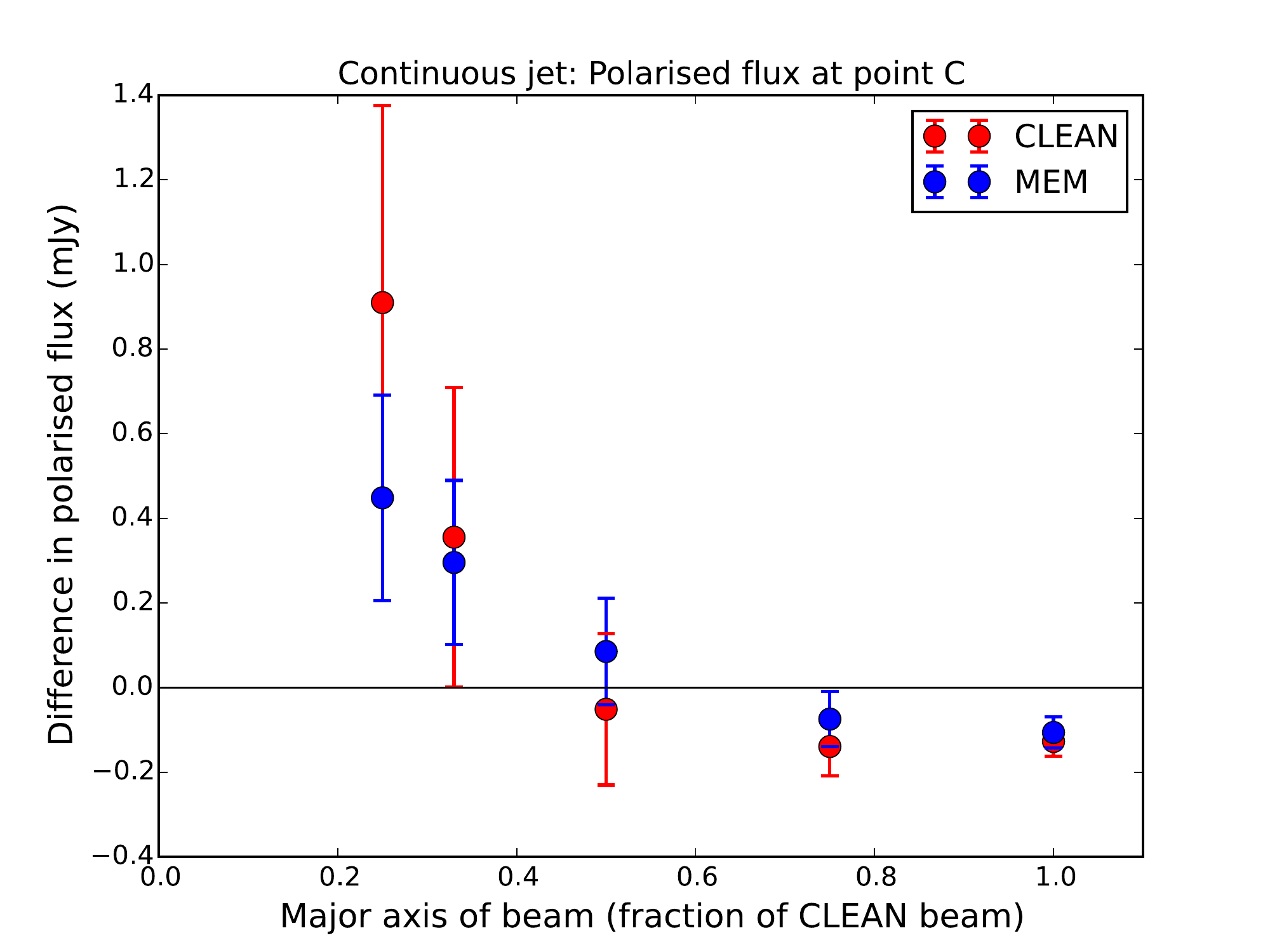}
}

\subfloat[Point B $m$]{
	\includegraphics[width=  1.0\columnwidth]{mem-sims-images/CTSJ/CTSJ_m_B-eps-converted-to.pdf}
}
\subfloat[Point C $m$]{
	\includegraphics[width=  1.0\columnwidth]{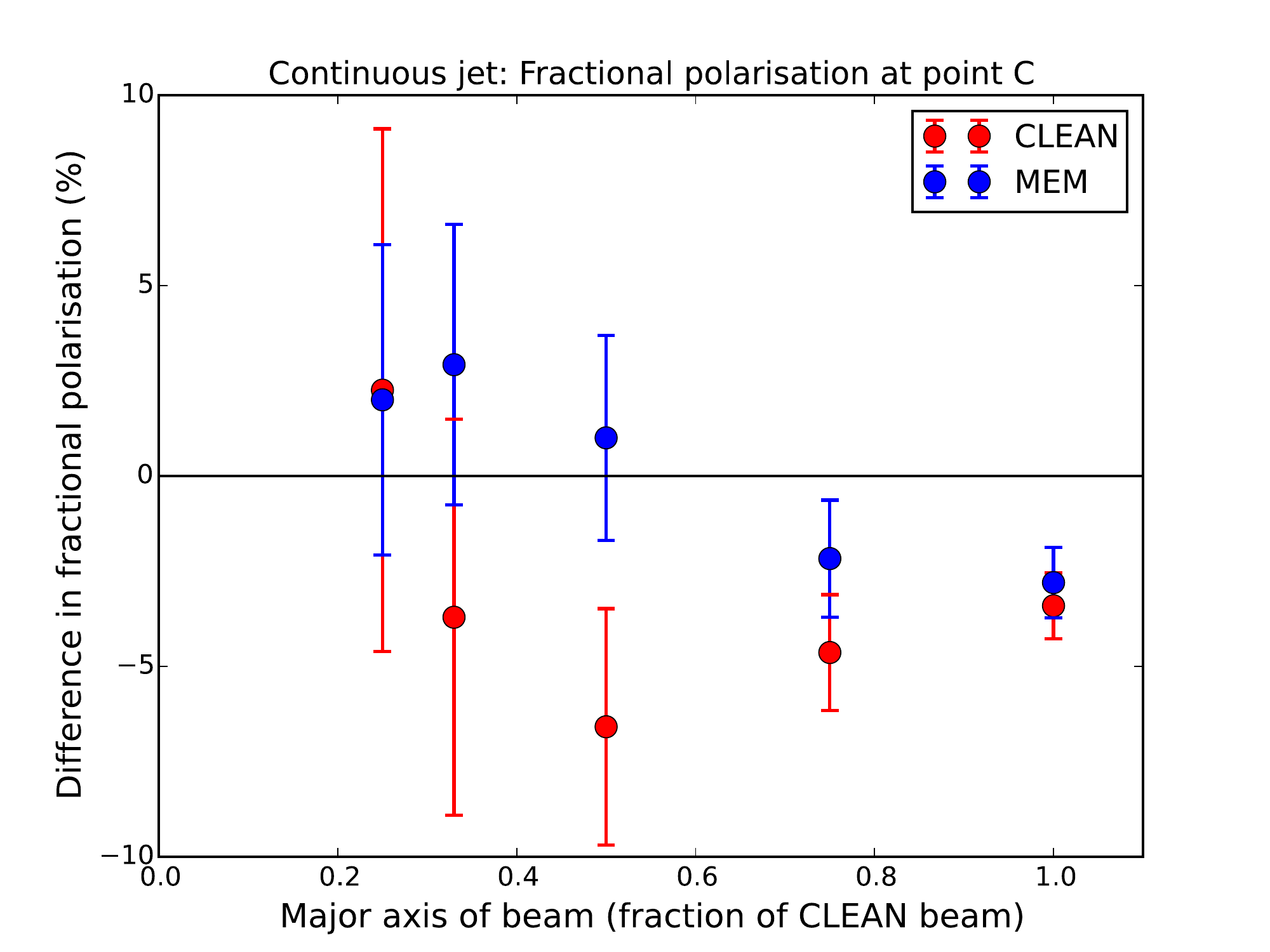}
}

\subfloat[Point B $\chi$]{
	\includegraphics[width=  1.0\columnwidth]{mem-sims-images/CTSJ/CTSJ_chi_B-eps-converted-to.pdf}
}
\subfloat[Point C $\chi$]{
	\includegraphics[width=  1.0\columnwidth]{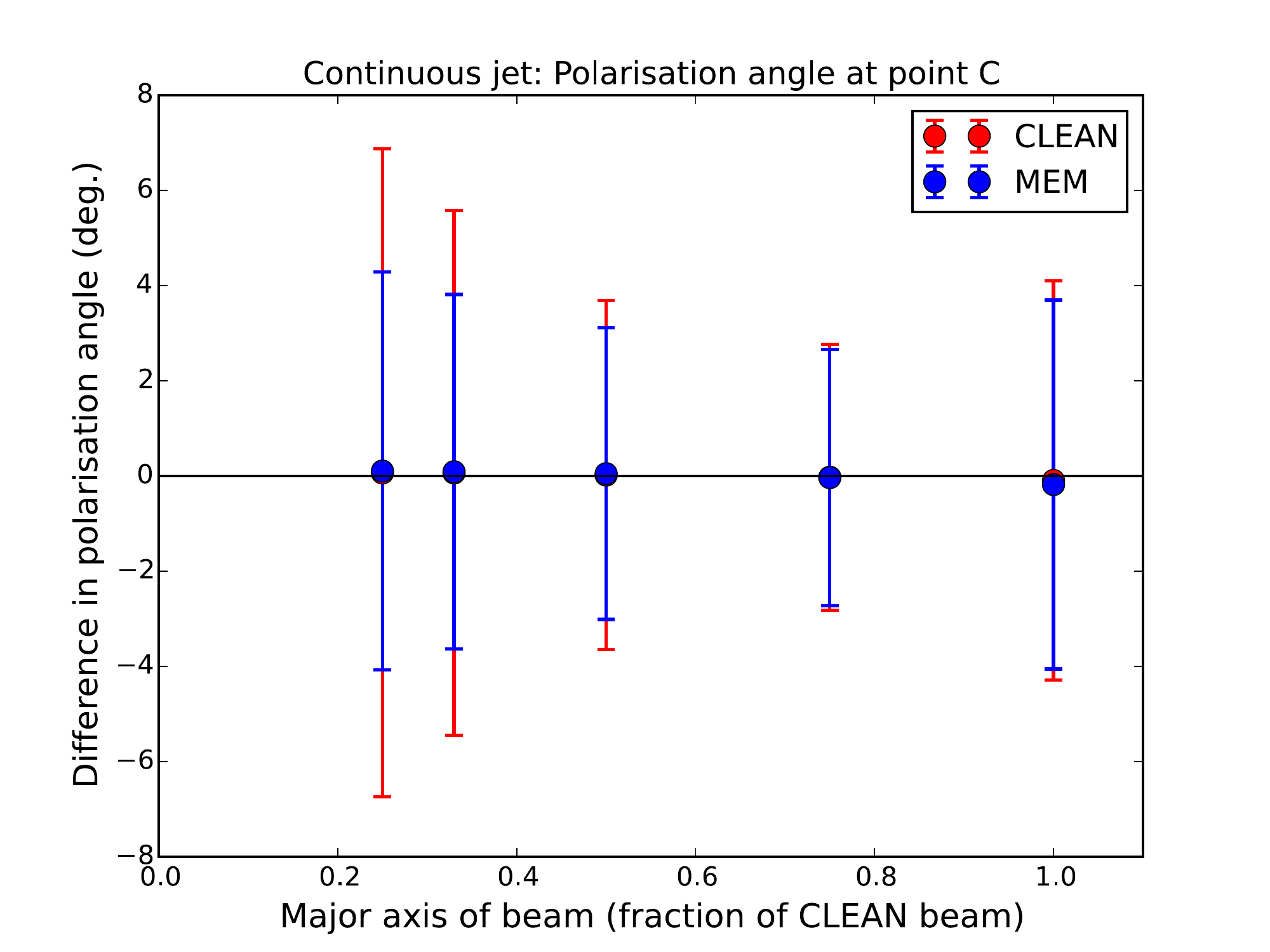}
}
\caption[Continuous, bent jet model. Error distributions for p, m, chi for B and C.]{Distribution 
of the errors in the (a, b) polarised flux $p$, (c, d) fractional 
polarisation $m$ and (e, f) polarisation angle $\chi$ for the 
positions of components (a, c, e) B and (b, d, f) C, for
the MEM and CLEAN maps of the bent continuous jet model convolved
with beams comprising various fractions of the full CLEAN beam. The size of the major axis of convolving beam used is indicted on the x-axis as a fraction of the standard CLEAN major axis (note that the beam area is the square of this factor).}
\label{fig-app-cts-pmchi-BC}
\end{center}
\end{figure*}

The same polarisation quantities are then measured at multiple regions in each source and presented in Figures \ref{mem-main-fig-pmchi-totA} to \ref{mem-main-fig-mchi-DEF} (Triple Gaussian: points A, B and C - with points D, E and F transversely across the jet) and Figures \ref{fig-app-cts-pmchi-totA} to \ref{fig-app-cts-pmchi-BC} (Continuous jet: points A, B and C).The polarisation quantities were derived by taking the average Stokes Q and U flux in the $0.3~\textrm{mas}\times 0.3~\textrm{mas}$ region centred at the point and deriving $P=\sqrt{Q^{2} + P^{2}}$, $m = \frac{P}{I}$ and $\chi = \frac{1}{2} \arctan{U/Q}$. In all cases both algorithms performed very well down at both CLEAN beam resolution and with a mild super-resolution to 0.75 of the CLEAN beam. Performance in both algorithms begins to deteriorate at higher resolutions, with MEM generally performing somewhat better than CLEAN - though CLEAN performs equally well in some cases (the polarisation angles of points A, B and C for the continuous bent jet source) and and does better than MEM in measuring the total polarisation at point B of the triple Gaussian.

The large variation in the errors of both algorithms across the points and sources considered mean that, short of conducting a Monte Carlo simulation with comparable UV coverage and a suitable model, determining the absolute and relative performance of each algorithm in at a location in a real map is difficult. However, these simulations support the general conclusion that mild super-resolution has little effect on the error, and that higher super-resolution can give good results in some cases - particularly using MEM.

\end{document}